\documentclass[paper=a4, fontsize=11pt]{scrartcl}
\usepackage[T1]{fontenc}
\usepackage{amsmath,amssymb,amsfonts,amscd,amsxtra,amsthm}
\usepackage{latexsym}
\usepackage{bm}
\usepackage{fourier}
\usepackage[english]{babel}								 
\usepackage{amssymb}
\usepackage[pdftex]{graphicx}	
\usepackage{url}
\usepackage[title,titletoc,page]{appendix} 
\usepackage{listings} 
\usepackage{color}
\usepackage{cite}
\usepackage{verbatim}
\usepackage{caption}
\usepackage{subcaption}

\numberwithin{equation}{section}		
\numberwithin{figure}{section}			
\numberwithin{table}{section}				

\newtheorem{thm}{Theorem}[section]
\newtheorem{defi}[thm]{Definition}
\newtheorem{lem}[thm]{Lemma}
\newtheorem{cor}[thm]{Corollary}

\newtheorem{prop}[thm]{Proposition}
\newtheorem{remark}{Remark}[section]

\newcommand{\babs}[1]{\Big|{#1}\Big|}
\newcommand{\btwonorm}[1]{\Big|\Big|{#1}\Big|\Big|_2}

\title{A mathematical theory of computational resolution limit in one dimension}
\author{
Ping Liu\thanks{\footnotesize Department of Mathematics, 
  HKUST,  Clear Water Bay, Kowloon, Hong Kong (pliuah@connect.ust.hk).}
 \; and Hai Zhang\thanks{\footnotesize 
 Department of Mathematics, 
  HKUST,  Clear Water Bay, Kowloon, Hong Kong (haizhang@ust.hk). Hai Zhang was partially supported by Hong Kong RGC grant GRF 16306318 and GRF 16305419.}}
\date{}

\begin{document}

\bibliographystyle{plain}
\maketitle

\begin{abstract}
Given an image generated by the convolution of point sources with a band-limited function, the deconvolution problem is to reconstruct the source number, positions, and amplitudes. This problem arises from many important applications in imaging and signal processing. It is well-known that it is impossible to resolve the sources when they are close enough in practice. Rayleigh investigated this problem and formulated a resolution limit, the so-called Rayleigh limit, for the case of two sources with identical amplitudes. 
On the other hand, many numerical experiments
demonstrate that a stable recovery of the sources is possible even if the sources are separated below the Rayleigh limit. 
In this paper, a mathematical theory 
for the deconvolution problem in one dimension is developed. The theory addresses the issue when one can recover the source number exactly from noisy data. The key is a new concept ``computational resolution limit'' which is defined to be the minimum separation distance between the sources such that exact recovery of the source number is possible. This new resolution limit is determined by the signal-to-noise ratio and the sparsity of sources, in addition to the cutoff frequency of the image. Quantitative bounds for this limit is derived, which 
reveal the importance of the sparsity as well as the signal-to-noise ratio to the recovery problem. The stability for recovering the source positions is also analyzed under a condition on their separation distances. Moreover, a singular-value-thresholding algorithm is proposed to recover the source number of a cluster of closely spaced point sources and to verify our theoretical results on the computational resolution limit. The results are based on a multipole expansion method and a non-linear approximation theory in Vandermonde space. 
\end{abstract}



\section{Introduction}
\subsection{Problem Setting}\label{sec-setting}
In numerous imaging and signal processing problems, an image is obtained by convoluting point sources with a band-limited function which is called the point spread function. The problem of recovering the source number, positions and amplitudes is called deconvolution. 
This paper mainly focuses on recovering the source number and positions from their image under a certain noise level in one dimension. 
%
More precisely, we assume the collection of point sources is a discrete measure
\[
\mu^*=\sum_{j=1}^{n}a_j^*\delta_{y_j^*},
\]
where $y_1^*,\cdots,y_n^*$ are the locations of point sources and $a_1^*,\cdots,a_n^*$ the amplitudes. We assume that the amplitudes are real numbers. 
We denote 
\[m_{\min}^*=\min_{j=1,\cdots,n}|a_{j}^*|,
\quad m^*=\|\mu^*\|_{TV}=\sum_{j=1}^{n}|a_{j}^*|.\] 

 
%
Let $f_{\Omega}$ be the band-limited point spread function with cut-off frequency $\Omega$. Throughout the paper, we restrict our investigation to the case $f_{\Omega}(x)=\frac{\sin \Omega x}{\sqrt{\Omega}x}$, although our methods are also applicable to other point spread functions.  
The corresponding Rayleigh limit is $\frac{\pi}{\Omega}$. We assume that the sources are located in $[\frac{-d}{\Omega},\frac{d}{\Omega}]$ with $d$ of order one.

The noiseless image is the convolution of $\mu^*$ and $f_{\Omega}$. We sample the image at evenly spaced points $x_1=\frac{-R}{\Omega}, x_2=\frac{-R+h}{\Omega},x_3=\frac{-R+2h}{\Omega},\cdots, x_{N}=\frac{R}{\Omega}$, where $\frac{R}{\Omega}$ is a large truncation number such that the image outside is negligible, and $\frac{h}{\Omega}$ is the spacing of the sample points. We obey the Shannon sampling criterion by assuming that $\frac{h}{\Omega}\leq \frac{\pi}{\Omega}$. The measurement is
\begin{equation}\label{introductionequ-1}
\mathbf Y(x_t)=\mu^{*}*f_{\Omega}(x_t)+\sqrt{\Omega}\mathbf W(x_t)=\sum_{j=1}^na_{j}^*f_{\Omega}(x_t-y_j^*)+\sqrt{\Omega}\mathbf W(x_t), \quad  t=1,\cdots,N
\end{equation}
where $\sqrt{\Omega}\mathbf W(x)$ is band-limited noise. More precisely, 
$\sqrt{\Omega}\mathbf W(x)= \frac{1}{2\pi}\int_{-\infty}^{+\infty} w(\omega)e^{i \omega x}d\omega$ with $w(\omega) =0$ for $|\omega| >\Omega$. We assume the noise level $||w||_2=\sqrt{\int_{-\infty}^{+\infty}|w(\omega)|^2d\omega}\leq \sqrt{2\pi}\sigma.$ 
We denote 
$$
\mathbf Y= \left(\mathbf Y(x_1), \cdots, \mathbf Y(x_N)\right)^T, \quad [\mu*f_{\Omega}]= \left(\mu*f_{\Omega}(x_1), ,\cdots, \mu*f_{\Omega}(x_N)\right)^T, \quad
\mathbf W=\left(\mathbf W(x_1),\cdots,\mathbf W(x_N)\right)^T.
$$
Then the following estimate holds
\begin{align}\label{noiselevel1}
\sqrt{h}||\mathbf W||_2=\sqrt{\frac{h}{\Omega}}\sqrt{\Omega}||\mathbf W||_2\leq\frac{1}{\sqrt{2\pi}}||w||_2\leq \sigma.
\end{align}

%

The deconvolution problem is to recover the source number $n$ and their locations $y_j^*$'s and amplitudes $a_j^*$'s from the measurements in (\ref{introductionequ-1}).
It is well-known that it is impossible to resolve the sources when they are close enough in practice. Rayleigh investigated this issue and formulated a resolution limit, the so-called Rayleigh limit, for the case of two sources. It is defined to be $\frac{\pi}{\Omega}$
for one dimensional images, where $\Omega$ is the cutoff frequency of the point spread function. The Rayleigh limit is empirical and only applies to instrumental imaging methods.
On the other hand, many numerical experiments demonstrate that a stable recovery of the sources is possible even if the sources are separated below the Rayleigh limit. 
Indeed, consider the noiseless deconvolution problem. Using the linear independence of the functions $f_{\Omega}(x-y_j^*)$ for different $y_j^*$'s, 
the locations and amplitudes can be uniquely determined and infinite resolution can be achieved at least theorectically. Therefore, from a data-processing point of view, a proper definition of resolution limit should take into account of the noise level in the measurement. In this paper, a new concept of computational resolution limit will be proposed to highlight the importance of the noise
level and the number of the sources on the resolvability of closely spaced point sources. Moreover, this resolution limit will be characterized quantitatively.

We note that the deconvolution problem has an equivalent formulation in the frequency domain. The available data is the following Fourier data
\begin{equation} \label{eq-fouriermeasure}
\hat{\mathbf Y} (\omega) = \sum_{j=1}^n a_j^* e^{iy_j^* \omega}, \quad |\omega|\leq \Omega.
\end{equation}
One may take equal-spaced sampling at $\omega_t = -\Omega + \tau(t-1)$ where $\tau = \frac{2 \Omega}{M}$ is the sampling spacing and $M$ is some integer to get the measurement. This then becomes the line spectral estimation problem.

\subsection{Literature Review}  
The precise characterization of limit of resolution has been a long-standing problem in signal processing and spectral estimation. The idea that the actual resolution limit is determined by the signal-to-noise ratio (SNR) was known at a very early age. To our knowledge, as early as 1980, it was pointed out in \cite{papoulis1979improvement} that, Rayleigh limit is adequate if one relies on the direct observation of the data for the determination of sources, however, it is not useful if the data is subjected to elaborate processing. 
This promoted many investigations from the perspective of statistical estimation and hypothesis testing, see for instance \cite{helstrom1964detection, helstrom1969detection, lucy1992statistical, lucy1992resolution, den1996model}. Most of the studies focus on the two-point resolution limit which is defined to be the minimum detectable distance between two point sources at a given SNR. 
Especially, in \cite{shahram2004imaging, shahram2004statistical, shahram2005resolvability}, by unifying and generalizing much of the literature on the topic which spanned the course of roughly four decades, the authors derived explicit formula for the minimum SNR that is required to discriminate two point sources separated by a distance smaller than the Rayleigh limit.

To understand the puzzle of resolution limit, Donoho developed a theory from the optimal recovery point of view to explain the possibility and difficulties of superresolution via sparsity constraint \cite{donoho1992superresolution}. He considered a grid setting where a discrete measure is supported on a lattice 
and the available measurement is its low frequency information. He derived bounds for the minimax error of the recovery of a special class of sparse measures. These bounds are given by noise amplification, and increase polynomially with the super-resolution factor which is defined to be the ratio between Rayleigh limit and the grid spacing. His results emphasize the importance of sparsity in superresolution. 
Further discussed in \cite{demanet2015recoverability}, the authors obtained sharper bounds using estimate of the minimum singular value for the measurement matrix.
See also similar results for multi-clumps case in \cite{batenkov2020conditioning, li2018stable}. In \cite{Moitra:2015:SEF:2746539.2746561}, the author demonstrated a sharp phase transition for the relationship between the cutoff frequency and the separation distance for off-the-grid sources.  We also note that in a series of papers \cite{akinshin2015accuracy, batenkov2018stability, batenkov2019super}, the authors considered the resolution limit for closely spaced point sources in the off-the-grid case. By carefully analyzing the associated ``Prony-type system'' with dedicated quantitative singularity theory, sharp minimax error rate for the reconstruction of the source supports was obtained. 
Moreover, they showed that the Matrix Pencil Method can achieve the accuracy bound. 

On the other hand, this work is also motivated by the recent developments of superresolution algorithms. It is demonstrated   \cite{candes2014towards} that when there is no noise, well-separated sources can be exactly recovered by the sparsity promoting convex optimization. Moreover, if the separation distance is beyond several Rayleigh limits, the recovery is stable. 
In the presence of noise, stability results are further established in \cite{candes2013super} for well-separated sources. Many interesting results are obtained in this research line, see, for instance,  
\cite{fernandez2013support, duval2015exact, cai2019fast, tang2014near, bernstein2019deconvolution, chi2020harnessing}. 
We remark that in order for most of the sparsity promoting convex optimization to work, it is necessary to assume that the sources are well-separated \cite{duval2015exact, Moitra:2015:SEF:2746539.2746561}. 
In \cite{duval2015exact, morgenshtern2016super, morgenshtern2020super}, the restriction on the separation distance is relaxed. However, the point sources considered therein are assumed to be additionally positive.
In addition to these sparsity promoting optimization algorithms,
a class of algorithms called subspace methods also have demonstrated great promise for achieving superresolution, see for instance MUltiple SIgnal Classification (MUSIC) \cite{schmidt1986multiple}, Estimation of Signal Parameters via Rotational Invariance Technique (ESPRIT) \cite{roy1989esprit}, and Matrix Pencil Method \cite{hua1990matrix}. 
These algorithms have root in the Prony's method \cite{Prony-1795}.  In \cite{stoica1989music}, a statistical analysis of MUSIC was provided along with the analysis on performance limits based on Cramer-Rao bound. 
Moreover, recently in the grid setting, a mathematical theory was developed in \cite{li2018stable} to explain the numerical superresolution observed in \cite{liao2016music}. Their theory is based on a sharp bound of the minimax error of recovery. 

%

\subsection{Main Results}
In this paper, we investigate the resolution limit problem for the deconvolution problem (\ref{introductionequ-1}) for a cluster of closely spaced point sources separated below the Rayleigh limit as is described in Section \ref{sec-setting}. 
We aim to answer the following question: what is the minimum separation distance between the sources so that one can recover their number from their noisy image. More precisely, for an image generated by $n$ point sources, what is the minimum separation distance between them so that it cannot be approximated within the noise level by one with less than $n$ point sources. From the approximation theory point of view, let $f$ be a general point spread function, we may view the functions $f(x-y)$ 
as a continuous family of dictionary indexed by the location parameter $y$'s. The coherence of elements in the dictionary is determined by the separation distance of their associated indices. Then the problem we are interested in amount to the following one: for a vector that is a linear combination of $n$ elements in the dictionary, what is the condition on the coherence of the elements so that it cannot be approximated by a linear combination of less than $n$ elements. To our knowledge, all the existing results in this direction only consider the case of two point sources, see for instance \cite{shahram2004imaging, shahram2004statistical, shahram2005resolvability}. The results developed in this paper seem to be the first attempt for the general case with multiple point sources. To achieve the goal, we developed a multiple expansion method and reduced the original problem to a non-linear approximation problem in the so-called Vandermonde space. We obtained a sharp bound to the approximation problem and this yields a quantitative characterization to the resolution limit. In what follows, we briefly present our main results. 
We first introduce the concept of admissible measures. 

\begin{defi} \label{definition-ad}
For given a priori noise level $\sigma$, interval size $\frac{d}{\Omega}$, and total-variation norm bound $M \geq m^*$, we say that $\mu=\sum_{j=1}^{k} a_j \delta_{y_j}$ is a $(d,\sigma,M)$-admissible discrete measure for the image $\mathbf Y$ only if $\mu$ is supported in $[\frac{-d}{\Omega}, \frac{d}{\Omega}]$ such that 
$\|\mu\|_{TV} =\sum_{j=1}^{k}|a_{j}|\leq M$ and
\[
\sqrt{\frac{h}{\Omega}}\btwonorm{[\mu*f_{\Omega}]-\mathbf Y}\leq \sigma.
\]
\end{defi}
The set of admissible measures of $\mathbf Y$ characterizes all possible solutions to the deconvolution problem with the given image $\mathbf Y$. 
If there exists one admissible measure with less than $n$ sources, then the image $\mathbf Y$ can be generated by less than $n$ sources, and it is impossible to detect the right number of sources without additional priori information. On the other hand, if all admissible measures have at least $n$ sources, then one can determine the source number $n$ correctly if one restricts to the sparsest admissible measures. This leads to the following definition of computational resolution limit for number detection.  
\begin{defi}\label{def:numbercomputresolutionlimit}
For an image $\mathbf Y$ generated by $n$ point sources $\mu^*=\sum_{j=1}^{n}a_j^* \delta_{y_j^*}$, the computational resolution limit is defined as the minimum nonnegative number $\mathcal D_{num}$ so that if
\[
\min_{p\neq j, 1\leq p,j\leq n}|y_p^*-y_j^*|\geq \mathcal D_{num}
\]
then there does not exist any $(d,\sigma,M)$-admissible measure for $\mathbf Y$ with less than $n$ supports.
\end{defi}


According to the definition, detection of the source number is impossible without additional priori information if the sources are separated below this limit. 
The main contribution of this work is the following bounds for the computational resolution limit (see Proposition \ref{thm:numberlowerboundthm0} and Theorem \ref{necessarythm1}): 
\[
\frac{0.9e^{-\frac{3}{2}}}{\Omega}\Big(\frac{\sigma}{m_{\min}^*}\Big)^{\frac{1}{2n-2}} \leq \mathcal D_{num} \leq \frac{4.7(1+d)}{\Omega}\Big(\frac{3}{\sigma_{\min}(s^*)}\frac{\sigma}{m_{\min}^*}\Big)^{\frac{1}{2n-2}},
\]
where $\Omega$ is the cutoff frequency, $\frac{d}{\Omega}$ is a priori estimate of the size of the interval where the sources are located in, $n$ is the source number, $\sigma_{\min}(s^*)$ is the minimum singular value of certain multipole matrix and $\frac{\sigma}{m_{\min}^*}$ is the noise-to-signal ratio. 
The factor $\sigma_{\min}(s^*)$ characterizes the correlation of multipoles and is determined by the multipole matrix which is further determined by the point spread function. A lower bound for $\sigma_{\min}(s^*)$ is given in section \ref{sec-sigmas}. It demonstrates that as the noise-to-signal ratio tends to zero, the upper bound for the computational resolution limit $\mathcal D_{num}$ also tends to zero. 
As a consequence, provided that SNR is sufficiently large, one can recover the source number exactly even if the sources are separated below the Rayleigh limit to achieve the so-called ``super-resolution''.  

Following the same methods we developed for the source number detection problem, we also considered the stability of recovering the source positions. We showed that when the separation distance exceeds
\[
\frac{6.24(1+d)}{\Omega}\Big({\frac{3}{\sigma_{\min}(s^*)}\frac{\sigma}{m_{\min}^*}}\Big)^{\frac{1}{2n-1}},
\]
we can stably recover the source positions from admissible measures (see Theorem \ref{reconaccuracythm2}). We remark that such result was also reported in the related work \cite{batenkov2019super} under a more general setting where some of the point sources (or nodes as is called therein) form a cluster while the rests are well separated. Their results use measurement in the Fourier space and  
are based on the analysis of ``Prony mapping'' and the ``quantitative inverse function theorem''. The techniques are very different from ours. 

Finally, following the multipole expansion method, we developed a singular-value-thresholding algorithm to recover the source number. Our numerical results show that exact source number can be recovered in the super-resolution regime when the source separation distance is comparable to the upper bound we derived for the computational resolution limit $\mathcal D_{num}$ and hence confirms the theory.

\subsection{Organization of this paper}
The rest of the paper is organized as follows. In Section 2, we introduce the multipole expansion method and derive the quantitative bounds for computational resolution limit. Moreover, we investigate the stability of support recovery from admissible measures. In Section 3, we introduce the nonlinear approximation theory in the Vandermonde space that are used in the proofs of results in Section 2. In Section 4, we propose a singular-value-thresholding algorithm for number detection. In Section 5, we perform numerical experiments that verify our theory on the computational resolution limit. 
Finally, some technical lemmas and their proofs are given in the appendix.

\section{Main results}
\subsection{Multipole expansion}\label{multi-pole expansion}
A nature way to solve the deconvolution problem (\ref{introductionequ-1}) is to consider the following linear problem
\begin{equation}\label{linearinverseproblem1}
\left (
\begin{array}{cc}
\mathbf Y(x_1) \\
\vdots\\
\mathbf Y(x_N)
\end{array}
\right )
= 
\begin{pmatrix}
f_{\Omega}(x_1-y_1) &\cdots& f_{\Omega}(x_1-y_K)\\
\vdots&\vdots&\vdots\\
f_{\Omega}(x_N-y_1)&\cdots& f_{\Omega}(x_N-y_K)
\end{pmatrix}
\begin{pmatrix}
a_{1}\\
\vdots\\
a_{K}
\end{pmatrix}
+
\begin{pmatrix}
\mathbf W(x_1)\\
\vdots\\
\mathbf W(x_N)
\end{pmatrix}
\end{equation}
where $y_1,\cdots,y_K$ are given grid points. In this setting, the sources are assumed to be lying on the grid points and one only need to reconstruct their amplitudes. This approach introduces model errors when the sources are off the grid \cite{chi2011sensitivity}. To avoid this issue, we propose a multipole expansion method. The key observation is that
\begin{align*}
&\sum_{j=1}^{n}a_{j}^*f_{\Omega}(x-y_j^*)=\sum_{j=1}^{n}a_{j}^*f_{\Omega}(x)+\sum_{j=1}^{n}a_{j}^*f_{\Omega}^{(1)}(x)(-y_{j}^*)+\sum_{j=1}^{n}a_{j}^*\frac{f_{\Omega}^{(2)}(x)(-y_{j}^*)^2}{2}+\cdots\\
=&\sqrt{\Omega}\Big(\sum_{j=1}^{n}a_{j}^*f(\Omega x)+\sum_{j=1}^{n}a_{j}^* f^{(1)}(\Omega x)d_{j}^*+\sum_{j=1}^{n}a_{j}^* \frac{f^{(2)}(\Omega x)(d_{j}^*)^2}{2}+\cdots \Big)\\
=&\sqrt{\Omega}\sum_{r=0}^{+\infty}c_r^* \sqrt{2r+1}f^{(r)}(\Omega x)
\end{align*}
where $d_{j}^*:=-\Omega y_{j}^*$, $f^{(r)}(x):=(\frac{\sin x}{x})^{(r)}, r=0,1,\cdots$ and 
\[
c_r^*:=\sum_{j=1}^{n}a_j^*\frac{(d_j^{*})^r}{r!\sqrt{2r+1}}.
\] 
Hence, the measurement $\mathbf Y$ has the following multipole expansion:
\begin{equation}\label{equ:measurementexpansion2} 
\mathbf Y = \sqrt{\Omega} \sum_{r=0}^{+\infty}c_r^*\mathbf h_{r}+\sqrt{\Omega} \mathbf W,
\end{equation}
where $c_r^*$'s are the multipole coefficients and $\mathbf h_r$'s are the multipoles defined as 
\begin{equation}\label{equ:multipole1}
\mathbf h_r=\sqrt{2r+1}(f^{(r)}(\Omega x_1),\cdots,f^{(r)}(\Omega x_N))^T,\ r=0,1,\cdots.
\end{equation}
Here $\sqrt{2r+1}$ is a normalization factor. Especially, We call 
$\mathbf h_0$
the monopole, $\mathbf h_1$
the dipole and $\mathbf h_2$
the quadrapole. We have the following $2$-norm estimate
\begin{equation}\label{2normofmultipole}
\sqrt{h}||\mathbf h_r||_2=\sqrt{2r+1}\Big(h\sum_{t=1}^{N}f^{(r)}(\Omega x_t)^2\Big)^{\frac{1}{2}}\leq \sqrt{\pi},
\end{equation}
which can be proved using the inequality 
\[
h\sum_{t=1}^{N}f^{(r)}(\Omega x_t)^2\leq h\sum_{j=-\infty}^{+\infty}f^{(r)}(jh)^2= \int_{-\infty}^{+\infty}|f^{(r)}|^2dx=\frac{\pi}{2r+1}.
\] 

The analysis of the resolution limit is based on the idea that for a certain level of SNR, we can only stably recover a finite number of low-order multipole coefficients from measurement (\ref{equ:measurementexpansion2}). We shall show that these partial multipole coefficients set a limit to the resolution. For the purpose, we introduce multipole matrix
\begin{equation} \label{eq-matrix}
\mathbf H(s)=\Big(\mathbf h_0,\mathbf h_1,\cdots,\mathbf h_{s-1}\Big).
\end{equation}
We denote the minimum singular value of $\sqrt{h}\mathbf H(s)$ as 
\begin{equation}\label{equ:sigmamin}
\sigma_{\min}(s)=\sigma_{\min}(\sqrt{h}\mathbf H(s)).
\end{equation}
We remark that $\sigma_{\min}(s)$ is determined by $s$ and the point spread function. Note that $\sigma_{\min}(\sqrt{h}\mathbf H(s))\leq ||\sqrt{h}\mathbf H(s)v||_2\leq \sqrt{\pi}$ by (\ref{2normofmultipole}) for $v=(1,0,\cdots,0)^T$.
We have the following estimate
\begin{equation}\label{minisingularvaluebound1}
\sigma_{\min}(s)\leq \sqrt{\pi}, \quad \forall \,\, s \geq 1.  
\end{equation}

\begin{remark}
\begin{enumerate}

\item
The multipole expansion method is a nature idea to solve the deconvolution problem for closely spaced point sources when the measurement is given in the form of (\ref{introductionequ-1}). The multipole functions form a natural quantized basis that can be used to efficiently represent functions which are linear combinations of the continuous dictionary functions $f_{\Omega}(x-y)$ (indexed by the parameter $y$) for all $y$'s that are close to the center of the multipole expansion.  
In an ongoing work, it is used to develop efficient algorithms to resolve point sources which have multiple scales.

\item
For ease of exposition, we only considered the case when the point spread function is the sinc function in this paper. The extension to other point spread functions that are smooth and are essentially supported on a bounded domain 
is straightforward. 
\end{enumerate}	
\end{remark}

\subsection{Main results on the computational resolution limit}\label{sufficientsection1}
In this section, we present two quantitative bounds for the computational resolution limit $\mathcal{D}_{num}$. 
we first show that $\mathcal{D}_{num}$
is at least of the order $O(\Big(\frac{\sigma}{m_{\min}^*}\Big)^{\frac{1}{2n-2}})$.

\vspace{0.2cm}
\begin{prop}\label{thm:numberlowerboundthm0}
For given integer $n\geq 2$ and $0<\sigma<m_{\min}^*$, let
\begin{equation}\label{sufficientthm2equ-1}
\tau=\frac{0.9e^{-\frac{3}{2}}}{\Omega}\Big(\frac{\sigma}{m_{\min}^*}\Big)^{\frac{1}{2n-2}}.
\end{equation}
Then there exist two measures $\mu^*=\sum_{j=1}^{n}a_j^*\delta_{y_j^*}$ with $n$ supports and $\mu$ with $n-1$ supports, both supported in $[\frac{-(n-1)\tau}{\Omega},\frac{(n-1)\tau}{\Omega}]$, such that $\sqrt{\frac{h}{\Omega}}\btwonorm{[\mu*f] -[\mu^**f]}\leq \sigma$. Moreover, 
\[
\min_{1\leq j\leq n}|a_j^*|=m_{\min}^*, \quad \min_{p\neq j}|y_p^*-y_j^*|= \tau.
\]
\end{prop}	
Proof: \textbf{Step 1}. Let $t_1=-(n-1)\tau,t_2=-(n-2)\tau, \cdots, t_{2n-2}=(n-2)\tau, t_{2n-1}=(n-1)\tau$. We consider the linear system  
\begin{equation}\label{sufficientthm2equ1}
Aa=0
\end{equation}
where $a = (a_1,\cdots, a_{2n-1})^T$ and $A=\big(\phi_{2n-3}(t_1), \cdots, \phi_{2n-3}(t_{2n-1})\big)$ with $\phi_{2n-3}(t)= (1, t, \cdots, t^{2n-3})^T$. 
Since $A$ is underdetermined, there exists a nonzero $a=(a_1,\cdots,a_{2n-1})^T$ satisfying (\ref{sufficientthm2equ1}). Using the linear independence of the column vectors in the matrix $A$, we can show that all $a_j$'s are not zero. By a scaling of $a$, we can assume that $\min_{1\leq j\leq 2n-1}|a_{j}|=m_{\min}^*$. We define
\[
\mu^*=\sum_{j=1}^{n}a_j \delta_{-t_j},\  \mu=\sum_{j=n+1}^{2n-1}-a_j\delta_{-t_j}, \quad 
\mbox{if \,\,$\min_{1\leq j\leq n}|a_{j}|=m_{\min}^*$};
\] 
and 
\[
\mu^*=\sum_{j=n}^{2n-1}a_j \delta_{-t_j},\  \mu=\sum_{j=1}^{n-1}-a_j\delta_{-t_j}, \quad \mbox{otherwise}. 
\]
We shall show that $\sqrt{\frac{h}{\Omega}}\btwonorm{[\mu^**f]-[\mu*f]}\leq \sigma$. In subsequent steps, we show this for the case $\mu^*=\sum_{j=1}^{n}a_j \delta_{-t_j},\  \mu=\sum_{j=n+1}^{2n-1}-a_j\delta_{-t_j}$. The other case can be proved in a similar way. \\
\textbf{Step 2.}	
We estimate $\sum_{j=1}^{2n-1}|a_{j}|$.
We begin by ordering $a_j$'s such that
\[
m_{\min}^*=|a_{j_1}|\leq |a_{j_2}|\leq  \cdots \leq |a_{j_{2n-1}}|.
\]
Then (\ref{sufficientthm2equ1}) implies that
\[
a_{j_1}\phi_{2n-3}(t_{j_1}) = \big(\phi_{2n-3}(t_{j_2}), \cdots, \phi_{2n-3}(t_{j_{2n-1}})\big)(-a_{j_2},\cdots, -a_{j_{2n-1}})^T,
\]
where $\phi_{2n-3}(t)=(1, t, \cdots, t^{2n-3})^T$. Hence
\[
a_{j_1} \left(\phi_{2n-3}(t_{j_2}), \cdots, \phi_{2n-3}(t_{j_{2n-1}})\right)^{-1}\phi_{2n-3}(t_{j_1}) = (-a_{j_2},\cdots, -a_{j_{2n-1}})^T.
\]
By Lemma \ref{lem:invervandermonde}, we have 
\[
a_{j_1}  \Pi_{2\leq q\leq 2n-2}\frac{t_{j_1}-t_{j_q}}{t_{2n-1}-t_{j_q}}= -a_{j_{2n-1}}.
\]
Therefore
\[
|a_{j_{2n-1}}|\leq \frac{(2n-2)!}{(n-1)!(n-2)!}|a_{j_1}|\leq \frac{2^{2n-2}\big((n-1)!\big)^2}{(n-1)!(n-2)!}|a_{j_1}|=2^{2n-2}(n-1)m_{\min}^*.
\]
It follows that
\begin{equation}\label{equ:numberlowerboundthm0equ2}
\sum_{j=1}^{2n-1}|a_j| = \sum_{q=1}^{2n-1}|a_{j_q}| \leq (2n-1) |a_{j_{2n-1}}|\leq (2n-1)(n-1)2^{2n-2}m_{\min}^*.
\end{equation}
\textbf{Step 3}. Let $d_{j}= \Omega t_j, 1\leq j \leq 2n-1$. Similar to (\ref{equ:measurementexpansion2}), we have the expansion 
\begin{align*}
[\mu^* * f]-[\mu * f]=\sqrt{\Omega}\sum_{r=0}^{2n-3}(c_r^*-c_r)\mathbf h_r+ \sqrt{\Omega}\mathbf{Res},
\end{align*}
where $c_r^* =\sum_{j=1}^{n}a_{j}\frac{(d_j)^{r}}{r!\sqrt{2r+1}}, \ c_r=\sum_{p=n+1}^{2n-1}a_{p}\frac{(d_p)^{r}}{r!\sqrt{2r+1}}$, and $\mathbf {Res}=\sum_{r=2n-2}^{+\infty}(c_r^*-c_r)\mathbf h_{r}$ with $\mathbf h_r$'s being the multipoles defined in (\ref{equ:multipole1}). By (\ref{sufficientthm2equ1}), $\sum_{r=0}^{2n-3}(c_r^*-c_r)\mathbf h_{r}=0$. On the other hand, 
\begin{align*}
&\sqrt{h}\Big|\Big|\mathbf {Res}\Big|\Big|_2=\sqrt{h}\Big|\Big|\sum_{r=2n-2}^{+\infty}(c_r^*-c_r)\mathbf h_{r}\Big|\Big|_2\leq  \sqrt{h}\sum_{r=2n-2}^{+\infty}\Big|\Big|(c_r^*-c_r)\mathbf h_{r}\Big|\Big|_2\nonumber \\
=&\sqrt{h}\sum_{r=2n-2}^{+\infty}\Big|\Big|\Big(\sum_{j=1}^{n}a_{j}\frac{(d_j)^{r}}{r!\sqrt{2r+1}}+\sum_{p=n+1}^{2n-1}a_{p}\frac{(d_p)^{r}}{r!\sqrt{2r+1}}\Big)\mathbf h_{r}\Big|\Big|_2 \nonumber\\
\leq& \sum_{r=2n-2}^{+\infty}\frac{(2n-1)(n-1)2^{2n-2}m_{\min}^*((n-1)\Omega\tau)^{r}}{r!\sqrt{2r+1}} \ \sqrt{h}\Big|\Big|\mathbf h_{r}\Big|\Big|_2 \nonumber  \quad \Big(\text{by (\ref{equ:numberlowerboundthm0equ2}) and $|d_j|\leq (n-1)\Omega\tau$} \Big)  \\
<& \frac{(2n-1)(n-1)2^{2n-2}m_{\min}^*((n-1)\Omega\tau)^{2n-2}}{(2n-2)!\sqrt{4n-3}}\Big(\sum_{r=2n-2}^{+\infty}\frac{((n-1)\Omega\tau)^{r-(2n-2)}}{(r-(2n-2))!}\Big) \sqrt{\pi} \qquad \Big(\text{by (\ref{2normofmultipole})}\Big) \nonumber \\
=& \frac{(2n-1)(n-1)2^{2n-2}m_{\min}^*((n-1)\Omega\tau)^{2n-2}\sqrt{\pi}e^{((n-1)\tau)}}{(2n-2)!\sqrt{4n-3}}\nonumber \\
=& \frac{(2n-1)(n-1)2^{2n-2}m_{\min}^*((n-1)\Omega\tau)^{2n-2}\sqrt{\pi}e^{(n-1)}}{(2n-2)!\sqrt{4n-3}} \quad \Big(\text{using $\tau<1$ which follows from $\sigma<m_{\min}^*$}\Big)\nonumber\\ 
\leq & 1.11^{2n-2} e^{3n-3}(\Omega\tau)^{2n-2}m_{\min}^*\leq \sigma. \quad (\text{by Lemma \ref{sufficienthm2estimate1} and (\ref{sufficientthm2equ-1})}) 
\end{align*}
Thus,
\[
\sqrt{h}\btwonorm{[\mu^**f]-[\mu*f]}\leq \sqrt{h}\btwonorm{\sum_{r=0}^{2n-3}(c_r^*-c_r)\mathbf h_r}+ \sqrt{h}\btwonorm{\mathbf{Res}}\leq \sigma. 
\]
This completes the proof.

\medskip
In what follows, we derive an upper bound for the computational resolution limit introduced in Definition \ref{def:numbercomputresolutionlimit}. 
For given $\sigma,M$ and $d$, we define 
\begin{equation}\label{polesrecovered2}
s^*:=\min\Big\{l\in \mathbb N: \frac{d^l}{l!\sqrt{2l+1}}\leq \frac{\sigma}{2e^d\sqrt{\pi}M}\Big\},
\end{equation}
which can be viewed as the maximum number of multipole coefficents one can expect to recover stably with the given SNR.

\begin{thm}{\label{necessarythm1}}
Let $n\geq 2$ and let $\mu^*=\sum_{j=1}^{n}a_j^*\delta_{y_j^*}$ be a measure supported in $[\frac{-d}{\Omega},\frac{d}{\Omega}]$. Assume that the following separation condition is satisfied
\begin{equation}\label{necessray1separationcond}
\min_{p\neq j}|y_p^*-y_j^*|\geq \frac{4.7(1+d)}{\Omega}\Big(\frac{3}{\sigma_{\min}(s^*)}\frac{\sigma}{m_{\min}^*}\Big)^{\frac{1}{2n-2}},
\end{equation}
 then any discrete measure $\mu$ with $k<n$ supports cannot be a $(d, \sigma, M)$-admissible measure.
\end{thm}
Proof: 
\textbf{Step 1}. 
We first show that $s^*\geq 2n-1$.
Let $d_{\min}^*=\min_{p\neq j}|y_p^*-y_j^*|$. Note that $y_j^*$'s are in $[\frac{-d}{\Omega},\frac{d}{\Omega}]$. We have
$$d\geq \frac{(n-1)\Omega d_{\min}^*}{2}.$$ 
Therefore 
\begin{align*} 
&\frac{d^{2n-2}}{(2n-2)!\sqrt{4n-3}}\geq \frac{(\frac{1}{2})^{2n-2}(n-1)^{2n-2}(\Omega d_{\min}^*)^{2n-2}}{(2n-2)!\sqrt{4n-3}}\\
\geq& \frac{(\frac{1}{2})^{2n-2}(n-1)^{2n-2}4.7^{2n-2}(1+d)^{2n-2}}{(2n-2)!\sqrt{4n-3}} \frac{3\sigma}{\sigma_{\min}(s^*)m_{\min}^*} \quad \Big(\mbox{by} \,\,(\ref{necessray1separationcond})\Big)\\
\geq& \frac{(n-1)^{2n-2}2.35^{2n-2}(1+d)^{2n-2}}{e(2n-2)^{2n-2+\frac{1}{2}}e^{-(2n-2)}\sqrt{4n-3}}  \frac{3\sigma}{\sigma_{\min}(s^*)\frac{M}{n}} \quad \Big(\text{since $m_{\min}^*\leq \frac{m^*}{n}\leq \frac{M}{n}$}\Big)\\
\geq & \frac{e^{2n-3}1.175^{2n-2}(1+d)^{2n-2}}{2\sqrt{2}} \frac{3}{\sigma_{\min}(s^*)}\frac{\sigma}{M} 
>   \frac{1}{2e^d\sigma_{\min}(s^*)} \frac{\sigma}{M} \\
\geq &  \frac{1}{2e^d \sqrt{\pi}} \frac{\sigma}{M}
,  \quad \Big(\text{by \,\,(\ref{minisingularvaluebound1})}\Big)
\end{align*}
which implies that $s^*\geq 2n-1$ by definition (\ref{polesrecovered2}).\\
\textbf{Step 2}. Let $\mu=\sum_{j=1}^{k}a_j\delta_{y_j}$ with $k<n$. By (\ref{equ:measurementexpansion2}), we have
\begin{equation}\label{necessarybaseeq0}\mathbf Y-[\mu *f_{\Omega}]=\sqrt{\Omega}\sum_{r=0}^{+\infty}(c_r^*-c_r)\mathbf h_{r}+\sqrt{\Omega}\mathbf W,
\end{equation}
where 
\[
c_r^*=\sum_{j=1}^{n}a_j^*\frac{(d_j^{*})^r}{r!\sqrt{2r+1}}, \quad c_r=\sum_{j=1}^{k}a_j \frac{(d_j)^r}{r!\sqrt{2r+1}}, \,\,\,\, d_j^{*}=-y_j^*\cdot \Omega, \,\,\,\,d_j=-y_j \cdot \Omega.
\]
It follows that
\begin{align}\label{necessarybaseeq1}
\frac{1}{\sqrt{\Omega}}\btwonorm{\mathbf Y-[\mu *f]}&\geq \btwonorm{\sum_{r=0}^{s^*-1}(c_r^*-c_r)\mathbf h_r}-\btwonorm{\mathbf W}-\btwonorm{\mathbf {Res}},
\end{align}
where $\mathbf {Res}=\sum_{r=s^*}^{\infty}(c_r^*-c_r)\mathbf h_r$ is the residual term. Note that $\sum_{j=1}^{k}|a_j|\leq M$, $\sum_{j=1}^{n}|a_j^*|=m^*\leq M$, and $|d_j|\leq d$. We have 
\begin{align}\label{necessarybaseeq2}
&\sqrt{h}\Big|\Big|\mathbf {Res}\Big|\Big|_2=\sqrt{h}\Big|\Big|\sum_{r=s^*}^{+\infty}(c_r^*-c_r)\mathbf h_{r}\Big|\Big|_2
=\sqrt{h}\Big|\Big|\sum_{r=s^*}^{+\infty}\Big(\sum_{j=1}^{n}a_{j}^*\frac{(d_j^*)^{r}}{r!\sqrt{2r+1}}-\sum_{p=1}^{k}a_{p}\frac{d_p^{r}}{r!\sqrt{2r+1}}\Big)\mathbf h_{r}\Big|\Big|_2\nonumber \\
\leq& \sum_{r=s^*}^{+\infty}\sqrt{h}\Big|\Big|\Big(\sum_{j=1}^{n}a_{j}^*\frac{(d_j^*)^{r}}{r!\sqrt{2r+1}}-\sum_{p=1}^{k}a_{p}\frac{d_p^{r}}{r!\sqrt{2r+1}}\Big)\mathbf h_{r}\Big|\Big|_2\leq \sum_{r=s^*}^{+\infty}\frac{2Md^{r}}{r!\sqrt{2r+1}}\sqrt{h}||\mathbf h_{r}||_2\nonumber \\
<& \frac{2Md^{s^*}}{s^*!\sqrt{2s^*+1}}\Big(\sum_{r=s^*}^{+\infty}\frac{d^{r-s^*}}{(r-s^*)!}\Big) \sqrt{\pi}\qquad \Big( \text{by (\ref{2normofmultipole})} \Big)\nonumber \\
=& \frac{2e^d\sqrt{\pi} Md^{s^*}}{s^*!\sqrt{2s^*+1}} \leq\sigma. \quad \Big(\text{since $\frac{d^{s^*}}{s^*!\sqrt{2s^*+1}}\leq \frac{\sigma}{2e^d\sqrt{\pi}M}$ }\Big)
\end{align}
\textbf{Step 3}. We estimate the term $\sqrt{h}\Big|\Big|\sum_{r=0}^{s^*-1}(c_r^*-c_r)\mathbf h_r\Big|\Big|_2$ in this step. Recall the multipole matrix
\begin{equation*}{\label{matrix1}}
\mathbf H(s^*)=\Big(
\mathbf h_0\ \mathbf h_1\ \cdots\ \mathbf h_{s^*-1}\Big).
\end{equation*}
Denote
\[b(s^*)=\Big(c_0^*-c_0,\ \cdots, c_{s^*-1}^*-c_{s^*-1} \Big)^{T}.
\]
We have 
\begin{equation*}
\begin{aligned}
\sqrt{h}\Big|\Big|\sum_{r=0}^{s^*-1}(c_r^*-c_r)\mathbf h_r\Big|\Big|_2
=\sqrt{h}\Big|\Big|\mathbf H(s^*) b(s^*)\Big|\Big|_2\geq  
\sigma_{\min}(s^*)\Big|\Big|b(s^*)\Big|\Big|_2,
\end{aligned}
\end{equation*}
where $\sigma_{\min}(s^*)=\sigma_{\min}(\sqrt{h}\mathbf H(s^*))$ is defined in (\ref{equ:sigmamin}). Note that $b(s^*)$ can be written in the form $\tilde A a -\tilde A^* a^* $ with $\tilde A, \tilde A^*, a, a^*$ being defined as in Corollary \ref{spaceapproxlowerbound2} and that
$ \min_{j\neq p}|d_j^*-d_p^*| = \Omega d_{\min}^*$. 
An application of Corollary \ref{spaceapproxlowerbound2} yields 
\begin{align*}
||b(2n-1)||_2\geq \frac{1.15m_{\min}^*(\Omega d_{\min}^*)^{2n-2}}{2^{8n-8}(n-1)(1+d)^{2n-2}}.
\end{align*}
Using the fact that $s^*\geq 2n-1$ (proved in Step 1) and
the separation condition (\ref{necessray1separationcond}), we further get
\[||b(s^*)||_2\geq ||b(2n-1)||_2 \geq \frac{1.15m_{\min}^*(\Omega d_{\min}^*)^{2n-2}}{2^{8n-8}(n-1)(1+d)^{2n-2}}\geq \frac{3\sigma}{\sigma_{\min}(s^*)}.\]
It follows that 
\begin{equation}\label{necessarybaseeq3}
\sqrt{h}\Big|\Big|\sum_{r=0}^{s^*-1}(c_r^*-c_r)\mathbf h_r\Big|\Big|_2\geq 3\sigma. 
\end{equation}
\textbf{Step 4}. Finally, combining (\ref{necessarybaseeq1})-(\ref{necessarybaseeq3}), we have  \begin{equation*}\sqrt{\frac{h}{\Omega}}\btwonorm{\mathbf Y-[\mu*f_{\Omega}]}> 3\sigma-2\sigma= \sigma.
\end{equation*}
This proves that any discrete measure $\mu$ with only $k<n$ supports cannot be a $(d, \sigma, M)$-admissible measure and thus completes the proof of the theorem.

\medskip

\subsection{Discussion of $\sigma_{\min}(s^*)$ and the upper bound for the computational resolution limit} \label{sec-sigmas}
Combining Proposition \ref{thm:numberlowerboundthm0} and Theorem \ref{necessarythm1}, we have the following bounds for the computational resolution limit $\mathcal{D}_{num}$: 
%
\[
 \frac{0.9e^{-\frac{3}{2}}}{\Omega}\Big(\frac{\sigma}{m_{\min}^*}\Big)^{\frac{1}{2n-2}} \leq \mathcal D_{num} \leq \frac{4.7(1+d)}{\Omega}\Big(\frac{3}{\sigma_{\min}(s^*)}\frac{\sigma}{m_{\min}^*}\Big)^{\frac{1}{2n-2}}.
\]
The major gap in the two bounds lies in the factor $\sigma_{\min}(s^*)$ which is determined by $s^*$ in (\ref{polesrecovered2}) and the multipole matrix (\ref{eq-matrix}). 
This gap is due to the correlation between the multipoles which amplifies the noise-to-signal ratio $\frac{\sigma}{m_{\min}^*}$ when one reconstructs the multipole coefficients. In this section, we derive estimate for the factor $\sigma_{\min}(s^*)$ and use it to demonstrate the dependence of the upper bound on the noise level. 
We first present a useful lemma. 
\begin{lem} \label{lem:sigmaminsestimate1}
	We have
	\[
	\min_{||a||_2\leq 1, a\in \mathbb C^{s}}\int_{-1}^1 \babs{a_1+a_2 x+\cdots+ a_{s} x^{s-1}}^2dx \geq \Big(\frac{8\sqrt{2}\pi }{27}\frac{1}{(2e)^{s-1}}\Big)^2.
	\] 
\end{lem}
Proof: Let $\tau = \frac{2}{s}$. We have
\begin{align*}
&\min_{||a||_2\leq 1, a\in \mathbb C^{s}}\int_{-1}^1 \babs{a_1+a_2 x+\cdots+ a_{s} x^{s-1}}^2dx\\
=& \min_{||a||_2\leq 1, a\in \mathbb C^{s}}\int_{-1}^{-1+\tau} \babs{a_1+a_2 x+\cdots+ a_{s} x^{s-1}}^2dx
+\int_{-1+\tau}^{-1+2\tau} \babs{a_1+a_2 x+\cdots+ a_{s} x^{s-1}}^2dx\\
&+\cdots+ \int_{-1+(s-1)\tau}^{1} \babs{a_1+a_2 x+\cdots+ a_{s} x^{s-1}}^2dx \quad \\
=&\min_{||a||_2\leq 1, a\in \mathbb C^{s}}\int_{-1}^{-1+\tau} \babs{a_1+a_2 x+\cdots+ a_{s} x^{s-1}}^2+  \babs{a_1+a_2 (x+\tau)+\cdots+ a_{s} (x+\tau)^{s-1}}^2\\
&+\cdots + \babs{a_1+a_2 (x+(s-1)\tau)+\cdots+ a_{s} (x+(s-1)\tau)^{s-1}}^2dx\\
=&\min_{||a||_2\leq 1, a\in \mathbb C^{s}} \int_{-1}^{-1+\tau} \btwonorm{
	\begin{pmatrix}
	1&x&\cdots &x^s\\
	1&x+\tau&\cdots &(x+\tau)^s\\
	\vdots&\vdots&\ddots&\vdots\\
	1&x+(s-1)\tau&\cdots &(x+(s-1)\tau)^s\\
	\end{pmatrix} a }^2dx\\
\geq &  \int_{-1}^{-1+\tau} \Big(\frac{\zeta(s)\tau^{s-1}}{2^{s-1}\sqrt{s}}\Big)^2dx   \quad \Big(\text{using Lemma \ref{singularvaluevandermonde2} and Corollary \ref{norminversevandermonde2}}\Big)\\
=&\Big(\frac{\zeta(s)\tau^{s-\frac{1}{2}}}{2^{s-1}\sqrt{s}}\Big)^2 =  \Big(\sqrt{2}\zeta(s)(\frac{1}{s})^{s}\Big)^2 \geq \Big(\frac{8\sqrt{2}\pi }{27}\frac{1}{(2e)^{s-1}}\Big)^2. \quad \Big(\text{using Lemma \ref{equ:sigmaminsestimate1}}\Big)
\end{align*} 

Now for the integer $s^*$ defined in (\ref{polesrecovered2}), we denote $\mathcal{H}(s^*)$ the $s^* \times s^*$ matrix with entries
\begin{equation}\label{equ:defineofmathcalH}
	\mathcal{H}_{p, j}(s^*) = \sqrt{2p-1}\sqrt{2j-1}\frac{\pi}{2} \int_{-1}^{1} (i\omega)^{p-1}(-i\omega)^{j-1}d\omega,  \quad  1\leq p, j \leq s^*.
\end{equation}
Note that $\mathcal{H}(s^*)$ is a real-valued matrix. We have the following estimate.
\begin{lem}\label{lem:lowerboundofsigmamins1}
\[
\Big(\min_{||a||_2\leq 1, a\in \mathbb R^{s^*}} a^T \mathcal H(s^*) a\Big)^{\frac{1}{2}} \geq  \frac{8\pi^{\frac{3}{2}}} {27}\frac{1}{(2e)^{s^*-1}}.
\]
\end{lem}
Proof:
\begin{align*}
&\Big(\min_{||a||_2\leq 1, a\in \mathbb R^{s^*}} a^T \mathcal H(s^*) a\Big)^{\frac{1}{2}} \nonumber \\
=& \Big(\min_{||a||_2\leq 1, a\in \mathbb R^{s^*}}\frac{\pi}{2}\int_{-1}^{1}|a_1 + a_2\sqrt{3}(i\omega)+\cdots+ a_{s^*} \sqrt{2s^*-1}(i\omega)^{s^*-1}|^2d\omega \Big)^{\frac{1}{2}} \quad \Big(\text{by (\ref{equ:defineofmathcalH})}\Big)\nonumber \\
\geq& \Big(\frac{\pi}{2}\min_{||a||_2\leq 1, a\in \mathbb C^{s^*}}\int_{-1}^{1}|a_1 + a_2\omega+\cdots+ a_{s^*}\omega^{s^*-1}|^2d\omega \Big)^{\frac{1}{2}} \nonumber \\
\geq & \frac{8\pi^{\frac{3}{2}}}{27}\frac{1}{(2e)^{s^*-1}}. \quad  \Big( \text{by Lemma \ref{lem:sigmaminsestimate1}}\Big)
\end{align*}

We are ready to estimate $\sigma_{\min}(s^*)$. Recall that
\begin{align*}
\sigma_{\min}(s^*)=\sigma_{\min}(\sqrt{h}\mathbf H(s^*))=\sqrt{\lambda_{\min}(h\mathbf H(s^*)^T\mathbf H(s^*))},
\end{align*} 
where $\mathbf H(s^*)=\Big(\mathbf h_0,\mathbf h_1,\cdots,\mathbf h_{s^*-1}\Big)$ with 
\[
\mathbf h_r=\sqrt{2r+1}(f^{(r)}(\Omega x_1),\cdots,f^{(r)}(\Omega x_N))^T,\ r=0,1,\cdots,
\] 
and $f(x)=\frac{\sin x}{x}$.
Consider the $(p,j)$-th entry $h\mathbf h_{p-1}^T\mathbf h_{j-1}$ of the matrix $h\mathbf H(s^*)^T \mathbf H(s^*)$. 
When the truncation number $R\to +\infty$, we have 
\begin{align*}
h\mathbf h_{p-1}^T\mathbf h_{j-1}\to &\sqrt{2p-1}\sqrt{2j-1}\int_{-\infty}^{+\infty}f^{(p-1)}(x)f^{(j-1)}(x)dx \\
=&\sqrt{2p-1}\sqrt{2j-1} \frac{1}{2\pi}\int_{-\infty}^{+\infty} \widehat {f^{(p-1)}}\overline{\widehat {f^{(j-1)}}}d\omega \\
=&\sqrt{2p-1}\sqrt{2j-1}\frac{\pi}{2} \int_{-1}^{1} (i\omega)^{p-1}(-i\omega)^{j-1}d\omega\\
=&\mathcal{H}_{p,j}(s^*),
\end{align*}
where the notation $\widehat \cdot$
denotes the Fourier transform. 
It follows that
\begin{align}\label{equ:lowerboundofsigmamins0}
\sigma_{\min}(s^*)= &\Big(\min_{||a||_2\leq 1, a\in \mathbb R^{s^*}} a^Th \mathbf H(s^*)^T \mathbf H(s^*) a\Big)^{\frac{1}{2}} \to \Big(\min_{||a||_2\leq 1, a\in \mathbb R^{s^*}} a^T \mathcal H(s^*) a\Big)^{\frac{1}{2}}, \nonumber
\end{align}	
as $R\to +\infty$. 
Therefore, using Lemma \ref{lem:lowerboundofsigmamins1}, we can obtain the following lower bound for $\sigma_{\min}(s^*)$, provided that $R$ is sufficiently large: 
%
\begin{equation}\label{equ:lowerboundofsigmamins1}
\sigma_{\min}(s^*) \gtrsim \frac{1}{(2e)^{s^*}}.
\end{equation}
We now reexamine the upper bound in Theorem \ref{necessarythm1}. Let $M$ and $m_{\min}^*$ be two fixed positive numbers that are of order one and let $\sigma \ll 1$. Recall that $s^*$ is defined to be the minimum integer so that $\frac{d^{s^*}}{s^*!\sqrt{2s^*+1}}\leq \frac{\sigma}{2e^d\sqrt{\pi}M}$ (see (\ref{polesrecovered2})). Thus  
\[
\sigma \leq \frac{d^{s^*-1}2e^d\sqrt{\pi}M}{(s^*-1)!\sqrt{2s^*-1}}.  
\]
Using formula (\ref{strilingformula}), we get
\begin{equation}\label{equ:orderofsigma}
 \sigma \lesssim  \frac{(ed)^{s^*}}{(s^*)^{s^*}}.
\end{equation}
Together with (\ref{equ:lowerboundofsigmamins1}), we have 
\[
\frac{\sigma}{\sigma_{\min}(s^*)} \lesssim \frac{(2e^{2}d)^{s^*}}{(s^*)^{s^*}}.
\]
Therefore as the noise level $\sigma\to 0$, we have $s^*\to \infty$ and the 
 the upper bound for the computational resolution limit,  $\frac{4.7(1+d)}{\Omega}\Big(\frac{3}{\sigma_{\min}(s^*)}\frac{\sigma}{m_{\min}^*}\Big)^{\frac{1}{2n-2}}$, tends to $0$ as well. 
On the other hand, 
using (\ref{equ:lowerboundofsigmamins1})-(\ref{equ:orderofsigma})
we can derive that
\[
 \frac{1}{\sigma_{\min}(s^*)} \lesssim \Big(\frac{1}{\sigma}\Big)^{\epsilon(\sigma)},
\]
 where $\epsilon(\sigma)= \frac{\ln(2e)}{\ln s^* - \ln (ed)}$ depends on the noise level $\sigma$. Therefore, the upper bound for computational resolution limit scales as
 \[
 \frac{1}{\Omega}\cdot \sigma^{\frac{1}{2n-2}(1-\epsilon(\sigma))}.
 \]
Especially in the limit $\sigma\to 0$, we have $\epsilon(\sigma) \to 0$, and the upper bound scales asymptotically as
 \[
 \frac{1}{\Omega}\cdot \sigma^{\frac{1}{2n-2}}.
 \]
 
\begin{remark}
We remark that the factor $\sigma_{\min}(s^*)$ in the upper bound of the computation resolution limit seems to be a disadvantage of the reconstruction using direct measurement of the convoluted image (\ref{introductionequ-1}). This unpleasant amplification of noise may be mitigated if one uses the Fourier measurement (\ref{eq-fouriermeasure}) and exploits the special data structure therein. This is indeed the case for the point spread function $f_{\Omega}$ as considered in the paper, see \cite{liu2020resolution}. 
On the other hand, most reconstruction methods using Fourier data are based on equal-spaced sampling and hence may not be practicable to the case when the support of the point spread function in the Fourier domain is supported on a union of disjoint intervals where equal-spaced sampling is not possible. Moreover, in some situation when the available measurement is only given by a truncated image, the Fourier data need to be calculated using the Fourier transform. The truncation in the image will inevitably induce error in the Fourier data which may increase the noise level. The issue is more severe if the point spread function does not decay fast at the infinity and the truncation number is not large enough.  
We expect that the reconstruction using the direct measurement (\ref{introductionequ-1}) has certain edge in these cases.    
\end{remark} 

\subsection{Stability of the support recovery problem} \label{sec-stability}
In this section, 
we study the stability of recovering source supports from admissible measures and quantitatively demonstrate the dependence on the super-resolution factor. In the grid setting, the super-resolution factor can be defined as the ratio between Rayleigh limit and the grid spacing, see for instance \cite{candes2014towards}. In our off-the-grid setting, recalling that the point spread function is $f_{\Omega}(x)=\frac{\sin \Omega x}{\sqrt{\Omega}x}$ and the associated Rayleigh limit is $\frac{\pi}{\Omega}$, we define the super-resolution factor as 
\[
SRF:= \frac{\pi}{d_{\min}^*\Omega}, 
\]
where $d_{\min}^* = \min_{p\neq j, 1\leq p, j\leq n}|y_p^*-y_j^*|$. We have the following stability result. 

\begin{thm}\label{reconaccuracythm2}
Let $n\geq 2$ and let $\mu^*=\sum_{j=1}^{n}a_j^*\delta_{y_j^*}$ be a measure such that $\frac{-d}{\Omega} \leq y_1^* < y_2^* < \cdots < y_n^* \leq \frac{d}{\Omega}$. Assume that the following separation condition holds
\begin{equation}\label{equ:supportupperboundequ0}
d_{\min}^*=\min_{p\neq j}\babs{y_p^*-y_j^*}\geq
\frac{6.24(1+d)}{\Omega}\Big(\frac{3}{\sigma_{\min}(s^*)}\frac{\sigma}{ m_{\min}^*}\Big)^{\frac{1}{2n-1}},
\end{equation}
where $s^*$ is defined in (\ref{polesrecovered2}). If $\mu=\sum_{j=1}^na_j\delta_{y_j}$ is a $(d, \sigma, M)$-admissible measure such that $y_j \in [\frac{-d}{\Omega}, \frac{d}{\Omega}], 1\leq j \leq n$, then after reordering $y_j$'s, we have
\begin{equation}\label{equ:supportupperboundequ1}
|y_j- y_j^*|< \frac{d_{\min}^*}{2},\quad  j=1, \cdots, n. 
\end{equation}
Moreover, 
\[ 
|y_j-y_j^*|< \frac{C(n,d)}{\Omega}SRF^{2n-2}\frac{3\sigma}{\sigma_{\min}(s^*)m_{\min}^*},
\]
where $C(n,d)=\frac{7.73\sqrt{4n-1}(2n-1)(4+4d)^{2n-1}}{4e\pi^{2n-\frac{1}{2}}}$.	
\end{thm}
Proof: 
\textbf{Step 1}. We show that $s^*\geq 2n$. 
Since $d\geq \frac{(n-1)\Omega d_{\min}^*}{2}$, we have 
\begin{align*}
	&\frac{d^{2n-1}}{(2n-1)!\sqrt{4n-1}}\geq \frac{(\frac{1}{2})^{2n-1}(n-1)^{2n-1}(\Omega d_{\min}^*)^{2n-1}}{(2n-1)!\sqrt{4n-1}}\\
	\geq& \frac{(\frac{1}{2})^{2n-1}(n-1)^{2n-1}6.24^{2n-1}(1+d)^{2n-1}}{(2n-1)!\sqrt{4n-1}} \frac{3\sigma}{\sigma_{\min}(s^*)m_{\min}^*}  \quad \Big(\text{by (\ref{equ:supportupperboundequ0})}\Big)\\
	\geq& \frac{(n-1)^{2n-1}3.12^{2n-1}(1+d)^{2n-1}}{e(2n-1)^{2n-1+\frac{1}{2}}e^{-(2n-1)}\sqrt{4n-1}} \frac{3\sigma}{\sigma_{\min}(s^*)\frac{M}{n}} \quad \Big(\text{since $m_{\min}^*\leq \frac{m^*}{n}\leq \frac{M}{n}$}\Big)\\
	\geq & \frac{e^{2n-3}1.56^{2n-1}(1+d)^{2n-1}}{2\sqrt{2}} \frac{3}{\sigma_{\min}(s^*)}\frac{\sigma}{M} >  \frac{\sigma}{2e^d\sigma_{\min}(s^*)M}  \\
	\geq & \frac{\sigma}{2e^d\sqrt{\pi}M}.  \quad \Big(\text{by (\ref{minisingularvaluebound1})}\Big)
\end{align*}
Recalling (\ref{polesrecovered2}), we get $s^*\geq 2n$.\\
\textbf{Step 2}. Let $\mu=\sum_{j=1}^na_j\delta_{y_j}$ be a $(d,\sigma, M)$-admissible measure such that
\begin{equation}\label{reconaccuracy1equ1}
\sqrt{\frac{h}{\Omega}}\btwonorm{\mathbf Y-[\mu*f_{\Omega}]}\leq\sigma.
\end{equation}
Based on the decomposition (\ref{equ:measurementexpansion2}),
\[
\mathbf Y-[\mu *f_{\Omega}]=\sqrt{\Omega}\sum_{r=0}^{+\infty}(c_r^*-c_r)\mathbf h_{r}+\sqrt{\Omega}\mathbf W,
\]
where $c_r^*=\sum_{j=1}^na_j^*\frac{(d_j^*)^r}{r!\sqrt{2r+1}}$ and $c_r=\sum_{j=1}^na_j\frac{d_j^r}{r!\sqrt{2r+1}}$. We have 
\begin{align}\label{reconaccuracy1equ2}
	\sqrt{\frac{h}{\Omega}}\btwonorm{\mathbf Y-[\mu *f]}&\geq \sqrt{h}\btwonorm{\sum_{r=0}^{s^*-1}(c_r^*-c_r)\mathbf h_r}-\sqrt{h}\btwonorm{\mathbf W}-\sqrt{h}\btwonorm{\mathbf {Res}}.
\end{align}
Here $\mathbf {Res}$ is the residual term and $s^*$ is defined in (\ref{polesrecovered2}). By a similar argument as in Step 2 and 3 of the proof of Theorem \ref{necessarythm1}, we have
\begin{align}\label{reconaccuracy1equ3}
\sqrt{h}\btwonorm{\mathbf {Res}}< \sigma,\ \text{and} \ \sqrt{h}\btwonorm{\sum_{r=0}^{s^*-1}(c_r^*-c_r)\mathbf h_r}\geq \btwonorm{b(s^*)}\sigma_{\min}(s^*),
\end{align}
where 
\[
b(s^*)=(c_0^*-c_0,\cdots,c_{s^*-1}^*-c_{s^*-1})^{T},\quad  \sigma_{\min}(s^*)=\sigma_{\min}(\sqrt{h}\mathbf H(s^*)).
\]
Here recall that $\mathbf H(s^*)=\Big(\mathbf h_0, \mathbf h_1, \cdots, \mathbf h_{s^*-1} \Big)$ is the multipole matrix. Therefore, by the constraint (\ref{reconaccuracy1equ1}) and the estimates (\ref{reconaccuracy1equ2})-(\ref{reconaccuracy1equ3}), we have 
\[
\btwonorm{b(s^*)}<\frac{3\sigma}{\sigma_{\min}(s^*)}. 
\]
Since $s^*\geq 2n$ (by result in Step 1), we have $||b(2n)||_{2}\leq||b(s^*)||_{2}<\frac{3\sigma}{\sigma_{\min}(s^*)}$. 
Note that $b(2n)$ can be written in the form $\tilde A a -\tilde A^* a^* $ with $\tilde A, \tilde A^*, a, a^*$ being defined as in Corollary \ref{spaceapproxlowerbound4}. 
An application of the result therein yields
\begin{equation}\label{stability2equ1}
||\eta_{n,n}(d_1^*,\cdots, d_{n}^*, d_1, \cdots, d_{n})||_{\infty}<\frac{(1+d)^{2n-1}}{\zeta(n)(\Omega d_{\min}^*)^{n-1}}\frac{\sqrt{4n-1}(2n-1)!3\sigma}{\sigma_{\min}(s^*)m_{\min}^*},
\end{equation}
where $\eta_{n,n}(d_1^*,\cdots, d_{n}^*, d_1, \cdots, d_{n})$ is defined as in (\ref{equ:defineofeta}).\\
\textbf{Step 3}. We apply Lemma \ref{lem:multiproductstability1} to estimate $|d_j-d_j^*|$'s. For the purpose, let 
$$
\epsilon =\frac{(1+d)^{2n-1}}{\zeta(n)(\Omega d_{\min}^*)^{n-1}}\frac{\sqrt{4n-1}(2n-1)!3\sigma}{\sigma_{\min}(s^*)m_{\min}^*}.
$$ 
We have $||\eta_{n,n}||_{\infty}<\epsilon$. We only need to check that the following condition holds: 
\begin{equation}\label{stability2equ4}
\Omega d_{\min}^* \geq \big(\frac{4\epsilon}{\lambda(n)}\big)^{\frac{1}{n}}, \ \text{or equivalently} \,\, (\Omega d_{\min}^*)^n \geq \frac{4\epsilon}{\lambda(n)}.
\end{equation}
Indeed, by the separation condition and Lemma \ref{stability2calculation1} in appendix C, we have
\begin{equation}\label{stability2equ2}
\Omega d_{\min}^*\geq \Big(\frac{4(1+d)^{2n-1}\sqrt{4n-1}(2n-1)!}{\zeta(n)\lambda(n)}\Big)^{\frac{1}{2n-1}} \Big(\frac{3}{\sigma_{\min}(s^*)}\frac{\sigma}{m_{\min}^*}\Big)^{\frac{1}{2n-1}},
\end{equation}
where $\lambda(n)$ is defined as (\ref{equ:lambda1}). Then
\[
(\Omega d_{\min}^*)^{2n-1}\geq\frac{4(1+d)^{2n-1}\sqrt{4n-1}(2n-1)!}{\zeta(n)\lambda(n)}\frac{3\sigma}{\sigma_{\min}(s^*)m_{\min}^*},
\]
whence we get (\ref{stability2equ4}). Therefore, we can apply Lemma \ref{lem:multiproductstability1} to get that, after reordering $d_j$'s,
\begin{equation*}
|d_{j}-d_j^*|< \frac{\Omega d_{\min}^*}{2}, \text{ and }|d_{j}-d_j^*|< \frac{(1+d)^{2n-1}2^{n-1}}{(n-2)!\zeta(n)(\Omega d_{\min}^*)^{2n-2}}\frac{\sqrt{4n-1}(2n-1)!3\sigma}{\sigma_{\min}(s^*)m_{\min}^*},\ 1\leq j\leq n.
\end{equation*}
The first inequality in above result proves (\ref{equ:supportupperboundequ1}) and the second one gives
\[ 
|d_j-d_j^*|<\frac{2^{n-1}(1+d)^{2n-1}(2n-1)!\sqrt{4n-1}}{\zeta(n)(n-2)! }(\frac{1}{\Omega d_{\min}^*})^{2n-2}\frac{3\sigma}{m_{\min}^*\sigma_{\min}(s^*)}.
\]
Finally, using Lemma \ref{stability2calculation2} in appendix C, we have 
\[ 
|d_j-d_j^*|< \frac{7.73\sqrt{4n-1}(2n-1)(4+4d)^{2n-1}}{4e\pi^{\frac{3}{2}}}(\frac{1}{\Omega d_{\min}^*})^{2n-2}\frac{3\sigma}{m_{\min}^*\sigma_{\min}(s^*)}.
\]
It follows that
\[ 
|y_j-y_j^*|< \frac{7.73\sqrt{4n-1}(2n-1)(4+4d)^{2n-1}}{\Omega 4e\pi^{\frac{3}{2}}}(\frac{1}{\Omega d_{\min}^*})^{2n-2}\frac{3\sigma}{m_{\min}^*\sigma_{\min}(s^*)}.
\]
This proves the theorem by substituting $SRF$ into the above inequality.

\medskip

The theorem above states that when the sources are separated sufficiently, 
one can stably recover the source positions from admissible measures.

\section{A Non-linear approximation theory in Vandermonde space}  
In this section, we introduce the main technique, a nonlinear approximation theory in Vandermonde space, 
that is used to derive the main results in the previous section.
For a given positive integer $s$ and $\omega\in \mathbb R$, we denote
\begin{equation}\label{equ:defineofphi}
\phi_s(\omega)=(1,\omega,\cdots,\omega^{s})^T. 
\end{equation}
and call it a Vandermonde vector. We define the $k$-dimensional Vandermonde space
\begin{equation}\label{equ:kdvandermondespace}
W_s^{k}(\omega_1,\cdots,\omega_k):=\text{span} \Big\{\ \phi_s(\omega_1), \ \cdots,\ \phi_s(\omega_k)\ \Big\}.
\end{equation}
At the heart of our theory is the  
following nonlinear approximation problem in the Vandermonde space 
\begin{equation}\label{vandermondeapprox}
\min_{a_j,d_j\in \mathbb{R},|d_j|\leq d,j=1,\cdots,k}\Big|\Big|\sum_{j=1}^k a_j\phi_s(d_j)-v\Big|\Big|_2,
\end{equation}
where $v=\sum_{j=1}^{k+1}a_j^*\phi_s(d_j^*)$ is a given vector. 
We shall derive a sharp lower bound for this problem. 
The lower bound shall lead to the upper bound for the computational resolution limit $\mathcal{D}_{num}$. In addition, we shall also investigate the stability of the approximation problem (\ref{vandermondeapprox}) for $v=\sum_{j=1}^{k}a_j^*\phi_s(d_j^*)$.

The material in this section can be viewed independently. It is origanized as follows. Section \ref{sec-vand-pre} introduces some notations and preliminaries. Section \ref{sec-vand-lowerbound} presents the results on the lower bound to the problem (\ref{vandermondeapprox}). The stability results are given in section 
\ref{sec-vand-stability}.


\subsection{Notations and Preliminaries} \label{sec-vand-pre}
We introduce some notations and preliminaries.  
We denote for integer $k\geq 1$, 
\begin{equation}\label{equ:zetaxiformula1}
	\zeta(k+1)= \left\{
	\begin{array}{cc}
		(\frac{k+1}{2})!(\frac{k-1}{2})!,& \text{$k$ is odd},\\
		(\frac{k}{2}!)^2,& \text{$k$ is even},
	\end{array} 
	\right.	  \quad \xi(k)=\left\{
	\begin{array}{cc}
		\frac{1}{2},  & k=1,\\
		\frac{(\frac{k-1}{2})!(\frac{k-3}{2})!}{4},& \text{$k$ is odd, $ k\geq 3$},\\
		\frac{(\frac{k-2}{2}!)^2}{4},& \text{$k$ is even},
	\end{array} 
	\right.	
\end{equation}
and for positive integers $p,q$ and real numbers $z_1^*, \cdots, z_p^*, z_1, \cdots, z_q$, 
\begin{equation}\label{equ:defineofeta}
\eta_{p,q}(z_1^*, \cdots, z_p^*, z_1, \cdots, z_q)= 
\begin{pmatrix}
(z_1^*-z_1)\cdots(z_1^*-z_q)\\
(z_2^*-z_1)\cdots(z_2^*-z_q)\\
\vdots\\
(z_p^*-z_1)\cdots(z_p^*-z_q)
\end{pmatrix}.
\end{equation}
We denote 
\begin{align} \label{eq-Vn}
V_{s}(k)=\begin{pmatrix}
1&\cdots&1\\
d_1&\cdots&d_{k}\\
\vdots&\ddots&\vdots\\
d_1^s&\cdots&d_{k}^{s}
\end{pmatrix}=
\Big(
\phi_s(d_1)\ \ \phi_s(d_2)\ \ \cdots\ \ \phi_s(d_k) 	
\Big).
\end{align}
For a real matrix or vector $A$, we denote by $A^T$ its transpose. 

\medskip
We first present some basic properties for Vandermonde matrices.
\begin{lem}{\label{norminversevandermonde1}}
Let $d_i\neq d_j$ for $i\neq j, i,j=1,\cdots,k$. For the Vandermonde matrix $V_{k-1}(k)$ defined as in (\ref{eq-Vn}), we have the following estimate
\[
||V_{k-1}(k)^{-1}||_{\infty}\leq \max_{1\leq i\leq k}\Pi_{1\leq p\leq k,p\neq i}\frac{1+|d_p|}{|d_i-d_p|}. 
\]
\end{lem}
Proof: see Theorem 1 in \cite{gautschi1962inverses}.

\vspace{0.2cm}
\begin{cor}{\label{norminversevandermonde2}}
Let $d_{\min}=\min_{i\neq j}|d_i-d_j|$ and assume that $\max_{i=1,\cdots,k}|d_i|\leq d$. Then 
\[
||V_{k-1}(k)^{-1}||_{\infty}\leq \frac{(1+d)^{k-1}}{\zeta(k) (d_{\min})^{k-1}},
\] 
where $\zeta(k)$ is defined in (\ref{equ:zetaxiformula1}).
\end{cor}
Proof: WLOG, we may arrange the $d_i$'s such $d_1<d_2<\cdots<d_k$. Then, $|d_i-d_p|\geq |i-p|d_{\min}$. By Lemma \ref{norminversevandermonde1} we have $||V_{k-1}(k)^{-1}||_{\infty}\leq \max_{1\leq i\leq k}\Pi_{1\leq p\leq k,p\neq i}\frac{1+|d_p|}{|d_i-d_p|}$. It follows that
\[
\max_{1\leq i\leq k}\Pi_{1\leq p\leq k,p\neq i}\frac{1+|d_p|}{|d_i-d_p|}\leq \max_{1\leq i\leq k}\Pi_{1\leq p\leq k,p\neq i}\frac{1+d}{|i-p|d_{\min}}\leq \frac{(1+d)^{k-1}}{\zeta(k)(d_{\min})^{k-1}},
\]
where $\zeta(k)$ is defined by (\ref{equ:zetaxiformula1}).

\vspace{0.2cm}
\begin{lem}{\label{lem:invervandermonde}}
Let $d_1, \cdots, d_k$ be $k$ different real numbers and let $t$ be a real number. We have
\[
\left(V_{k-1}(k)^{-1}\phi_{k-1}(t)\right)_{j}=\Pi_{1\leq q\leq k,q\neq j}\frac{t- d_q}{d_j- d_q},
\]
where $V_{k-1}(k)=  \big(\phi_{k-1}(d_1),\cdots,\phi_{k-1}(d_k)\big)$. 
\end{lem}
Proof: 	
We denote $\left(V_{k-1}(k)^{-1}\right)_{jq}=b_{jq}$. Observe that $\left(V_{k-1}(k)^{-1}\phi_{k-1}(t)\right)_{j}=\sum_{q=1}^{k}b_{jq}t^{q-1}$.
We have 
\[
\sum_{q=1}^{k}b_{jq}(t_p)^{q-1}=\delta_{jp},\ \forall j,p=1,\cdots,k,
\]
where $\delta_{jp}$ is the Kronecker delta function. We consider the polynomial $P_j(x)=\sum_{q=1}^{k}b_{jq}x^{q-1}$. It is clear that $P_j(x)$ satisfies $P_{j}(d_1)=0,\cdots,P_j(d_{j-1})=0,P_j(d_j)=1,P_j(d_{j+1})=0,\cdots,P_j(d_k)=0$. Therefore, it must be the Lagrange polynomial, i.e.,
\[
P_j(x)=\Pi_{1\leq q\leq k,q\neq j}\frac{x-d_q}{d_j-d_q}.
\]
It follows that $\left(V_{k-1}(k)^{-1}\phi_{k-1}(t)\right)_{j}=\Pi_{1\leq q\leq k,q\neq j}\frac{ t- d_q}{d_j-d_q}$.

\vspace{0.2cm}
\begin{lem}\label{singularvaluevandermonde2}
For distinct $d_1,\cdots, d_k \in \mathbb R$, define the Vandermonde matrices $V_{k-1}(k), V_s(k)$ as in (\ref{eq-Vn}) with $s\geq k-1$. Then the following estimate on their singular values holds:
\[
\frac{1}{\sqrt{k}}\min_{1\leq i\leq k}\Pi_{1\leq p\leq k,p\neq i}\frac{|d_i-d_p|}{1+|d_p|} \leq \frac{1}{||V_{k-1}(k)^{-1}||_{2}}\leq  \sigma_{\min}(V_{k-1}(k))\leq \sigma_{\min}(V_{s}(k)).
\]
\end{lem}
Proof: By Lemma \ref{norminversevandermonde1}, $||V_{k-1}(k)^{-1}||_{\infty}\leq \max_{1\leq i\leq k}\Pi_{1\leq p\leq k,p\neq i}\frac{1+|d_p|}{|d_i-d_p|}$.
Then 
\[
||V_{k-1}(k)^{-1}||_2\leq \sqrt{k}||V_{k-1}(k)^{-1}||_{\infty}\leq \sqrt{k}\max_{1\leq i\leq k}\Pi_{1\leq p\leq k,p\neq i}\frac{1+|d_p|}{|d_i-d_p|}.
\]
It follows that
\[
\sigma_{\min}(V_{k-1}(k))\geq \frac{1}{||V_{k-1}(k)^{-1}||_{2}}\geq \frac{1}{\sqrt{k}}\min_{1\leq i\leq k}\Pi_{1\leq p\leq k,p\neq i}\frac{|d_i-d_p|}{1+|d_p|}.
\]
Therefore, we have 
\[
\frac{1}{\sqrt{k}}\min_{1\leq i\leq k}\Pi_{1\leq p\leq k,p\neq i}\frac{|d_i-d_p|}{1+|d_p|}\leq \sigma_{\min}(V_{k-1}(k))\leq \sigma_{\min}(V_{s}(k)).
\]

\medskip
A key step in our approximation theory for Vandermonde vectors is to calculate the projection of $\phi_{s}(\omega)$ onto the Vandermonde subspace $W_{s}^{k}(\omega_1,\cdots,\omega_{k})$ 
(see Lemma \ref{lem:vandemondevolumeratio3}). 
For this, we  
first recall the following definition of volume for parallelotopes. 
\begin{defi}
	For $s\geq k-1$, the $k$-dimensional volume of the parallelotope spanned by the vectors $\phi_s(\omega_j),\omega_j\in \mathbb R, j=1,\cdots,k$ is $\sqrt{\det(A^TA)}$
where $A=\big(\phi_s(\omega_1),\cdots,\phi_s(\omega_k)\big)$.
\end{defi}

\begin{lem}\label{lem:vandemondevolumeratio3}
Let $V=W_k^{k}(\omega_1,\cdots,\omega_{k})$ be defined as in (\ref{equ:kdvandermondespace}) where $\omega_1, \cdots, \omega_k$ are $k$ different real numbers, and let $V^{\perp}$ be the orthogonal complement of $V$ in $\mathbb {R}^{k+1}$. Let $P_{V^{\perp}}$ be the orthogonal projection onto $V^{\perp}$, then 
\begin{equation} \label{eq-matrix}
||P_{V^{\perp}}(v)||_2=\sqrt{\frac{\det(\hat A^T\hat A)}{\det(A^TA)}},
\end{equation}
where
\[
A=\Big(\begin{array}{cccc}\phi_k(\omega_1)&\phi_k(\omega_2)&\cdots&\phi_k(\omega_k)\end{array}\Big)\ \text{ and }\ \hat A=\Big(A, v\Big).
\]
\end{lem}
Proof: The conclusion follows from the observation that the $k+1$-dimensional volume of the parallelotope spanned by the vectors $\phi_k(\omega_1), \cdots \phi_k(\omega_k)$ and $v$ can be computed as the product of $||P_{V^{\perp}}(v)||_2$ and the $k$-dimensional volume of the parallelotope spanned by the $k$ vectors $\phi_k(\omega_1), \cdots, \phi_k(\omega_k)$.

\medskip

We now present a key result that is used to calculate of the right hand side of (\ref{eq-matrix}).  
We denote 
\[
S_{1n}^j:=\{(\tau_1,\cdots,\tau_j): \text{$\tau_p\in \{1,\cdots,n\},p=1,\cdots,j$ and $\tau_p\neq \tau_q$, for $p\neq q$}\}.
\]

\begin{prop}\label{vandermondegaussianelimiate1}
The matrix $V_{n}(n)$ (defined as in (\ref{eq-Vn})) can be reduced to the following form by using elementary column-addition operations, i.e.,
\begin{align}\label{equ:vandermondegaussianelimiate1}
V_{n}(n)G(1)\cdots G(n-1)DQ(1)\cdots Q(n-1)=
\begin{pmatrix}
1&0&\cdots&0\\
0&1&\cdots&0\\
\vdots&\vdots&\ddots&\vdots\\
0&0&\cdots&1\\
v_{(n+1)1}&v_{(n+1)2}&\cdots&v_{(n+1)n}
\end{pmatrix}
\end{align}
where $G(1),\ \cdots,\ G(n-1),\ Q(1),\ \cdots,\ Q(n-1)$ are elementary column-addition matrices, $D=\text{diag}(1,\frac{1}{(d_2-d_1)},\cdots,\frac{1}{\Pi_{p=1}^{n-1}(d_n-d_p)})$ and
\begin{equation}\label{equ:vandermondegaussianelimiate2}
v_{(n+1)j}=(-1)^{n-j}\sum_{(\tau_1,\cdots,\tau_{n+1-j})\in S_{1n}^{n+1-j}}d_{\tau_1}\cdots d_{\tau_{n+1-j}}.
\end{equation}
\end{prop}	
Proof: See Appendix B.\\

Following Proposition \ref{vandermondegaussianelimiate1}, we have the following  result.  
\begin{lem}\label{vandemondevolumeratio2}
We have 
\begin{equation} \label{eq-deter}
\sqrt{\frac{\det(V_{n}(n)^TV_n(n))}{\det(V_{n-1}(n)^TV_{n-1}(n))}}= \sqrt{\sum_{j=0}^{n}v_{j}^2},
\end{equation}
where $V_{s}(k)$ is defined as in (\ref{eq-Vn}) and $v_{j}=\sum_{(\tau_1,\cdots,\tau_j)\in S_{1n}^j}d_{\tau_1}\cdots d_{\tau_j}$. Especially, if $|d_j|<d,j=1,\cdots,n$, then
\begin{equation}\label{equ:projectiondisestimate1}
\sqrt{\frac{\det(V_{n}(n)^TV_n(n))}{\det(V_{n-1}(n)^TV_{n-1}(n))}}\leq (1+d)^n.
\end{equation}

\end{lem}
Proof: Note that in Proposition \ref{vandermondegaussianelimiate1}, 
all the elementary column-addition matrices have unit determinant. As a result,
$ \det(V_{n}(n)^TV_n(n)) = \det(F^TF) \cdot \frac{1}{(\det D)^2}$, where $F$ is the matrix in the right hand side of (\ref{equ:vandermondegaussianelimiate1}), and $D$ is the diagonal matrix in Proposition \ref{vandermondegaussianelimiate1}. A direct calculation shows that $\det(F^TF)= \sum_{j=0}^{n}v_j^2$ where we use (\ref{equ:vandermondegaussianelimiate2}). On the other hand, $V_{n-1}(n)$ is a standard Vandermonde matrix and we have $\det(V_{n-1}(n)) = \det(V_{n-1}(n)^T) =\frac{1}{\det D}$. Combining these results, 
(\ref{eq-deter}) follows.  The last statement can be derived from (\ref{eq-deter}) and the estimate that 
\[
\sqrt{\sum_{j=0}^{n}v_j^2}\leq \sum_{j=0}^{n}|v_j|\leq \sum_{j=0}^{n}
\begin{pmatrix}
n\\
j
\end{pmatrix}
d^j=(1+d)^n.
\]


\medskip
We finally present two lemmas that are used in establishing the stability result in section \ref{sec-stability}. 
\begin{lem}\label{lem:multiproductlowerbound0}
	Let $k\geq 1$. Assume that $d_1^*<d_2^*<\cdots<d_{k+1}^*$ and let $d_{\min}^*=\min_{i\neq j}|d_i^*-d_j^*|$. Then for real numbers $d_1\leq  d_2 \leq \cdots \leq d_k$, we have the following estimate
	\[
	||\eta_{k+1,k}(d_1^*, \cdots, d_{k+1}^*, d_1, \cdots, d_k)||_{\infty}\geq \xi(k)(d_{\min}^*)^k,
	\]
	where $\eta_{k+1,k}(d_1^*, \cdots, d_{k+1}^*, d_1, \cdots, d_k)$ is defined as (\ref{equ:defineofeta}).	
\end{lem}
Proof: See Appendix A.

\begin{lem}\label{lem:multiproductstability1}
Let $-d\leq d_1^*<d_2^*<\cdots<d_k^*\leq d$ and $d_1, \cdots, d_k \in [-d, d]$. Assume that 
\begin{equation}\label{equ:satblemultiproductlemma1equ1}
||\eta_{k,k}(d_1^*, \cdots, d_k^*, d_1, \cdots, d_k)||_{\infty}< \epsilon, 
\end{equation}
where $\eta_{k,k}(\cdots)$ is defined as in (\ref{equ:defineofeta}), and that 
\begin{equation}\label{equ:satblemultiproductlemma1equ2}
d_{\min}^* = \min_{p\neq j}|d_p^*-d_j^*| \geq \big(\frac{4\epsilon}{\lambda(k)}\big)^{\frac{1}{k}},
\end{equation}
where 
\begin{equation}\label{equ:lambda1}
\lambda(k)=\left\{
\begin{array}{ll}
1,  & k=2,\\
\xi(k-2),& k\geq 3.
\end{array} 
\right.	
\end{equation}
	Then after reordering $d_j$'s, we have
\begin{equation}\label{equ:satblemultiproductlemma1equ4}
	|d_j -d_j^*|< \frac{d_{\min}^*}{2},  \quad j=1,\cdots,k,
\end{equation}
and moreover
\begin{equation}\label{equ:satblemultiproductlemma1equ5}
	|d_j -d_j^*|\leq \frac{2^{k-1}\epsilon}{(k-2)!(d_{\min}^*)^{k-1}}, \quad j=1,\cdots, k.
\end{equation}
\end{lem}
Proof: See Appendix A.

\subsection{Lower bound of the approximation problem (\ref{vandermondeapprox})}\label{sec-vand-lowerbound}
In this section we derive a sharp lower bound for the non-linear approximation problem (\ref{vandermondeapprox}). We first consider the special case when $v$ is a Vandermonde vector. 

\begin{thm}\label{thm:spaceapproxlowerbound0}
Let $k\geq 1$ and $d_1,\cdots, d_{k}\in [-d, d]$ be $k$ distinct real number. Denote $A = \big(\phi_{k}(d_1), \cdots, \phi_k(d_k)\big)$ where $\phi_k(d_j)$'s are defined as in (\ref{equ:defineofphi}). Let $V$ be the $k$-dimensional real space spanned by the column vectors of $A$, and $V^{\perp}$ be the one-dimensional orthogonal complement of $V$ in $\mathbb R^{k+1}$. Let $P_{V^{\perp}}$ be the orthogonal projection onto $V^{\perp}$ in $\mathbb R^{k+1}$, we have 
\[
\min_{a\in \mathbb R^k}||Aa- \phi_{k}(d^*)||_2 = ||P_{V^{\perp}}(\phi_k(d^*))||_2 =|v^{T}\phi_{k}(d^*)|\geq \frac{1}{(1+d)^k}|\Pi_{j=1}^k (d^*- d_j)|,
\]
where $v$ is a unit vector in $V^{\perp}$ and $v^{T}$ is its transpose.
\end{thm}
Proof: By Lemma \ref{lem:vandemondevolumeratio3}, 
\[
\min_{a\in \mathbb R^k}||Aa- \phi_{k}(d^*)||_2 = \sqrt{\frac{\text{det}(D^T D)}{\text{det}(A^TA)}},
\]
where $D=\big(\phi_{k}(d_1),\cdots, \phi_{k}(d_k), \phi_{k}(d^*)\big)$. Denote $\tilde{A} = \big(\phi_{k-1}(d_1),\cdots, \phi_{k-1}(d_k)\big)$. By (\ref{equ:projectiondisestimate1}), we have
\[
\sqrt{\frac{\text{det}(A^TA)}{\text{det}(\tilde{A}^T\tilde{A})}} \leq (1+d)^k.
\]
Therefore, 
\[
\min_{a\in \mathbb R^k}||Aa- \phi_{k}(d^*)||_2 \geq \frac{1}{(1+d)^{k}}\sqrt{\frac{\text{det}(D^TD)}{\text{det}(\tilde{A}^T\tilde{A})}}.
\]
Note that $D$ and $\tilde{A}$ are square Vandermonde matrices. We can use the determinant formula to derive that 
\[
\min_{a\in \mathbb R^k}||Aa- \phi_{k}(d^*)||_2 \geq \frac{1}{(1+d)^{k}} \frac{|\Pi_{1\leq t<p\leq k} (d_t - d_p)\Pi_{q=1}^k(d^*-d_q)|}{|\Pi_{1\leq t<p\leq k} (d_t - d_p)|} = \frac{1}{(1+d)^k}|\Pi_{j=1}^k (d^*- d_j)|.
\]
This completes the proof of the theorem.

\medskip

We now consider the approximation problem (\ref{vandermondeapprox})
for the general case when $v$ is a linear combination of Vandermonde vectors. 

\begin{thm}\label{spaceapproxlowerbound1}
Let $k\geq 1$. Assume $-d \leq d_1^*< d_2^* < \cdots <d_{k+1}^* \leq d$ and $|a_j^*|\geq m_{\min}^*,1\leq j\leq k+1$. Let $d_{\min}^*:=\min_{i\neq j}|d_i^*-d_j^*|$.  For $q\leq k$, let $a(q)=(a_1, a_2, \cdots, a_{q})^T, a^*=(a_1^*, a_2^*, \cdots, a_{k+1}^{*})^T$ and 
\[
 A(q) = \big(\phi_{2k}(d_1),\ \cdots,\ \phi_{2k}(d_q)\big),\  A^* = \big(\phi_{2k}(d_1^*),\ \cdots,\ \phi_{2k}(d_{k+1}^*)\big)
\]
where $\phi_{2k}(z)$ is defined as in (\ref{equ:defineofphi}). Then
\begin{align*}
\min_{a_p,d_p\in \mathbb R, |d_p|\leq d, p=1,\cdots,q}||A(q)a(q)-A^*a^*||_2\geq \frac{\zeta(k+1)\xi(k)m_{\min}^*(d_{\min}^*)^{2k}}{(1+d)^{2k}}.
\end{align*}

\end{thm}
Proof: \textbf{Step 1}. Note that for $q<k$, we have 
\[\min_{a_p,d_p\in \mathbb R, |d_p|\leq d, p=1,\cdots,q}||A(q)a(q)-A^*a^*||_2\geq \min_{a_p,d_p\in \mathbb R, |d_p|\leq d, p=1,\cdots,k}||A(k)a(k)-A^*a^*||_2.\]
So we need only to consider the case when $q=k$. It suffices to show that for any given $-d \leq d_1 < d_2 < \cdots  < d_k \leq d$, the following holds
\begin{equation}\label{equ:spaceapproxlowerboundequ-4}
\min_{a_p\in \mathbb R, p=1,\cdots,k}||A(k)a(k)-A^*a^*||_2\geq \frac{\zeta(k+1)\xi(k)m_{\min}^*(d_{\min}^*)^{2k}}{(1+d)^{2k}}.
\end{equation}
So we fix $d_1,\cdots,d_k$ in our subsequent argument.\\
\textbf{Step 2}. For $l=0, \cdots, k$, we define the following partial matrices 
\[
A_{l}=
\left(\begin{array}{ccc}
	d_1^l&\cdots&d_k^l\\
	d_1^{l+1}&\cdots &d_k^{l+1}\\
	\vdots &\vdots &\vdots\\ 
	d_1^{l+k}&\cdots&d_k^{l+k}
\end{array}
\right),
\quad 
A_{l}^*=
\left(\begin{array}{ccc}
	(d_1^*)^l&\cdots&(d_{k+1}^*)^l\\
	(d_1^*)^{l+1}&\cdots &(d_{k+1}^*)^{l+1}\\
	\vdots &\vdots &\vdots\\ 
	(d_1^*)^{l+k}&\cdots& (d_{k+1}^*)^{l+k}
\end{array}
\right).
\]
It is clear that for all $l$, 
\begin{equation}\label{equ:spaceapproxlowerboundequ-3}
\min_{a\in \mathbb R^{k}}||A_la-A_l^*a^*||_2 \geq \min_{a\in \mathbb R^k}||A_l a- A_l^* a^*||_2.
\end{equation}
\textbf{Step 3}. For each $l$, observe that $A_l = A_0 \text{diag}(d_1^l,\ \cdots,\ d_k^l), A_l^*=A_0^*\text{diag}\left((d_1^*)^l, \cdots, (d_{k+1}^*)^l\right)$, we have
\begin{equation}\label{equ:spaceapproxlowerboundequ-2}
\min_{a\in \mathbb R^{k}}||A_la-A_l^*a^*||_2\geq \min_{\alpha_l\in \mathbb R^{k}}||A_0\alpha_l-A_0^*\alpha_l^*||_2,
\end{equation}
where $\alpha_l^*=\left(a_1^*(d_1^*)^l,\cdots,a_{k+1}(d_{k+1}^*)^l\right)^T$. Let $V$ be the space spanned by column vectors of $A_0$. Then the dimension of $V$ is $k$, and the dimension of $V^\perp$, the orthogonal complement of $V$ in $\mathbb R^{k+1}$, is one.  Let 
$P_{V^{\perp}}$ be the orthogonal projection onto $V^{\perp}$. Note that $||P_{V^{\perp}}u||_2=|v^{T}u|$ for $u\in \mathbb{R}^{k+1}$ where $v$ is a unit vector in $V^{\perp}$ and $v^{T}$ is its transpose. We have
\begin{align}\label{equ:spaceapproxlowerboundequ-1}
\min_{\alpha_l\in \mathbb R^k}||A_0\alpha_l-A_0^*\alpha_l^*||_2=||P_{V^{\perp}}(A_0^*\alpha_l^*)||_2=|v^TA_0^*\alpha_l^* |=\Big|\sum_{j=1}^{k+1}a_j^*(d_j^*)^l v^T \phi_{k}(d_j^*)\Big| = |\beta_l|,
\end{align} 
where
\[
\beta_l = \sum_{j=1}^{k+1}a_j^*(d_j^*)^l v^T \phi_{k}(d_j^*), \quad \text{for $l=0, 1, \cdots, k.$}
\] 
\textbf{Step 4}. Denote $\beta = (\beta_0, \cdots, \beta_k)^T$. We have $B\hat\eta=\beta$, where
\[B=\left(\begin{array}{cccc}
	a_1^*&a_2^*&\cdots&a_{k+1}^*\\
	a_1^*d_1^*&a_2^*d_2^*&\cdots&a_{k+1}^*d_{k+1}^*\\
	\vdots&\vdots&\vdots&\vdots\\
	a_1^*(d_1^*)^{k}&a_2^*(d_2^*)^{k}&\cdots&a_{k+1}^*(d_{k+1}^*)^k
\end{array}\right),\quad \hat \eta=
\left(\begin{array}{c}
v^T \phi_{k}(d_1^*)\\
v^T \phi_{k}(d_2^*)\\
\vdots\\
v^T \phi_{k}(d_{k+1}^*)
\end{array}\right).
\]
By Corollary \ref{norminversevandermonde2}, we have 
\begin{align*}
||\hat \eta||_{\infty}=||B^{-1}\beta||_{\infty}\leq ||B^{-1}||_{\infty}||\beta||_{\infty}\leq \frac{(1+d)^{k}}{ \zeta(k+1)m_{\min}^*(d_{\min}^*)^{k}}||\beta||_{\infty}.
\end{align*}
On the other hand, applying Theorem \ref{thm:spaceapproxlowerbound0} to each term $|v^T \phi_{k}(d_j^*)|$, $j=1, 2, \cdots k+1$,  we have 
\[
||\hat \eta||_{\infty}\geq \frac{1}{(1+d)^k} ||\eta_{k+1, k}(d_1^*, \cdots, d_{k+1}^*, d_1, \cdots, d_{k} )||_{\infty},
\]
where $\eta_{k+1, k}(\cdots)$ is defined as in (\ref{equ:defineofeta}). Combing this inquality with Lemma \ref{lem:multiproductlowerbound0}, we get 
\[
||\hat \eta||_{\infty}\geq \frac{\xi(k)(d_{\min}^*)^{k}}{(1+d)^k}.
\]
It follows that 
\[
||\beta||_{\infty} \geq \frac{\zeta(k+1)\xi(k)m_{\min}^*(d_{\min}^*)^{2k}}{(1+d)^{2k}}.
\]
Therefore, recalling (\ref{equ:spaceapproxlowerboundequ-3})--(\ref{equ:spaceapproxlowerboundequ-1}), we have
\[
\min_{a(k)\in \mathbb R^{k}}||A(k)a(k)-A^*a^*||_2\geq \max_{0\leq l\leq k} \min_{a\in \mathbb R^k}||A_la-A_l^*a^*||_2= \max_{0\leq l\leq k} |\beta_l|=||\beta||_{\infty}\geq \frac{\zeta(k+1)\xi(k)m_{\min}^*(d_{\min}^*)^{2k}}{(1+d)^{2k}}.
\]
This proves (\ref{equ:spaceapproxlowerboundequ-4}) and hence the theorem.

\medskip
We next demonstrate the sharpness of the lower bound above.
\begin{prop}
Let $k\geq 1$. There exist $-d \leq d_1^*< d_2^* < \cdots <d_{k+1}^* \leq d$ and nonzero real numbers $a_1^*,\cdots,a_{k+1}^*$ with $\min_{1\leq j\leq k+1}|a_j^*|=m_{\min}^*$ such that 
\begin{align*}
\min_{a_p,d_p\in \mathbb R, |d_p|\leq d, p=1,\cdots,k}||A(k)a(k)-A^*a^*||_2< (k+1)(2k)^{2k+1}m_{\min}^*(d_{\min}^*)^{2k},
\end{align*}
where $A(k),a(k),A^*,a^*,d_{\min}^*$ are defined in Theorem \ref{spaceapproxlowerbound1}. 
\end{prop}
Proof:
Let $\hat d_1=-d,\ \hat d_2=-d+\frac{d}{k},\ \cdots,\ \hat d_{2k}=d-\frac{d}{k},\ \hat d_{2k+1}=d$. There exists nonzero $\hat a$ such that  
\begin{equation}\label{equ:sharpnessapproximation1}
V_{2k}(2k+1)\hat a=\gamma,
\end{equation}
where $V_{2k}(2k+1)= \big(\phi_{2k}(\hat d_1), \cdots, \phi_{2k}(\hat d_{2k+1})\big)$, $\hat a = (\hat a_1, \hat a_2, \cdots, \hat a_{2k+1})^T$ and 
\begin{equation*}
\gamma=(0, \cdots, 0, \sum_{j=1}^{2k+1}\hat a_j (\hat d_j)^{2k})^T.
\end{equation*}
Using the linear independence of the column vectors in the matrix $V_{2k}(2k+1)$, we can show that all $\hat a_j$'s are nonzero. Denote $\min_{1\leq j\leq 2k+1}|\hat a_j| = m_{\min}^*$. We define
\[
\begin{cases}
d_1^*=\hat d_1,\ \cdots,\ d_{k+1}^*=\hat d_{k+1},\ a_1^*=\hat a_1,\ \cdots,\ a_{k+1}^*=\hat a_{k+1}, \quad \mbox{if}\,\,\,  \min_{1\leq j\leq n}|a_{j}|=m_{\min},\\
d_1^*=\hat d_{k+1},\ \cdots,\ d_{k+1}^*=\hat d_{2k+1},\  a_1^*=\hat a_{k+1},\ \cdots,\ a_{k+1}^*=\hat a_{2k+1}, \quad  \mbox{otherwise}.
\end{cases}
\]
On the other hand, we let $V_{2k-1}(2k+1)= \big(\phi_{2k-1}(\hat d_1), \cdots, \phi_{2k-1}(\hat d_{2k+1})\big)$. Then (\ref{equ:sharpnessapproximation1}) implies that $V_{2k-1}(2k+1) \hat a = 0$. Following the same argument in Step 2 of proof of Proposition \ref{thm:numberlowerboundthm0}, we can derive that
\[
\sum_{j}^{2k+1}|\hat a_j|\leq (2k+1)k2^{2k}m_{\min}^*.
\]
Note that $d_{\min}^*=\min_{i\neq j}|d_i^*-d_j^*| \geq\frac{d}{k}$.
We have 
\begin{align*}
&||A^*a^*-A(k)a(k)||_2=||V_{2k}(2k+1)\hat a||_2=||\gamma||_2=\Big|\sum_{j=1}^{2k+1}\hat a_j (\hat d_j)^{2k}\Big|\\
\leq& (2k+1)k2^{2k}m_{\min}^*d^{2k}\leq (2k+1)k2^{2k}m_{\min}^*(kd_{\min}^*)^{2k}< (k+1)(2k)^{2k+1}m_{\min}^*(d_{\min}^*)^{2k}.
\end{align*}
It follows that
\begin{align*}
	\min_{a_p,d_p\in \mathbb R, |d_p|\leq d, p=1,\cdots,k}||A(k)a(k)-A^*a^*||_2< (k+1)(2k)^{2k+1}m_{\min}^*(d_{\min}^*)^{2k}.
\end{align*}


\medskip
Finally, as a consequence of Theorem \ref{spaceapproxlowerbound1}, we have the following corollary. 

\begin{cor}\label{spaceapproxlowerbound2}
Let $k\geq 1$. Assume that $-d \leq d_1^*< d_2^* < \cdots <d_{k+1}^* \leq d$ and $a_1^*,\cdots,a_{k+1}^*\in \mathbb R$ with $|a_j^*|\geq m_{\min}^*,j=1,\cdots,k+1$. For $q\leq k$, we have 
\begin{align*}
\min_{a_p,d_p\in \mathbb R, |d_p|\leq d, p=1,\cdots,q}||\tilde{A}(q)a(q)-\tilde{A}^*a^*||_2\geq  \frac{1.15m_{\min}^*(d_{\min}^*)^{2k}}{2^{4k}k(1+d)^{2k}},
\end{align*}
where $d_{\min}^*:=\min_{i\neq j}|d_i^*-d_j^*|$ and 
\begin{equation}\label{equ:spaceapproxlowerbound2equ1}
\tilde{A}(q)=\text{diag}(1,\frac{1}{\sqrt{3}},\cdots, \frac{1}{(2k)!\sqrt{4k+1}})A(q), \quad \tilde{A}^*=\text{diag}(1,\frac{1}{\sqrt{3}},\cdots, \frac{1}{(2k)!\sqrt{4k+1}}) A^*
\end{equation}
with $A(q),a(q),A^*,a^*$ being defined in Theorem \ref{spaceapproxlowerbound1}.
\end{cor}
Proof: By (\ref{equ:spaceapproxlowerbound2equ1}), we have 
 \begin{align*}
 \min_{a_p,d_p\in \mathbb R, |d_p|\leq d, p=1,\cdots,q}||\tilde{A}(q)a(q)-\tilde{A}^*a^*||_2&\geq\frac{1}{(2k)!\sqrt{4k+1}}\min_{a_p,d_p\in \mathbb R, |d_p|\leq d, p=1,\cdots,q}||A(q)a(q)-A^*a^*||_2\\
 &\geq  \frac{\zeta(k+1)\xi(k)m_{\min}^*(d_{\min}^*)^{2k}}{(1+d)^{2k}(2k)!\sqrt{4k+1}}, \quad \Big(\text{by Theorem \ref{spaceapproxlowerbound1}}\Big)
 \end{align*}
 where $\zeta(k+1), \xi(k)$ is defined in (\ref{equ:zetaxiformula1}). Combining this with Lemma \ref{spaceapproxcalculate1} yields the corollary.

\subsection{Stability of the approximation problem (\ref{vandermondeapprox})}\label{sec-vand-stability}
In the section we present stability results for the approximation problem (\ref{vandermondeapprox}).

\begin{thm}\label{spaceapproxlowerbound3}
Let $k\geq 2$. Assume that $-d \leq d_1^*< d_2^* < \cdots <d_{k}^* \leq d$ and $|a_j^*|\geq m_{\min}^*,1\leq j\leq k$. Let $d_{\min}^*:=\min_{i\neq j}|d_i^*-d_j^*|$. Assume that $-d \leq d_1 <d_2 <\cdots <d_k \leq d$ satisfy
\[
||Aa-A^*a^*||_2< \sigma, 
\]
where $a=(a_1,\cdots, a_k)^T$, $a^* = (a_1^*,\cdots, a_k^*)^T$ and 
\[
A = \big(\phi_{2k-1}(d_1),\ \cdots,\ \phi_{2k-1}(d_k)\big),\  A^* = \big(\phi_{2k-1}(d_1^*),\ \cdots,\ \phi_{2k-1}(d_{k}^*)\big).
\]
Then 
\[
\Big|\Big|\eta_{k,k}(d_1^*,\cdots, d_k^*, d_1, \cdots, d_k)\Big|\Big|_{\infty}<\frac{(1+d)^{2k-1}}{ \zeta(k)(d_{\min}^*)^{k-1}}\frac{\sigma}{m_{\min}^*}.
\]
\end{thm}
Proof: Since $||Aa-A^*a^*||_2<\sigma$, we have 
\begin{equation*}\label{spaceapproxlowerbound3equ0}
\min_{\alpha \in \mathbb R^k}||A\alpha-A^*a^*||_2<\sigma. 
\end{equation*}
and hence
\begin{equation}\label{spaceapproxlowerbound3equ1}
\max_{0\leq l\leq k-1}\min_{\alpha \in \mathbb R^k}||A_l \alpha -A_l^*a^*||_2\leq \min_{\alpha \in \mathbb R^k}||A\alpha-A^*a^*||_2<\sigma,
\end{equation}
where 
\[
A_{l}=
\left(\begin{array}{ccc}
d_1^l&\cdots&d_k^l\\
d_1^{l+1}&\cdots &d_k^{l+1}\\
\vdots &\vdots &\vdots\\ 
d_1^{l+k}&\cdots&d_k^{l+k}
\end{array}
\right),
\quad 
A_{l}^*=
\left(\begin{array}{ccc}
(d_1^*)^l&\cdots&(d_{k}^*)^l\\
(d_1^*)^{l+1}&\cdots &(d_{k}^*)^{l+1}\\
\vdots &\vdots &\vdots\\ 
(d_1^*)^{l+k}&\cdots& (d_{k}^*)^{l+k}
\end{array}
\right).
\]
For each $l$, from the decomposition $A_l = A_0 \text{diag}(d_1^l,\ \cdots,\ d_k^l), A_l^*=A_0^*\text{diag}((d_1^*)^l, \cdots, (d_{k}^*)^l)$, we have
\begin{equation}\label{spaceapproxlowerbound3equ2}
\min_{\alpha \in \mathbb R^{k}}||A_l\alpha-A_l^*a^*||_2\geq \min_{\alpha_l\in \mathbb R^{k}}||A_0\alpha_l-A_0^*\alpha_l^*||_2,
\end{equation}
where $\alpha_l^*=(a_1^*(d_1^*)^l,\cdots,a_{k}(d_{k}^*)^l)^T$. Let $V$ be the space spanned by column vectors of $A_0$. Then the dimension of $V$ is $k$, and $V^\perp$, the orthogonal complement of $V$ in $\mathbb R^{k+1}$ is of dimension one. We let $v$ be a unit vector in $V^{\perp}$ and let 
$P_{V^{\perp}}$ be the orthogonal projection onto $V^{\perp}$. Similar to (\ref{equ:spaceapproxlowerboundequ-1}), we have
\begin{align}\label{spaceapproxlowerbound3equ3}
\min_{\alpha_l\in \mathbb R^k}||A_0\alpha_l-A_0^*\alpha_l^*||_2=||P_{V^{\perp}}(A_0^*\alpha_l^*)||_2=|v^TA_0^* \alpha_l^*|=\Big|\sum_{j=1}^{k}a_j^*(d_j^*)^lv^T\phi_{k}(d_j^*)\Big|=|\beta_l|,
\end{align} 
where $\beta_l = \sum_{j=1}^{k}a_j^*(d_j^*)^lv^T\phi_{k}(d_j^*)$.  Let $\beta = (\beta_0,\cdots, \beta_{k-1})^T$. Similar to Step 4 in the proof of Theorem \ref{spaceapproxlowerbound1}, we have 
\[
\Big|\Big|\eta_{k,k}(d_1^*,\cdots, d_k^*, d_1, \cdots, d_k)\Big|\Big|_{\infty} \leq \frac{(1+d)^{2k-1}}{\zeta(k)m_{\min}^*(d_{\min}^*)^{k-1}} ||\beta||_{\infty}.
\]
On the other hand, Equations (\ref{spaceapproxlowerbound3equ1})-(\ref{spaceapproxlowerbound3equ3}) indicate that $||\beta||_{\infty}<\sigma$. Hence we have 
\[
\Big|\Big|\eta_{k,k}(d_1^*,\cdots, d_k^*, d_1, \cdots, d_k)\Big|\Big|_{\infty} \leq \frac{(1+d)^{2k-1}}{\zeta(k)(d_{\min}^*)^{k-1}}\frac{\sigma}{m_{\min}^*}.
\]

\begin{cor}\label{spaceapproxlowerbound4}
Let $k\geq 2$. Assume that $-d \leq d_1^*< d_2^* < \cdots <d_{k}^* \leq d$ and $|a_j^*|\geq m_{\min}^*,j=1,\cdots,k$. Let $-d \leq d_1 <d_2 <\cdots <d_k \leq d$ and assume that 
\begin{equation}
||\tilde{A}a-\tilde{A^*}a^*||_2< \sigma, 
\end{equation}
where 
\begin{equation}\label{equ:spaceapproxlowerbound4equ1}
	\tilde{A}=\text{diag}(1,\frac{1}{\sqrt{3}},\cdots, \frac{1}{(2k-1)!\sqrt{4k-1}})A, \quad \tilde{A}^*=\text{diag}(1,\frac{1}{\sqrt{3}},\cdots, \frac{1}{(2k-1)!\sqrt{4k-1}}) A^*
\end{equation}
with $A, a, A^*, a^*$ being defined as in Theorem \ref{spaceapproxlowerbound3}. Then 
\[
\Big|\Big|\eta_{k,k}(d_1^*,\cdots, d_k^*, d_1, \cdots, d_k)\Big|\Big|_{\infty}<\frac{(2k-1)!\sqrt{4k-1}(1+d)^{2k-1}}{ \zeta(k)(d_{\min}^*)^{k-1}}\frac{\sigma}{m_{\min}^*},
\]
with $\eta_{k,k}(d_1^*,\cdots, d_k^*, d_1, \cdots, d_k), d_{\min}^*,\zeta(k)$ being defined as in Theorem \ref{spaceapproxlowerbound3}.
\end{cor}
Proof: By Theorem \ref{spaceapproxlowerbound3} and a similar argument as in the proof of Corollary \ref{spaceapproxlowerbound2}.

\section{A Singular-value-thresholding algorithm for number detection}\label{section:numberdetectionalgorithm}
In this section, we propose a singular-value-thresholding algorithm to detect the source number for a cluster of closely spaced point sources. We shall show that the algorithm can detect the source number correctly in the regime where the minimum separation distance is comparable to the upper bound of the computational resolution limit $\mathcal{D}_{num}$. 

We note that previous number detection algorithms usually deal with  multiple measurements (multiple $\mathbf Y$'s) with random noise of known statistical properties. Here we deal with a single measurement with a deterministic noise, see (\ref{introductionequ-1}). In general, in a statistical setting, there are two approaches to detect the source number (or model order). One approach selects the model which includes the model order using some generic information theoretic criteria by minimizing the summation of a log-likelihood function and a regularization term for the free parameters in the model, see for instance AIC \cite{akaike1998information, akaike1974new, wax1985detection}, BIC/MDL \cite{schwarz1978estimating, rissanen1978modeling, wax1989detection}. The other approach determines the model order by thresholding the eigenvalues of the covariance matrix of the data (the so-called eigen-thresholding method), see for instance \cite{lawley1956tests, chen1991detection, he2010detecting, han2013improved}. We shall follow the idea of eigen-thresholding to develop an algorithm to detect the source number. More precisely, we form a Hankel matrix from the multipole coefficients that are recovered from the measurement and derive a deterministic threshold to its singular values to estimate the source number (see Corollary \ref{MUSICnumberthm1}).

\subsection{Recovery of multipole coefficients}
We first recover the multipole coefficients. For given $d,\sigma,M\geq m^*$, we define $s^*$ as in (\ref{polesrecovered2}) and write the measurement as 
\begin{equation}\label{equ:measurementexpansion3}
\begin{aligned}
\begin{pmatrix}
\mathbf Y(x_1) \\
\vdots\\
\mathbf Y(x_N)
\end{pmatrix}
&= 
\begin{pmatrix}
f_{\Omega}(x_1-y_1^*) &\cdots& f_{\Omega}(x_1-y_n^*)\\
\vdots&\vdots&\vdots\\
f_{\Omega}(x_N-y_1^*)&\cdots& f_{\Omega}(x_N-y_n^*)\\
\end{pmatrix}
\begin{pmatrix}
a_{1}^*\\
\vdots\\
a_{n}^*
\end{pmatrix}
+
\sqrt{\Omega}
\begin{pmatrix}
\mathbf W(x_1)\\
\vdots\\
\mathbf W(x_N)\\
\end{pmatrix}\\
&=\sqrt{\Omega} \begin{pmatrix}
\mathbf h_0&\mathbf h_1& \cdots&\mathbf h_{s^*-1}
\end{pmatrix}
\begin{pmatrix}
c_{0}^*\\
c_{1}^*\\
\vdots\\
c_{s^*-1}^{*}
\end{pmatrix}
+\sqrt{\Omega}\begin{pmatrix}
\mathbf W(x_1)\\
\vdots\\
\mathbf W(x_N)\\
\end{pmatrix}
+\sqrt{\Omega}\mathbf {Res},
\end{aligned}
\end{equation}
where $c_{r}^*=\sum_{j=1}^{n}a_{j}^*\frac{(d_{j}^*)^r}{r!\sqrt{2r+1}}$ with $d_j^* = \Omega (-y_j^*)$, and $\mathbf {Res}$ is the residual term. We have 
\begin{equation}\label{reconalgorithmequ0}
\begin{aligned}
\sqrt{h}\Big|\Big|\mathbf {Res}\Big|\Big|_2&=\sqrt{h}\Big|\Big|\sum_{r=s^*}^{+\infty}\sum_{j=1}^{n}\frac{a_j^*(d_j^*)^r}{r!\sqrt{2r+1}}\mathbf h_{r}\Big|\Big|_2\leq \sum_{r=s^*}^{+\infty}\sum_{j=1}^{n}\frac{|a_j^*|d^r}{r!\sqrt{2r+1}}\sqrt{h}\Big|\Big|\mathbf h_{r}\Big|\Big|_2\\
&\leq  m^*\sum_{r=s^*}^{+\infty}\frac{d^r}{r!\sqrt{2r+1}}\sqrt{\pi}\qquad \Big(\text{by (\ref{2normofmultipole})} \Big)\\
&\leq m^*\sqrt{\pi}\frac{d^{s^*}}{s^*!\sqrt{2s^*+1}}\sum_{r=0}^{+\infty}\frac{d^r}{r!}=\frac{e^dm^*\sqrt{\pi}d^{s^*}}{s^*!\sqrt{2s^*+1}}
\leq\sigma. \quad \Big(\text{by (\ref{polesrecovered2})}\Big)
\end{aligned}
\end{equation}
We find the multipole coefficients by solving the following linear system  
\begin{equation}\label{equ:measurementexpansion4}
\left (
\begin{array}{c}
\mathbf Y(x_1) \\
\vdots\\
\mathbf Y(x_N)
\end{array}
\right)
= 
\sqrt{\Omega}
\begin{pmatrix}
\mathbf h_0&\mathbf h_1&\cdots&\mathbf h_{s^*-1}
\end{pmatrix}
\begin{pmatrix}
c_{0}\\
\vdots\\
c_{s^*-1}
\end{pmatrix}
.
\end{equation}
Recall that $\mathbf H(s^*)=\big(\mathbf h_{0},\ \mathbf h_{2},\ \cdots,\ \mathbf h_{s^*-1}
\big)$. Denote $(c_0^*,c_1^*,\cdots,c_{s^*-1}^*)^T=\theta^*, (c_{0},c_1,\cdots,c_{s^*-1})^T= \theta$. By (\ref{equ:measurementexpansion3}) and (\ref{equ:measurementexpansion4}), we have
\begin{align*}
&\mathbf H(s^*)(\theta-\theta^*)=\mathbf W + \mathbf {Res}\\
\implies &\sqrt{h}||\mathbf H(s^*)(\theta-\theta^*)||_2\leq \sqrt{h}||\mathbf W||_2+\sqrt{h}||\mathbf {Res}||_2\\
\implies & \sqrt{h}||\mathbf H(s^*)(\theta-\theta^*)||_2\leq 2\sigma. \qquad \Big(\text{by  (\ref{reconalgorithmequ0})}\Big)
\end{align*}
On the other hand, recall that $\sigma_{\min}(s^*)$ is the minimum singular value of the matrix $\sqrt{h}\mathbf H(s^*)$. We have $\sqrt{h}|| \mathbf H(s^*)(\theta-\theta^*)||_2\geq \sigma_{\min}(s^*)||\theta-\theta^*||_2$. It follows that
\begin{equation}\label{reconalgorithmequ2}
||\theta-\theta^*||_{2}\leq \frac{2\sigma}{\sigma_{\min}(s^*)}. \end{equation}
This gives a bound for the matching error of multipole coefficients.

\subsection{Determination of the source number}
In this section, we determine the source number from the recovered multipole coefficients. Note that for $n$ sources, we have $n$ positions and $n$ amplitudes to recover. But our available data is the $s^*$ multipole coefficients. So we assume that the source number $n$ satisfies the following condition
$$n\leq \frac{s^*-1}{2}.$$
Note that in general $2n$ multipole coefficients are enough to uniquely determine the $n$ sources. We shall only use the first $s$ multipole coefficients with $2n+1\leq s \leq s^*$ since the other higher-order multipole coefficients are less reliable.
From (\ref{reconalgorithmequ2}), we have the following equations for the  first $s$ multipole coefficients:
\begin{align}\label{reconalgorithmequ3}c_r=c_r^*+\delta_{r}, \quad r=0,\cdots,s-1,
\end{align}
where $\delta_r$ is the perturbation caused by noise. We need to recover the source number $n$ from (\ref{reconalgorithmequ3}). 
We rewrite (\ref{reconalgorithmequ3}) as 
\[r!\sqrt{2r+1}c_r=\sum_{j=1}^{n}a_{j}^*(d_{j}^*)^r+r!\sqrt{2r+1}\delta_{r} \quad \text{for $r=0,\cdots,s-1$}.\]
For convenience, we assume that $s$ is odd and form the following data matrix 
\[	X=\left (
\begin{array}{cccc}
c_0 &\sqrt{3}c_1&\cdots& (\frac{s-1}{2})!\sqrt{s}c_{\frac{s-1}{2}}\\
\sqrt{3}c_1 &2!\sqrt{5}c_2&\cdots& (\frac{s+1}{2})!\sqrt{s+2}c_{\frac{s+1}{2}}\\
\vdots&\vdots&\vdots&\vdots\\
(\frac{s-1}{2})!\sqrt{s}c_{\frac{s-1}{2}}&(\frac{s+1}{2})!\sqrt{s+2}c_{\frac{s+1}{2}}&\cdots&(s-1)!\sqrt{2s-1}c_{s-1}
\end{array}
\right). \]
We observe that the data matrix $X$ has the following decomposition 
\begin{equation}\label{MUSICnumber1equ0}
X=DAD^T+\Delta, 
\end{equation}
where $A=\text{diag}(a_1^*, \cdots, a_n^*)$, $D= \big(\phi_{\frac{s-1}{2}}(d_1^*), \cdots, \phi_{\frac{s-1}{2}}(d_n^*)\big)$ with $\phi_{\frac{s-1}{2}}(\cdot)$ being defined by (\ref{equ:defineofphi}), and 
\[ 
\Delta = \left (
\begin{array}{cccc}
\delta_0 &\sqrt{3}\delta_1&\cdots& (\frac{s-1}{2})!\sqrt{s}\delta_{\frac{s-1}{2}}\\
\sqrt{3}\delta_1 &2!\sqrt{5}\delta_2&\cdots& (\frac{s+1}{2})!\sqrt{s+2}\delta_{\frac{s+1}{2}}\\
\vdots&\vdots&\vdots&\vdots\\
(\frac{s-1}{2})!\sqrt{s}\delta_{\frac{s-1}{2}}&(\frac{s+1}{2})!\sqrt{s+2}\delta_{\frac{s+1}{2}}&\cdots&(s-1)!\sqrt{2s-1}\delta_{s-1}
\end{array}
\right).
\]
Recall that $s\geq 2n+1$, so $\frac{s-1}{2}+1\geq n+1$. We denote the singular value decomposition of $X$ as 
\[X=\hat U\hat \Sigma \hat U^*,\]
where 
$\hat \Sigma=\text{diag}(\hat \sigma_1,\hat \sigma_2,\cdots,\hat \sigma_n,\hat \sigma_{n+1},\cdots,\hat \sigma_{\frac{s+1}{2}})$ with the singular values $\hat \sigma_j$'s ordered in a decreasing manner. 
Note that $X=DAD^T$ if there is no noise. We have the following estimate for the singular values of $DAD^T$. 

\begin{thm}\label{MUSICnumberthm0}
Let $n\geq 2$ and let $U\Sigma U^*$ be the singular value decomposition of the matrix $DAD^T$. 
Let $\Sigma =\text{diag}(\sigma_1,\cdots,\sigma_n,0,\cdots,0)$.
Then the following estimate holds
\begin{align*}
\sigma_n \geq\frac{m_{\min}^*\zeta(n)^2(\Omega d_{\min}^*)^{2n-2}}{n(1+d)^{2n-2}},
\end{align*}
where $\Omega d_{\min}^*= \min_{p\neq j} |d_p^*-d_j^*|$ and $\zeta(n)$ is defined in (\ref{equ:zetaxiformula1}).
\end{thm}
Proof: Note that $\sigma_n$ is the minimum nonzero singular value of the matrix $DAD^T$. Let $S(D^T)$ be the kernel space of $D^T$ and $S^{\perp}(D^T)$ be its orthogonal complement, we have 
\begin{align*}
\sigma_n=\min_{||x||_2=1,x\in S^{\perp}(D^T)}||DAD^Tx||_2\geq \sigma_{\min}(DA)\sigma_n(D^T)\geq \sigma_{\min}(D)\sigma_{\min}(A)\sigma_{\min}(D).
\end{align*}
Since $s\geq 2n+1$, by Lemma \ref{singularvaluevandermonde2} and Corollary \ref{norminversevandermonde2}, we have
\begin{align*}
\sigma_{\min}(D)\geq \frac{1}{\sqrt{n}}\frac{\zeta(n)(\Omega d_{\min}^*)^{n-1}}{(1+d)^{n-1}}.
\end{align*}
Thus  
\[
\sigma_n\geq \sigma_{\min}(A)\Big(\frac{1}{\sqrt{n}}\frac{\zeta(n)(\Omega d_{\min}^*)^{n-1}}{(1+d)^{n-1}}\Big)^2.
\]
On the other hand, noting that $\sigma_{\min}(A)=m_{\min}^*$, we have 
\[
\sigma_n\geq \frac{m_{\min}^*\zeta(n)^2(\Omega d_{\min}^*)^{2n-2}}{n(1+d)^{2n-2}},
\]
which completes the proof of the theorem. 

\medskip

Theorem \ref{MUSICnumberthm0} illustrates that the minimum singular value of $DAD^T$ will increase as the minimum separation distance increases. With presence of noise, $X=DAD^T+\Delta$. For a given noise level, 
we expect to be able to distinguish the singular values that are associated with the real sources and those with the noise when the separation distance of sources are large enough. Precisely, we have the following result.
\begin{cor}\label{MUSICnumberthm1}
Let $n\geq 2$ and let $s^*$ be defined in (\ref{polesrecovered2}). Assume that $2n+1\leq s \leq s^*$ and that the minimum separation distance of the measure $\mu^*=\sum_{j=1}^{n}a_j^*\delta_{y_j^*}$ satisfies the following condition
\begin{equation}\label{MUSICseparationcondition1}
\min_{p\neq j}\babs{y_p^*-y_j^*}> \frac{(1+d)}{\Omega} \Big(\frac{4\pi n(s-1)!\sqrt{2s-1}}{\sqrt{6}\zeta(n)^2}\Big)^{\frac{1}{2n-2}}\Big(\frac{\sigma}{ \sigma_{\min}(s^*)m_{\min}^*}\Big)^{\frac{1}{2n-2}}.
\end{equation}
Then the following estimate on the singular values of the data matrix $X$ holds 
\[\hat \sigma_n> \frac{2\pi(s-1)!\sqrt{2s-1}\sigma}{\sqrt{6}\sigma_{\min}(s^*)},\quad \hat \sigma_{j}\leq \frac{2\pi (s-1)!\sqrt{2s-1}\sigma}{\sqrt{6}\sigma_{\min}(s^*)},\quad j=n+1,\cdots,\frac{s+1}{2}.\]
\end{cor}
Proof: First, by (\ref{reconalgorithmequ2}), we have
\begin{align*}
&||\Delta||_2\leq  ||\Delta||_F\leq \sqrt{\Big([(s-1)!]^2(2s-1)+[(s-2)!]^2(2s-3)+\cdots+[(\frac{s-1}{2})!]^2s\Big)(\frac{2\sigma}{\sigma_{\min}(s^*)})^2}\\
=&(s-1)!\sqrt{2s-1}\frac{2\sigma}{\sigma_{\min}(s^*)}\sqrt{1+\frac{2s-3}{(s-1)^2(2s-1)}+\cdots} \leq \frac{2\pi(s-1)!\sqrt{2s-1}\sigma}{\sqrt{6}\sigma_{\min}(s^*)}. \quad  \Big(\text{using $\sum_{j=1}^{+\infty}\frac{1}{j^2}= \frac{\pi^2}{6}$}\Big)
\end{align*}
Let $d_{\min}^* = \min_{p\neq j}|y_p^*-y_j^*|$. By Theorem \ref{MUSICnumberthm0} and the separation condition (\ref{MUSICseparationcondition1}), we have 
\begin{align*}
\sigma_n \geq\frac{m_{\min}^*\zeta(n)^2(\Omega d_{\min}^*)^{2n-2}}{n(1+d)^{2n-2}}>\frac{4\pi(s-1)!\sqrt{2s-1}\sigma}{\sqrt{6}\sigma_{\min}(s^*)}\geq 2||\Delta||_2.
\end{align*}
On the other hand, Weyl's theorem implies that $|\hat \sigma_n-\sigma_n|\leq ||\Delta||_2$. Thus 
\[\hat \sigma_n>||\Delta||_2\geq \frac{2\pi (s-1)!\sqrt{2s-1}\sigma}{\sqrt{6}\sigma_{\min}(s^*)}.\]
In the same fashion, we have that for $j=n+1,\cdots,\frac{s+1}{2}$, 
\[
|\hat \sigma_j|\leq ||\Delta||_2\leq \frac{2\pi (s-1)!\sqrt{2s-1}\sigma}{\sqrt{6}\sigma_{\min}(s^*)}.
\]
This completes the proof. 

\medskip

We note that the minimum separation distance in Corollary \ref{MUSICnumberthm1} is comparable to the upper bound we derived in Theorem \ref{necessarythm1}. Indeed, for $s=2n+1$, the required separation distance is 
\[
\frac{(1+d)}{\Omega} \Big(\frac{4\pi n(2n)!\sqrt{4n+1}}{\sqrt{6}\zeta(n)^2})^{\frac{1}{2n-2}}\Big(\frac{\sigma}{ \sigma_{\min}(s^*)m_{\min}^*}\Big)^{\frac{1}{2n-2}}
\leq \frac{4.75(1+d)}{\Omega}\Big(\frac{(n+1)^{3.2}}{\sigma_{\min}(s^*)}\frac{\sigma}{m_{\min}^*}\Big)^{\frac{1}{2n-2}}.
\]
In comparison, the upper bound in Theorem \ref{necessarythm1} is $
\frac{4.7(1+d)}{\Omega}\Big(\frac{3}{\sigma_{\min}(s^*)}\frac{\sigma}{m_{\min}^*}\Big)^{\frac{1}{2n-2}}$.
Under the separation condition, the lower bound of the first $n$ singular values $\frac{2\pi(s-1)!\sqrt{2s-1}\sigma}{\sqrt{6}\sigma_{\min}(s^*)}$ gives a natural threshold to differentiate the singular values that are associated with the real sources and those with noise. 

\medskip
In conclusion, our singular-value-thresholding algorithm can be summarized as follows. First, we choose a priori estimate of $(d, \sigma, M)$ for the unknown measure, and compute $s^*$ which is the number of multiple coefficients that can be stably recovered from the given image. Second, we solve a linear system to get the first $s^*$ multipole coefficients; Third, we choose $s$ multipole coefficients to form the data matrix and compute its singular values. Finally, we determine the source number by counting the number of singular values that exceed the threshold in Corollary \ref{MUSICnumberthm1}. Numerical examples of this algorithm will be given in next section. 
\section{Numerical experiments}
We perform numerical experiments using our singular-value-thresholding algorithm in this section. For ease of explanation, we set the cut-off frequency $\Omega = 1$ and the corresponding point spread is $f(x)=\frac{\sin x}{x}$.

\vspace{0.3cm}
\noindent\textbf{Experiment 1:} 
We consider the recovery of $3$ sources and the discrete measure 
\[\mu^*=\delta_{y_1^*}+\delta_{y_2^*}+\delta_{y_3^*}\]
where $y_1^*=0.4, y_2^*=0,y_3^*=-0.4$. Then the source number $n=3$, $m^*=3, m_{\min}^*=1$.  
We set $d=0.5, \sigma=1.38\times 10^{-9}$ and $M=4$. We sample the image evenly in $[-100,100]$ with $101$ sample points (with $R=100, h=2$) as follows:
\begin{equation}\label{equ:algorithmexample1equ1}
\mathbf Y(x_t)=\mu^**f(x_t)+\mathbf W(x_t),\qquad t=1,\cdots,101,
\end{equation}
where $\mathbf W(x_t)$'s are uniformly distributed random numbers in $(0,\frac{\sigma}{\sqrt{2R}})$.

\begin{figure}
	\centering
	\begin{subfigure}[b]{0.47\textwidth}
		\centering
		\includegraphics[width=\textwidth]{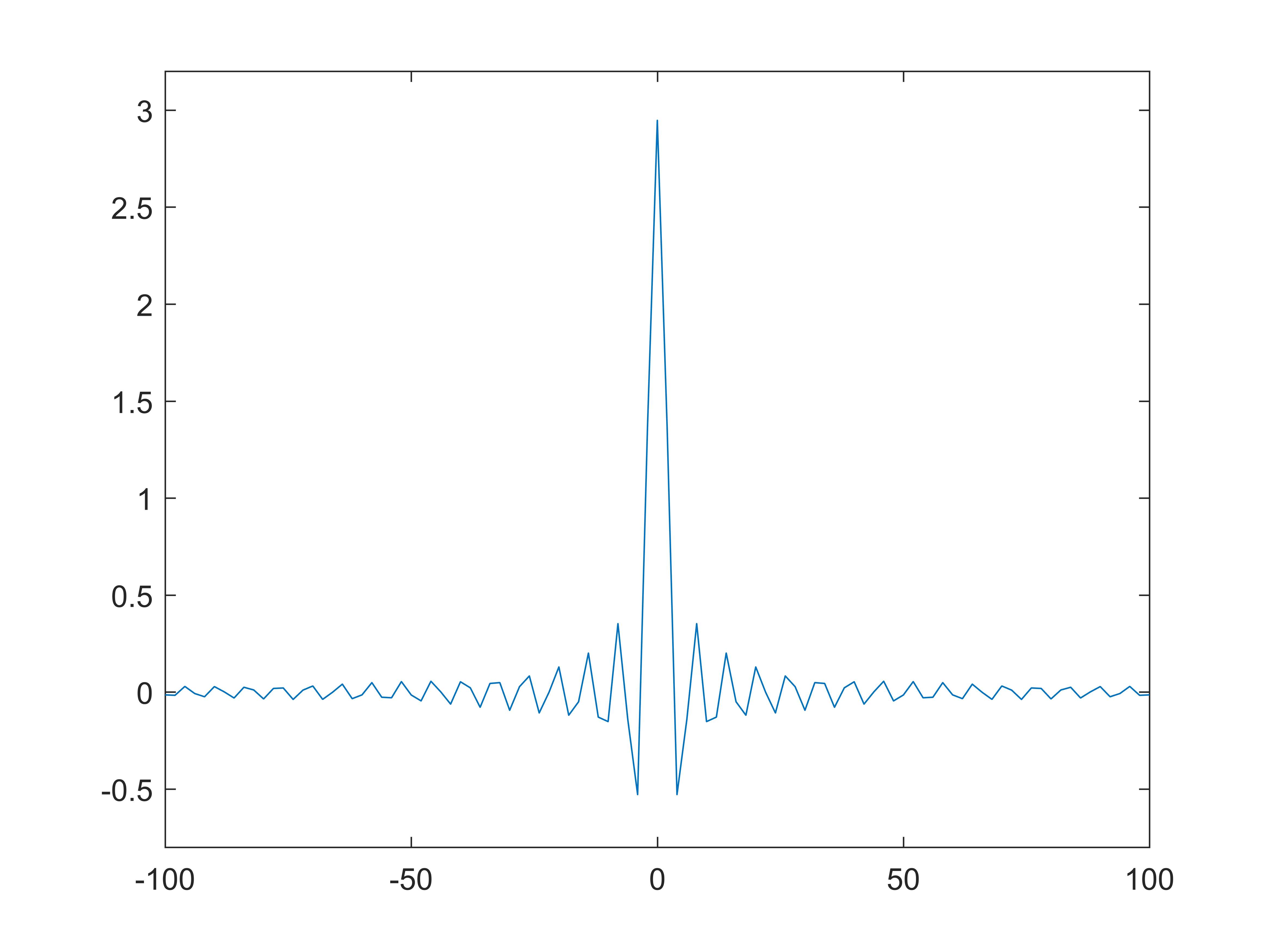}
		\caption{}
	\end{subfigure}
	\hfill
	\begin{subfigure}[b]{0.47\textwidth}
		\centering
		\includegraphics[width=\textwidth]{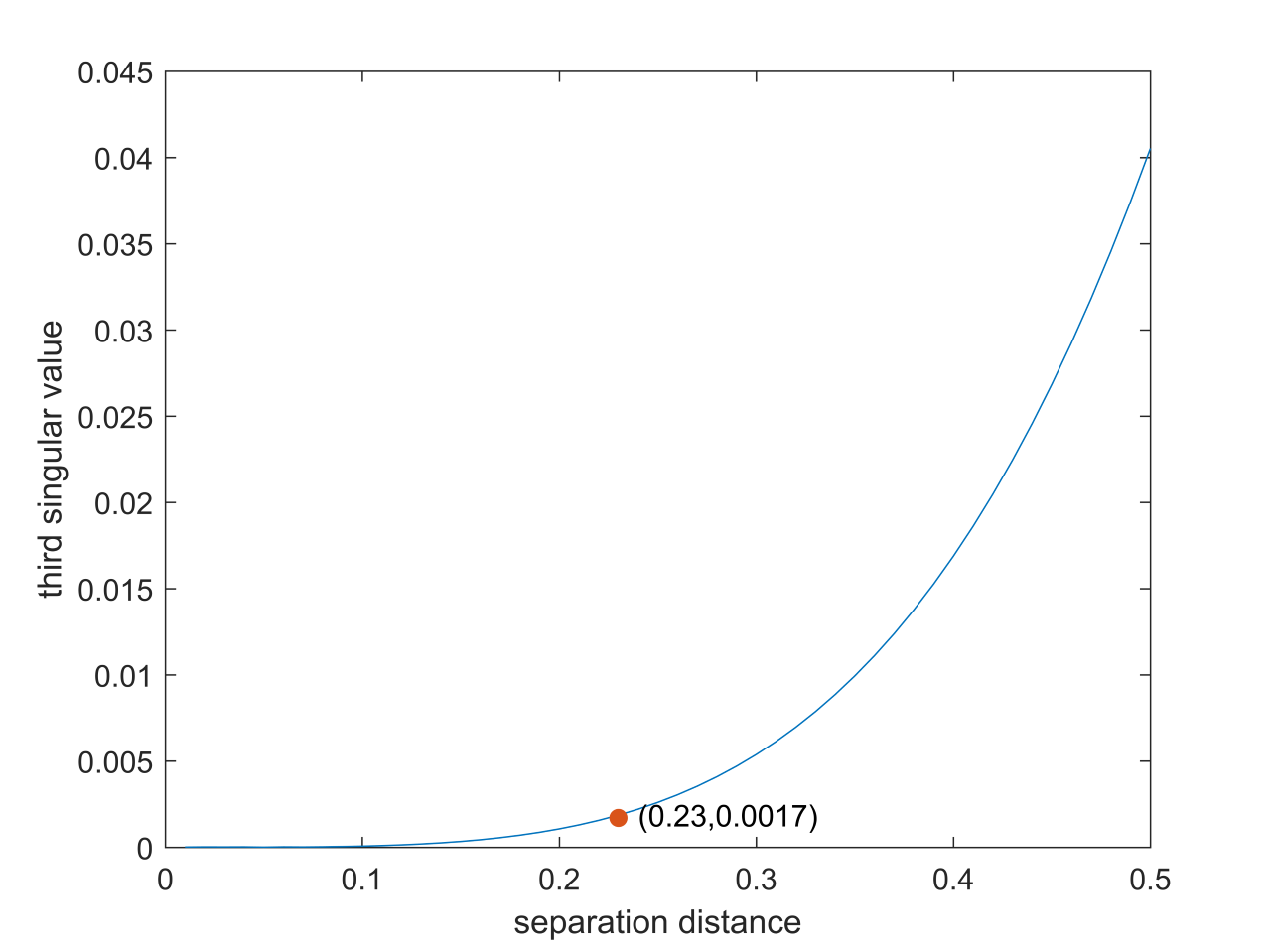}
		\caption{}
	\end{subfigure}
	\caption{a is the measurement in (\ref{equ:algorithmexample1equ1}). b is the behavior of the third singular value of the formulated Hankel matrix as the source separation distance increases.}
	\label{fig:MUSIC2image1}
\end{figure}

The measurements are shown in Figure \ref{fig:MUSIC2image1}:a. It is impossible to  discern visually that the source number is three. 
Note that the number of multipole coefficients that we can stably recover is 
$s^*=\min\Big\{l\in \mathbb N: \frac{d^l}{l!\sqrt{2l+1}}\leq \frac{\sigma}{2Me^d\sqrt{\pi}}\Big\}=10$. We consider the 
multipole matrix $\mathbf H(10)=(\mathbf h_0\cdots\mathbf h_9)$ and solve the following linear equations
\[\mathbf H(10) \theta = \mathbf Y.\]
We have $\theta=(3.0, -1.6171\times 10^{-10}, 0.32,-1.8209\times 10^{-9}, 0.0512, 1.8293 \times 10^{-7},0.0082,2.1694\times 10^{-5},0.0016,0.001)^T$. We use the first $7$ multipole coefficients to recover the source number. Consider the singular value decomposition of the data matrix
\[
X=\left(
\begin{array}{cccc}
\theta(1)&\sqrt{3}\theta(2)&2!\sqrt{5}\theta(3)&3!\sqrt{7}\theta(4)\\
\sqrt{3}\theta(2)&2!\sqrt{5}\theta(3)&3!\sqrt{7}\theta(4)&4!\sqrt{9}\theta(5)\\
2!\sqrt{5}\theta(3)&3!\sqrt{7}\theta(4)&4!\sqrt{9}\theta(5)&5!\sqrt{11}\theta(6)\\
3!\sqrt{7}\theta(4)&4!\sqrt{9}\theta(5)&5!\sqrt{11}\theta(6)&6!\sqrt{13}\theta(7)
\end{array}
\right)
 =\hat U\hat \Sigma \hat U^*.
 \]
We have $\hat \Sigma=\text{diag}(3.0343,0.3282,0.0169,1.2608\times 10^{-5})$ and  the threshold in Corollary \ref{MUSICnumberthm1} is $\frac{2\pi(s-1)!\sqrt{2s-1}\sigma}{\sqrt{6}\sigma_{\min}(s^*)}=0.0017$. Thus we can determine exactly the source number $3$ by the singular-value-thresholding algorithm. 

We now investigate the minimum separation distance required in this example beyond which one can recover the source number by this singular-value-thresholding algorithm. For the purpose, we draw a graph on the relation between $\hat \sigma_3$ and the separation distance between $y_j^*$'s (Figure \ref{fig:MUSIC2image1}:b). For simplicity, we consider $y_1^*=-y_3^*$ and $y_2^*=0$.  The threshold in Corollary \ref{MUSICnumberthm1} is 
\[\frac{2\pi(s-1)!\sqrt{2s-1}\sigma}{\sqrt{6}\sigma_{\min}(s^*)}=0.0017.\]
It is shown in Figure \ref{fig:MUSIC2image1}:b that, when the sources are separated beyond $0.23$, we are able determine the source number by the singular-value-thresholding algorithm. To show the efficiency of the algorithm, we calculate the upper bound for the computational resolution limit, which is equal to 
\begin{equation*}
\frac{4.7(1+d)}{\Omega}\Big(\frac{3}{\sigma_{\min}(s^*)}\frac{\sigma}{ m_{\min}^*}\Big)^{\frac{1}{2n-2}}=0.2083.
\end{equation*}
It is comparable to the separation distance required for our singular-value-thresholding algorithm. 

\vspace{0.3cm}
\noindent\textbf{Experiment 2:} We give an example of $4$ sources. 
We consider the recovery of the measure 
\[\mu^*=\delta_{y_1^*}-\delta_{y_2^*}+\delta_{y_3^*}+ \delta_{y_4^*}, \]
where $y_1^*=0.54, y_2^*=0.18,y_3^*=-0.18, y_4=-0.54$. Then the source number $n=4$, $m^*=4, m_{\min}^*=1$.  
We set $d=0.6, \sigma= 5\times 10^{-11}$ and $M=4$. We sample the image evenly in $[-100,100]$ with $101$ sample points (with $R=100, h=2$) as follows:
\begin{equation}\label{equ:algorithmexample2equ1}
\mathbf Y(x_t)=\mu^**f(x_t)+\mathbf W(x_t),\qquad t=1,\cdots,101,
\end{equation}
where $\mathbf W(x_t)$'s are uniformly distributed random numbers in $(0,\frac{\sigma}{\sqrt{2R}})$.
\begin{figure}
	\centering
	\begin{subfigure}[b]{0.47\textwidth}
		\centering
		\includegraphics[width=\textwidth]{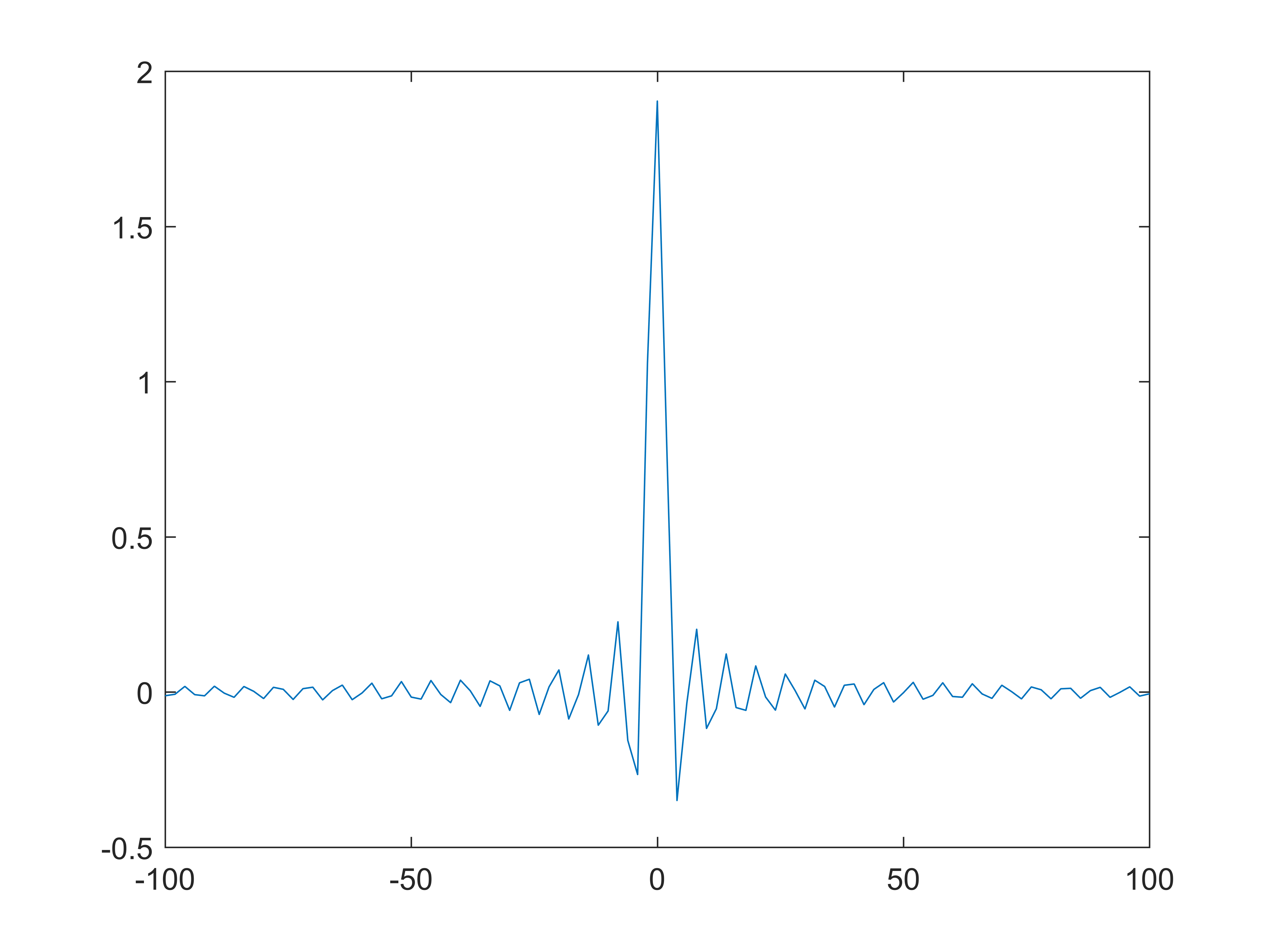}
		\caption{}
	\end{subfigure}
	\hfill
	\begin{subfigure}[b]{0.47\textwidth}
		\centering
		\includegraphics[width=\textwidth]{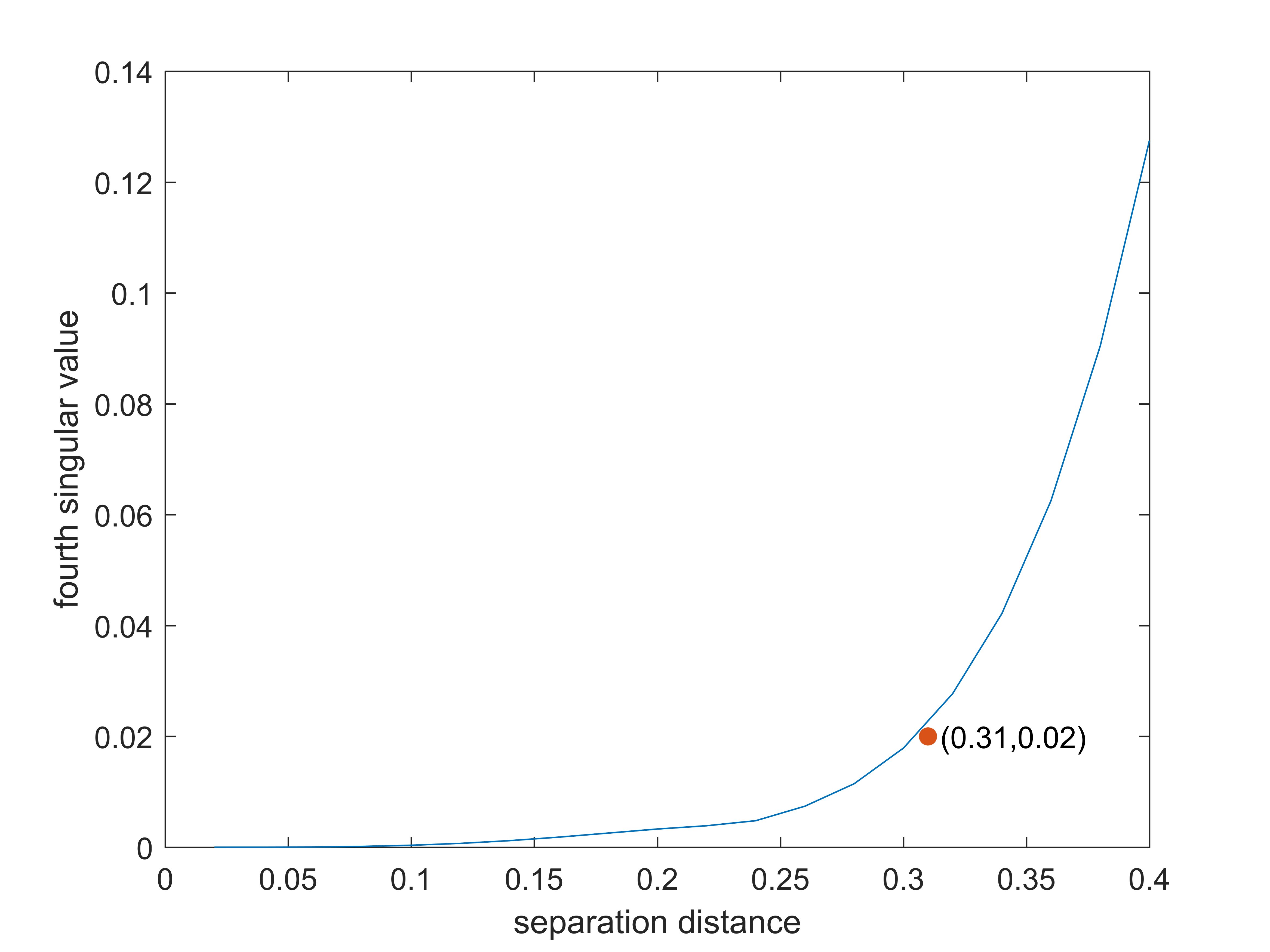}
		\caption{}
	\end{subfigure}
	\caption{a is the measurement in (\ref{equ:algorithmexample2equ1}). b is the behavior of the fourth singular value of the formulated Hankel matrix as the source separation distance increases.}
	\label{fig:MUSICsourceno4img1}
\end{figure}
The measurements are shown in Figure \ref{fig:MUSICsourceno4img1}:a. It is impossible to  discern visually that the source number is four. 
Note that the number of multipole coefficients that we can stably recover is 
$s^*=\min\Big\{l\in \mathbb N: \frac{d^l}{l!\sqrt{2l+1}}\leq \frac{\sigma}{2Me^d\sqrt{\pi}}\Big\}=12$. We consider the 
multipole matrix $\mathbf H(12)=(\mathbf h_0, \cdots, \mathbf h_{11})$ and solve the following linear equations
\[\mathbf H(12) \theta = \mathbf Y.\]
We have $\theta=(2.0000,
0.207846,
0.130406,
0.000735,
0.00235,
9.3393\times10^{-7},
-1.8379\times10^{-6},
-2.9499\times10^{-8}, -2.1253\times10^{-5}, -2.6784\times10^{-8}, -7.8629\times10^{-6}, -8.9346\times 10^{-9})^T$. We use the first $9$ multipole coefficients to recover the source number. Consider the singular value decomposition of the data matrix
\[
X=
\begin{pmatrix}
\theta(1)&\sqrt{3}\theta(2)&2!\sqrt{5}\theta(3)&3!\sqrt{7}\theta(4)&4!\sqrt{9}\theta(5)\\
\sqrt{3}\theta(2)&2!\sqrt{5}\theta(3)&3!\sqrt{7}\theta(4)&4!\sqrt{9}\theta(5)&5!\sqrt{11}\theta(6)\\
2!\sqrt{5}\theta(3)&3!\sqrt{7}\theta(4)&4!\sqrt{9}\theta(5)&5!\sqrt{11}\theta(6)&6!\sqrt{13}\theta(7)\\
3!\sqrt{7}\theta(4)&4!\sqrt{9}\theta(5)&5!\sqrt{11}\theta(6)&6!\sqrt{13}\theta(7)&7!\sqrt{15}\theta(9)\\
4!\sqrt{9}\theta(5)&5!\sqrt{11}\theta(6)&6!\sqrt{13}\theta(7)&7!\sqrt{15}\theta(9)&8!\sqrt{17}\theta(10)
\end{pmatrix}
=\hat U\hat \Sigma \hat U^*.
\]
We have $\hat \Sigma=\text{diag}(3.5385,2.2487,0.5701,0.0625, 0.00312)$ and  the threshold in Corollary \ref{MUSICnumberthm1} is $\frac{2\pi(s-1)!\sqrt{2s-1}\sigma}{\sqrt{6}\sigma_{\min}(s^*)}=0.02$. Thus we can determine exactly the source number $4$ by the singular-value-thresholding algorithm. 

We next investigate the minimum separation distance required to determine the exact source number. Figure \ref{fig:MUSICsourceno4img1}:b illustrates the relation between $\hat \sigma_4$ and the minimum separation distance of $y_1^*,y_2^*,y_3^*, y_4^*$. For simplicity, we consider $y_2^*=-y_3^*=\frac{d_{\min}^*}{2}$ and $y_1^*=-y_{4}^*=\frac{3}{2}d_{\min}^*$. The threshold in Corollary \ref{MUSICnumberthm1} is 
\[\frac{2\pi(s-1)!\sqrt{2s-1}\sigma}{\sqrt{6}\sigma_{\min}(s^*)}=0.02.\]
It is shown in Figure \ref{fig:MUSICsourceno4img1}:b that, when the sources are separated beyond $0.31$, we are able determine the source number by the singular-value-thresholding algorithm.

\section{Appendices}
\subsection{Appendix A: Proof of Lemma \ref{lem:multiproductlowerbound0} and Lemma \ref{lem:multiproductstability1}}
\textbf{Proof of Lemma \ref{lem:multiproductlowerbound0}}

\textbf{Step 1}. To prove the lemma, we only need to show that 
\begin{equation}\label{multiproducteq0}
\min_{d_1,\cdots,d_k \in \mathbb R} ||\eta_{k+1,k}(d_1^*, \cdots, d_{k+1}^*, d_1,\cdots,d_k)||_{\infty}\geq \xi(k)(d_{\min}^*)^k.
\end{equation}
It is easy to verify the result for $k=1$. For $k\geq 2$, we argue as follows. It is clear that the minimizer to (\ref{multiproducteq0}) exists (may not be unique) and $\min_{d_1,\cdots,d_k} ||\eta_{k+1, k}(d_1^*, \cdots, d_{k+1}^*, d_1,\cdots,d_k)||_{\infty}>0$. Let $(\hat d_1,\cdots,\hat d_k)$ be a minimizer with $\hat d_1 \leq \hat d_2\cdots \leq \hat d_{k}$.  We aim to estimate $||\eta_{k+1,k}(d_1^*, \cdots, d_{k+1}^*, \hat d_1,\cdots, \hat d_k)||_{\infty}$. For ease of notation, we define
\begin{equation}\label{equ:multiproducteq1}
	\eta_j(\hat d_1,\cdots, \hat d_k)=\Pi_{q=1}^{k}(d_j^*- \hat d_q).
\end{equation} 

\textbf{Step 2}. We prove that $\hat d_1,\cdots,\hat d_k \in [d_1^*,d_{k+1}^*]$. By contradiction, if $\hat d_p<d_1^*$ for some $p$, then for those $j$'s such that $\eta_{j}(\hat d_1,\cdots,\hat d_k)\neq 0$, we have
\begin{align*}
&|(d_j^*-\hat d_1)\cdots(d_j^*-\hat d_{p-1})(d_j^*-\hat d_{p})(d_j^*-\hat d_{p+1})\cdots(d_j^*-\hat d_{k})|\\
>&|(d_j^*-\hat d_1)\cdots(d_j^*-\hat d_{p-1})(d_j^*-d_{1}^*)(d_j^*-\hat d_{p+1})\cdots(d_j^*-\hat d_{k})|,
\end{align*}
i.e., 
\[
|\eta_j(\hat d_1,\cdots,\hat d_{p-1},\hat d_p,\hat d_{p+1},\cdots,\hat d_k)|>|\eta_j(\hat d_1,\cdots,\hat d_{p-1},d_1^*,\hat d_{p+1},\cdots,\hat d_k)|.
\]
While for those $j$'s such that $\eta_j(\hat d_1,\cdots,\hat d_k)=0$, we have $\eta_j(\hat d_1,\cdots,\hat d_{p-1},d_1^*,\hat d_{p+1},\hat d_k)=0$. This contradicts to the assumption that $(\hat d_1,\cdots,\hat d_k)$ is a minimizer and hence proves that 
$\hat d_p\geq d_1^*$ for all $p=1,\cdots,k$. Similarly, we can prove that $\hat d_p\leq d_{k+1}^*$ for all $p=1,\cdots,k$.  

\textbf{Step 3}. We show that each interval $[d_j^*, d_{j+1}^*)$ contains only one $\hat d_q$ for some $1\leq q \leq k$. 
We prove this by excluding the following three cases.

\textbf{Case 1:} There exist $j_0,p$ such that $d_p^* < \hat d_{j_0}< \hat d_{j_0+1}< d_{p+1}^*$. \\
Let $j_1$ be the integer such that 
\begin{equation}\label{multiproducteq2} 
||\eta_{k+1, k}(d_1^*, \cdots, d_{k+1}^*, \hat d_1,\cdots,\hat d_k)||_{\infty}=|\eta_{j_1}(\hat d_1,\cdots, \hat d_k)|>0,\end{equation}
where $\eta_j(\hat d_1,\cdots, \hat d_k)$ is defined in (\ref{equ:multiproducteq1}). If $1\leq j_1\leq  p$, then for $\Delta>0$ sufficiently small, we have 
\begin{equation}\label{multiproducteq3}
\begin{aligned}
&|\eta_{j_1}(\hat d_1,\cdots,\hat d_{j_0}-\Delta ,\hat d_{j_0+1}+\Delta ,\hat d_{j_0+2},\cdots,\hat d_{k})|-|\eta_{j_1}(\hat d_1,\cdots,\hat d_{k})|\\
=&\Big[(d_{j_1}^*-\hat d_{j_0}+\Delta)(d_{j_1}^*-\hat d_{j_0+1}-\Delta)-(d_{j_1}^*-\hat d_{j_0})(d_{j_1}^*-\hat d_{j_0+1})\Big]\Big|\Pi_{q=1,q\neq j_0,j_0+1}^k (d_{j_1}^*-\hat d_q)\Big| \\
=&\Big[\Delta(\hat d_{j_0}-\hat d_{j_0+1})-\Delta^2\Big]\Big|\Pi_{q=1,q\neq j_0,j_0+1}^k (d_{j_1}^*-\hat d_q)\Big|\\
<&0. \quad \Big(\text{using the inequality $\hat d_{j_0}<\hat d_{j_0+1}$ and (\ref{multiproducteq2})}\Big)
\end{aligned}
\end{equation}
On the other hand, if $p+1 \leq j_1\leq k+1$, then in the same fashion, for $\Delta$ sufficiently small, we have
\begin{equation*}|\eta_{j_1}(\hat d_1,\cdots,\hat d_{j_0}-\Delta ,\hat d_{j_0+1}+\Delta ,\hat d_{j_0+2},\cdots,\hat d_{k})|-|\eta_{j_1}(\hat d_1,\cdots,\hat d_{k})|<0.
\end{equation*}
Thus in all cases we have
\[
||\eta_{k+1, k}(d_1^*, \cdots, d_{k+1}^*, \hat d_1,\cdots,\hat d_{j_0}-\Delta, \hat d_{j_0+1}+\Delta,\cdots,\hat d_k)||_{\infty}<||\eta_{k+1,k}(d_1^*, \cdots, d_{k+1}^*, \hat d_1,\cdots,\hat d_k)||_{\infty}.
\]
This contradicts the assumption that $(\hat d_1,\cdots,\hat d_k)$ is a minimizer.\\
\textbf{Case 2:} There exist $j_0,p$ such that $\hat d_{j_0}=d_{p}^*<\hat d_{j_0+1}< d_{p+1}^*$.\\
We still denote $j_1$ the integer in $\{1,\cdots,k+1\}$ satisfying (\ref{multiproducteq2}). Since $\eta_p=0$, $j_1\neq p$.
Let $\Delta>0$ be sufficiently small. Similar to (\ref{multiproducteq3}), we can show that in both cases  $1\leq j_1\leq p-1$ and $p+1 \leq j_1 \leq k+1$,  
\begin{align*}
|\eta_{j_1}(\hat d_1,\cdots,\hat d_{j_0}-\Delta ,\hat d_{j_0+1}+\Delta ,\hat d_{j_0+2},\cdots,\hat d_{k})|-|\eta_{j_1}(\hat d_1,\cdots,\hat d_{k})|<0.
\end{align*}
Thus,
\[
||\eta_{k+1, k}(d_1^*, \cdots, d_{k+1}^*, \hat d_1,\cdots,\hat d_{j_0}-\Delta, \hat d_{j_0+1}+\Delta,\cdots,\hat d_k)||_{\infty}<||\eta_{k+1,k}(d_1^*, \cdots, d_{k+1}^*, \hat d_1,\cdots,\hat d_k)||_{\infty}.
\]
This contradicts the assumption that $(\hat d_1,\cdots,\hat d_k)$ is a minimizer.\\
\textbf{Case 3:} There exist $j_0,p$ such that $\hat d_{j_0}=\hat d_{j_0+1}=d_{p}^*$. \\
Denote $j_1$ the integer in $\{1,\cdots,k+1\}$ satisfying (\ref{multiproducteq2}). Since $\eta_p=0$, $j_1\neq p$. Let $\Delta>0$ be sufficiently small, we have for $1\leq j_1\leq p-1$, 
\begin{equation*}
\begin{aligned}
&|\eta_{j_1}(\hat d_1,\cdots,\hat d_{j_0}-\Delta ,\hat d_{j_0+1}+\Delta ,\hat d_{j_0+2},\cdots,\hat d_{k})|-|\eta_{j_1}(\hat d_1,\cdots,\hat d_{k})|\\
=&\Big[(d_{j_1}^*-\hat d_{j_0}+\Delta)(d_{j_1}^*-\hat d_{j_0+1}-\Delta)-(d_{j_1}^*-\hat d_{j_0})(d_{j_1}^*-\hat d_{j_0+1})\Big]\Big|\Pi_{q=1,q\neq j_0,j_0+1}^k (d_{j_1}^*-\hat d_q)\Big|\\
=&-\Delta^2|\Pi_{q=1,q\neq j_0,j_0+1}^k (d_{j_1}^*-\hat d_q)|<0.
\end{aligned}
\end{equation*}
On the other hand, for $p+1 \leq j_1 \leq k+1$, in the same fashion, we have
\begin{equation}|\eta_{j_1}(\hat d_1,\cdots,\hat d_{j_0}-\Delta ,\hat d_{j_0+1}+\Delta ,\hat d_{j_0+2},\cdots,\hat d_{k})|-|\eta_{j_1}(\hat d_1,\cdots,\hat d_{k})|<0.
\end{equation} 
Thus,
\[
||\eta_{k+1, k}(d_1^*, \cdots, d_{k+1}^*, \hat d_1,\cdots,\hat d_{j_0}-\Delta, \hat d_{j_0+1}+\Delta,\cdots,\hat d_k)||_{\infty}<||\eta_{k+1,k}(d_1^*, \cdots, d_{k+1}^*, \hat d_1,\cdots,\hat d_k)||_{\infty}.
\]
This contradicts the assumption that $(\hat d_1,\cdots,\hat d_k)$ is a minimizer. 

Finally, combining the results in the above three cases proves the claim.\\
\textbf{Step 4}. By the result in Step 3, we have for $j=1,\cdots,k,$
\begin{align}\label{multiproducteq1}d_{j}^*\leq\hat d_{j}<d_{j+1}^*.\end{align}
We then prove the lemma by considering the following two cases.\\
\textbf{Case 1:} For all $1\leq j \leq k$, $\hat d_j-d_{j}^*<\frac{d_{\min}^*}{2}$.\\ 
In this case, it is clearly that $|d_{k+1}^*-\hat d_{k}|>\frac{d_{\min}^*}{2}$. Thus
\begin{align*}
&|\Pi_{q=1}^{k}(d_{k+1}^*-\hat d_{q})|=|\Pi_{q=1}^{k-1}(d_{k+1}^*-\hat d_{q})||d_{k+1}^*-\hat d_{k}|\\
\geq &|\Pi_{q=1}^{k-1}(d_{k+1}^*-d_{q+1}^*)||d_{k+1}^*-\hat d_{k}| \quad \Big(\text{by (\ref{multiproducteq1})}\Big)\\
\geq &(k-1)!(d_{\min}^*)^{k-1}\frac{d_{\min}^*}{2}\geq \xi(k)(d_{\min}^*)^k.
\end{align*}
\textbf{Case 2:} There exist some $j$ such that $\hat d_j-d_{j}^*\geq \frac{d_{\min}^*}{2}$.\\
We let $j_0$ be the smallest integer such that $\hat d_{j_0}\geq d_{j_0}^*+\frac{d_{\min}^*}{2}$. Then
\begin{equation}\label{multiproducteq4}
|d_{j_0}^*-\hat d_{j_0-1}|\geq \frac{d_{\min}^*}{2}, \quad |d_{j_0}^*-\hat d_{j_0}|\geq \frac{d_{\min}^*}{2}.
\end{equation}
It follows that
\begin{align*}
&|\Pi_{q=1}^{k}(d_{j_0}^*-\hat d_{q})|=|\Pi_{q=1}^{j_0-2}(d_{j_0}^*-\hat d_{q})|\ |d_{j_0}^*-\hat d_{j_0-1}| |d_{j_0}^*-\hat d_{j_0}|\ |\Pi_{q=j_0+1}^{k}(d_{j_0}^*-\hat d_{q})|\\
\geq & |\Pi_{q=1}^{j_0-2}(d_{j_0}^*-d_{q+1}^*)||\Pi_{q=j_0+1}^{k}(d_{j_0}^*-d_{q}^*)||d_{j_0}^*-\hat d_{j_0-1}||d_{j_0}^*-\hat d_{j_0}| \quad (\text{by (\ref{multiproducteq1})})\\
\geq & (j_0-2)!(d_{\min}^*)^{j_0-2}(k-j_0)!(d_{\min}^*)^{k-j_0}|d_{j_0}^*-\hat d_{j_0-1}||d_{j_0}^*-\hat d_{j_0}|\\
\geq & (j_0-2)!(k-j_0)!(d_{\min}^*)^{k-2}(\frac{d_{\min}^*}{2})^2 \quad \Big(\text{by (\ref{multiproducteq4})}\Big)\\
=& \frac{(j_0-2)!(k-j_0)!}{4}(d_{\min}^*)^k.
\end{align*}
Minimizing $\frac{(j_0-2)!(k-j_0)!}{4}(d_{\min}^*)^k$ over $j_0=1,\cdots,k$ gives 
\begin{equation}
\min_{j_0=1,\cdots,k}\frac{(j_0-2)!(k-j_0)!}{4}(d_{\min}^*)^k\geq\left\{
\begin{array}{cc}
\frac{(\frac{k-1}{2})!(\frac{k-3}{2})!}{4}(d_{\min}^*)^k,& \text{$k$ is odd},\\
\frac{(\frac{k-2}{2}!)^2}{4}(d_{\min}^*)^k,& \text{$k$ is even}.
\end{array} 
\right.
\end{equation}
Therefore, in both cases, we have $||\eta_{k+1,k}(d_1^*, \cdots, d_{k+1}^*, \hat d_1,\cdots, \hat d_k)||_{\infty}\geq \xi(k) (d_{\min}^*)^k$. This completes the proof.

\vspace{0.8cm}
\textbf{Proof of Lemma \ref{lem:multiproductstability1}}

\textbf{Step 0.} We only prove the lemma for $k\geq 3$ and the case $k=2$ can be deduced in a similar manner. 

\textbf{Step 1.} We claim that 
for each $d_p, 1\leq p\leq k$, there exists one $d_j^*$ such that $|d_p-d_j^*|<\frac{d_{\min}^*}{2}$.
By contradiction, suppose there exists $p_0$ such that $|d_j^* - d_{p_0}|\geq \frac{d_{\min}^*}{2}$ for all $1\leq j\leq k$. Observe that 
\[
\eta_{k,k}(d_1^*,\cdots,d_k^*, d_1,\cdots, d_k)
=\text{diag}\left((d_1^*- d_{p_0}),\cdots,(d_{k}^*-d_{p_0})\right)\eta_{k,k-1}(d_1^*,\cdots,d_{k}^*, d_1,\cdots,d_{p_0-1}, d_{p_0+1},\cdots,d_k).
\]
We write $\eta_{k,k}=\eta_{k,k}(d_1^*,\cdots,d_k^*, d_1,\cdots, d_k)$ and $\eta_{k,k-1}=\eta_{k,k-1}(d_1^*,\cdots,d_{k}^*, d_1,\cdots,d_{p_0-1}, d_{p_0+1},\cdots,d_k)$. Using Lemma \ref{lem:multiproductlowerbound0}, we have
\[
||\eta_{k,k}||_{\infty}\geq \frac{d_{\min}^*}{2}||\eta_{k, k-1}||_{\infty} \geq  \frac{\xi(k-1)}{2}(d_{\min}^*)^{k}.  
\]	
By the formula of $\xi(k)$ in (\ref{equ:zetaxiformula1}), we can verify directly that $\frac{\xi(k-1)}{2}\geq \frac{\xi(k-2)}{4}$. 
Therefore, 
\[
||\eta_{k,k}||_{\infty}\geq \frac{\xi(k-2)}{4} (d_{\min}^*)^{k} 
\geq\epsilon,
\]
where we used (\ref{equ:satblemultiproductlemma1equ2}) in the last inequality above. This contradicts to (\ref{equ:satblemultiproductlemma1equ1}) and hence proves our claim.

\textbf{Step 2.} We claim that for each $d_j^*, 1\leq j\leq k$, there exists one and only one $d_p$ such that $|d_j^* -d_p|< \frac{d_{\min}^*}{2}$. It suffices to show that for each $d_j^*, 1\leq j\leq k$, there is only one $d_p$ such that $|d_j^*-d_p|<\frac{d_{\min}^*}{2}$. By contradiction, suppose there exist $p_1,p_2,$ and $j_0$ such that $|d_{j_0}^*-d_{p_1}^*|<\frac{d_{\min}^*}{2}, |d_{j_0}^*-d_{p_2}|<\frac{d_{\min}^*}{2}$. Then for all $j \neq j_0$, we have 
\begin{equation}\label{equ:satblemultiproductlemma1equ3}
	\Big|(d_j^*- d_{p_1}) (d_j^*-d_{p_2})\Big|\geq \frac{(d_{\min}^*)^2}{4}.
\end{equation}
Similar to the argument in Step 1, we separate the factors involving $d_{p_1}, d_{p_2}, d_{j_0}^*$ from $\eta_{k,k}$ and  consider
\[
\eta_{k-1,k-2}=	\eta_{k-1,k-2}(d_1^*,\cdots,d_{j_0-1}^*,d_{j_0+1}^*,\cdots,d_k^*, d_1,\cdots, d_{p_1-1}, d_{p_1+1}, \cdots, d_{p_2-1}, d_{p_2+1},\cdots, d_k).
\]
Note that the components of $\eta_{k-1,k-2}$ differ from those of $\eta_{k,k}$ only by the factors $|(d_j^*-d_{p_1})(d_{j}^*-d_{p_2})|$ for $j=1,\cdots,j_0-1,j_0+1,\cdots,k$. We can show that
$$	
||\eta_{k,k}||_{\infty}\geq \frac{(d_{\min}^*)^2}{4} ||\eta_{k-1,k-2}||_{\infty}. 
$$
Using Lemma \ref{lem:multiproductlowerbound0} and (\ref{equ:satblemultiproductlemma1equ2}), we further get 
\begin{align*}
	||\eta_{k,k}||_{\infty}\geq \frac{\xi(k-2)}{4}(d_{\min}^*)^{k}\geq \epsilon, 
\end{align*}
which contradicts to (\ref{equ:satblemultiproductlemma1equ1}). This contradiction proves our claim. 

\textbf{Step 3.} By the result in Step 2,  we can reorder $d_j$'s to get
\[
|d_j -d_j^*|< \frac{d_{\min}^*}{2}, \quad j=1,\cdots,k.
\]

\textbf{Step 4.} We prove (\ref{equ:satblemultiproductlemma1equ5}). By (\ref{equ:satblemultiproductlemma1equ4}), it is clear that
\begin{equation}\label{equ:satblemultiproductlemma1equ6}
	|d_{j}^* - d_{p}|\geq \left\{
	\begin{array}{cc}
		(j-p-\frac{1}{2})d_{\min}^* & p<j, \\
		(p-j-\frac{1}{2})d_{\min}^* & p>j.
	\end{array}
	\right.
\end{equation} 
We next show that 
\begin{equation} \label{eq-222}
	|(d_j^*-d_1)\cdots(d_j^*-d_k)|\geq |d_{j}^*-d_j|(\frac{d_{\min}^*}{2})^{k-1} (k-2)!, \quad j=1, 2, \cdots, k. 
\end{equation}  
Indeed, for $2\leq j\leq k-1$, we have
\begin{align*}
	&|(d_j^*-d_1)\cdots(d_j^*-d_k)|=|(d_j^*-d_j)| \Pi_{1\leq p\leq j-1}|(d_j^*-d_p)|\Pi_{j+1\leq p\leq k}|(d_j^*-d_p)|\\
	\geq &|(d_j^*-d_j)|\ \Big(\Pi_{1\leq p\leq j-1}\frac{2(j-p)-1}{2}\theta_{\min}\Big) \Big(\Pi_{j+1\leq p\leq k} \frac{2(p-j)-1}{2}\theta_{\min}\Big) \quad \Big(\text{by (\ref{equ:satblemultiproductlemma1equ6})}\Big)\\
	=& |d_{j}^*-d_j|(\frac{\theta_{\min}}{2})^{k-1}(2j-3)!!(2(k-j)-1)!!\quad  \,\, \\
	\geq& |d_{j}^*-d_j|(\frac{\theta_{\min}}{2})^{k-1} (k-2)!. \quad \quad \Big(\mbox{since $(2j-3)!!(2(k-j)-1)!!\geq (k-2)!$}\Big)
\end{align*}
Similarly, we can prove (\ref{eq-222}) for $j=1$ and $j=k$. Combining  (\ref{eq-222}) and (\ref{equ:satblemultiproductlemma1equ1}), we get 
\[
|d_{j}^*- d_j|(\frac{d_{\min}^*}{2})^{k-1}(k-2)!< \epsilon, \quad j=1, 2, \cdots, k,
\]
whence (\ref{equ:satblemultiproductlemma1equ5}) follows. This completes the proof of the lemma.

\subsection{Appendix B: Proof of Proposition \ref{vandermondegaussianelimiate1}}
We prove Proposition \ref{vandermondegaussianelimiate1} in this section. Denote
\begin{align*}
S_{ln}^j=\Big\{(\tau_1,\cdots,\tau_j): \tau_p\in \{l,l+1,\cdots,n\},p=1,\cdots,j, \tau_p\neq \tau_q  \mbox{\,\,for}\,\, p\neq q\Big\}.
\end{align*}
and  
$$
D_{ln}^j =\Big\{(\tau_1,\cdots,\tau_j): \tau_p\in \{l,l+1,\cdots,n\},p=1,\cdots,j \Big\}.
$$  
Let $M_{t}(d_1,\cdots,d_j)$ be the sum of all monomials of degree $t$ ($t\geq 0$) in $d_1,\cdots,d_j$. More precisely, 
\begin{equation}\label{vandermondegaussequ1}
	M_t(d_1,\cdots,d_j)=\sum_{(\tau_1,\cdots,\tau_j)\in D_{0t}^j,\ \sum_{p=1}^{j}\tau_{p}=t}(d_1)^{\tau_1}\cdots(d_j)^{\tau_j}, \quad 0\leq t\leq n.
\end{equation}

We first note that by a result in \cite{orucc2000explicit}, the Vandermonde matrix $V_n(n)$ defined in (\ref{eq-Vn}) can be reduced to the following form by applying a sequence of elementary column-addition matrices $G(1), G(2), \cdots, G(n-1)$,
\begin{equation}\label{vandermondegauss2}
	\begin{aligned}
	W=\left(\begin{array}{ccccc}
	w_{11}&0&0&\cdots&0\\
	w_{21}&w_{22}&0&\cdots&0\\
	w_{31}&w_{32}&w_{33}&\cdots&0\\
	\vdots&\vdots&\vdots&\ddots&\vdots\\
	w_{n1}&w_{n2}&w_{n3}&\cdots&w_{nn}\\
	w_{(n+1)1}&w_{(n+1)2}&w_{(n+1)3}&\cdots&w_{(n+1)n}
	\end{array}
	\right),
	\end{aligned}
\end{equation}
where 
\begin{equation}\label{vandermondeLUentry1}
w_{ij}=M_{i-j}(d_1,\cdots,d_j)\Pi_{p=1}^{j-1}(d_j-d_{p}), \quad i\geq j. 
\end{equation}
Therefore, after extracting the common factors, we get
\[
V(0)=
\begin{pmatrix}
1&0&0&\cdots&0\\
M_{1}(d_1)&1&0&\cdots&0\\
M_{2}(d_1)&M_{1}(d_1, d_2)&1&\cdots&0\\
\vdots&\vdots&\vdots&\ddots&\vdots\\
M_{n-1}(d_1)&M_{n-2}(d_1, d_2)&M_{n-3}(d_1, d_2, d_3)&\cdots&1\\
v_{(n+1)1}(0)&v_{(n+1)2}(0)&v_{(n+1)3}(0)&\cdots&v_{(n+1)n}(0)
\end{pmatrix}
=WD,
\]
where $D=\text{diag}(1,\frac{1}{(d_2-d_1)},\cdots,\frac{1}{\Pi_{p=1}^{n-1}(d_n-d_p)})$ and 
\begin{equation}\label{equ:vandermondeprocessequ0}
v_{(n+1)j}(0)= M_{n+1-j}(d_1, \cdots, d_j), \qquad 1\leq j \leq n.
\end{equation} 

We now perform further Gaussian eliminations to the matrix $V(0)$, using only elementary column-addition matrices. 

\textbf{Step 1:} Reduce the second to the last row in $V(0)$ by an elementary column operation  
to get
\begin{align*}
V(1)= 
\begin{pmatrix}
1&0&0&\cdots&0&0\\
M_{1}(d_1)&1&0&\cdots&0&0\\
M_{2}(d_1)&M_{1}(d_1, d_2)&1&\cdots&0&0\\
\vdots&\vdots&\vdots&\ddots&\vdots&\vdots\\
M_{n-2}(d_1)&M_{n-3}(d_1, d_2)&M_{n-4}(d_1,d_2,d_3)&\cdots&1&0\\
0&0&0&\cdots&0&1\\
v_{(n+1)1}(1)&v_{(n+1)2}(1)&v_{(n+1)3}(1)&\cdots&v_{(n+1)(n-1)}(1)&v_{(n+1)n}(1)
\end{pmatrix}
=V(0)Q(1).
\end{align*}
We have
\begin{align*}
v_{(n+1)j}(1) &= v_{(n+1)j}(0), \quad j = n, \\
v_{(n+1)j}(1) &= v_{(n+1)j}(0)-M_{n-j}(d_1,\cdots, d_j)v_{(n+1)n}(0),\quad  j \leq n-1.
\end{align*}
%

\textbf{Step 2:} Reduce the third to the last row in $V(1)$ by an elementary column operation to get $V(2)=V(1)Q(2)$. We have 
\begin{align*}
v_{(n+1)j}(2) &= v_{(n+1)j}(1), \quad j = n, n-1,\\
v_{(n+1)j}(2) &= v_{(n+1)j}(1)-M_{n-1-j}(d_1,\cdots, d_j) v_{(n+1)(n-1)}(1), \quad j \leq n-2.
\end{align*}
\textbf{Step t:} Reduec the $(t+1)$-th to the last row in $V(t-1)$ by an elementary column operation to get $V(t)=V(t-1)Q(t)$. We have 
\begin{align}
v_{(n+1)j}(t) &= v_{(n+1)j}(t-1), \quad j = n, n-1, \cdots, n-t+1,   \label{equ:vandermondeprocessequ1}\\
v_{(n+1)j}(t) &= v_{(n+1)j}(t-1)-M_{n-t+1-j}(d_1,\cdots, d_j) v_{(n+1)(n-t+1)}(t-1), \quad j \leq n-t. \label{equ:vandermondeprocessequ2}
\end{align}
$\cdots\cdots$\\
\textbf{Step n-1:} Reduce the second row in $V(n-2)$ by an elementary column operation to get
\begin{align*}
V(n-1)=\begin{pmatrix}
1&0&\cdots&0&0\\
0&1&\cdots&0&0\\
\vdots&\vdots&\ddots&\vdots&\vdots\\
0&0&\cdots&1&0\\
0&0&\cdots&0&1\\
v_{(n+1)1}(n-1)&v_{(n+1)2}(n-1)&\cdots&v_{(n+1)(n-1)}(n-1)&v_{(n+1)n}(n-1)
\end{pmatrix}
=V(n-2)Q(n-1).
\end{align*}
We have
\begin{align*}
v_{(n+1)j}(n-1) &= v_{(n+1)j}(n-2),  \quad j=n,n-1,\cdots,2 \\
v_{(n+1)j}(n-2) &= v_{(n+1)j}(n-2)-M_{1}(d_1)v_{(n+1)2}(n-2), \quad  j=1.
\end{align*}
Before we prove Proposition \ref{vandermondegaussianelimiate1}, we first present a useful equality: 
\begin{align}\label{equ:vandermondeprocessequ3}
	v_{(n+1)j}(n-1)=&M_{n-j+1}(d_1,\cdots,d_j)-v_{(n+1)n}(n-1)M_{n-j}(d_1,\cdots,d_j)-v_{(n+1)(n-1)}(n-1)M_{n-j-1}(d_1,\cdots,d_j) \nonumber \\
	&-\cdots-v_{(n+1)(j+1)}(n-1)M_{1}(d_1,\cdots,d_j), \quad j=1,\cdots,n.
\end{align}
To prove (\ref{equ:vandermondeprocessequ3}), we first observe that (\ref{equ:vandermondeprocessequ1}) implies
\[
v_{(n+1)q}(t) = v_{(n+1)q}(t-1), \quad   \text{for $n-1 \geq t \geq n-q+1$},
\]
and hence, 
\begin{equation}\label{equ:vandermondeprocessequ4}
v_{(n+1)q}(n-1) = 	v_{(n+1)q}(n-q), \quad q=1,\cdots, n.
\end{equation}
It follows that
\begin{align*}
&v_{(n+1)j}(n-1) = v_{(n+1)j}(n-j) \quad  (\text{by (\ref{equ:vandermondeprocessequ4}) for $q=j$})\\
=&v_{(n+1)j}(n-j-1)-M_{1}(d_1,\cdots, d_j) v_{(n+1)(j+1)}(n-j-1) \quad (\text{by (\ref{equ:vandermondeprocessequ2}) for $t=n-j$})\\
=&v_{(n+1)j}(n-j-1)-M_{1}(d_1,\cdots, d_j) v_{(n+1)(j+1)}(n-1) \quad (\text{by (\ref{equ:vandermondeprocessequ4}) for $q=j+1$})\\
=&v_{(n+1)j}(n-j-2)-M_{2}(d_1,\cdots,d_j)v_{(n+1)(j+2)}(n-j-2)-M_{1}(d_1,\cdots,d_j)v_{(n+1)(j+1)}(n-1)\\
&\ (\text{ by (\ref{equ:vandermondeprocessequ2}) for $t=n-j-1$})\\
=&v_{(n+1)j}(n-j-2)-M_{2}(d_1,\cdots,d_j)v_{(n+1)(j+2)}(n-1)-M_{1}(d_1,\cdots,d_j)v_{(n+1)(j+1)}(n-1)\ (\text{by (\ref{equ:vandermondeprocessequ4}) for $q=j+2$})\\
=&\cdots\cdots \quad  (\text{using (\ref{equ:vandermondeprocessequ2}) and (\ref{equ:vandermondeprocessequ4}) repeatedly})\\
=&M_{n-j+1}(d_1,\cdots,d_j)-M_{n-j}(d_1,\cdots,d_j)v_{(n+1)n}(n-1)-M_{n-j-1}(d_1,\cdots,d_j)v_{(n+1)(n-1)}(n-1)\\
-&\cdots-M_{1}(d_1,\cdots,d_j)v_{(n+1)(j+1)}(n-1),
\end{align*}
whence (\ref{equ:vandermondeprocessequ3}) follows. 

We are ready to prove Proposition \ref{vandermondegaussianelimiate1}. We need only to show that 
\begin{equation} \label{equ:vandermondeprocessequ5}
v_{(n+1)j}(n-1)=(-1)^{n-j}\sum_{(\tau_1,\cdots,\tau_{n+1-j})\in S_{1n}^{n+1-j}}d_{\tau_1}\cdots d_{\tau_{n+1-j}}, \quad j=1, \cdots, n.
\end{equation}
We first show (\ref{equ:vandermondeprocessequ5}) for $j=n$. Indeed, by (\ref{equ:vandermondeprocessequ4}) and (\ref{equ:vandermondeprocessequ0}), we have
$$
v_{(n+1)n}(n-1)=v_{(n+1)n}(0)=M_{1}(d_1, \cdots, d_{n})=\sum_{p=1}^{n}d_p.
$$ 
We next prove (\ref{equ:vandermondeprocessequ5}) for $j=n-1$. By (\ref{equ:vandermondeprocessequ3}), we have
\begin{align*}
&v_{(n+1)(n-1)}(n-1)=M_{2}(d_1,\cdots,d_{n-1})-v_{(n+1)n}(n-1)M_{1}(d_1,\cdots,d_{n-1})\\
=&M_{2}(d_1,\cdots,d_{n-1})-(\sum_{p=1}^{n}d_p)M_{1}(d_1,\cdots,d_{n-1})\quad \Big(\text{by (\ref{equ:vandermondeprocessequ5}) holds for $j=n$}\Big)\\
=&-\sum_{(\tau_1,\tau_2)\in S_{1n}^2}d_{\tau_1}d_{\tau_2}. \quad \Big(\text{by (\ref{vandermondedeterminantequ1})},\Big) 
\end{align*}
whence the case $j=n-1$ follows. 
Continuing the procedure (using (\ref{equ:vandermondeprocessequ3}) and (\ref{vandermondedeterminantequ1}) repeatedly), we can show that 
(\ref{equ:vandermondeprocessequ5}) holds for $j=n, n-1, \cdots, 1$ and this completes the proof of Proposition \ref{vandermondegaussianelimiate1}. What remains to show is the identity (\ref{vandermondedeterminantequ1}) which we prove below. 

\begin{lem}	\label{lem:vandermondeprocesslem1}
The following identity holds for $0\leq t\leq n-1$,
\begin{align}\label{vandermondedeterminantequ1}
&M_{t+1}(d_1,\cdots,d_{n-t})-\Big(\sum_{p=1}^{n}d_p\Big)M_{t}(d_1,\cdots,d_{n-t})-(-1)\Big(\sum_{(\tau_1,\tau_2)\in S_{1n}^2}d_{\tau_1}d_{\tau_2}\Big)M_{t-1}(d_1,\cdots,d_{n-t})\nonumber \\
&-\cdots-(-1)^{t-1}\Big(\sum_{(\tau_1,\cdots,\tau_{t})\in S_{1n}^t}d_{\tau_1}\cdots d_{\tau_t}\Big)M_{1}(d_1,\cdots,d_{n-t}).\nonumber \\
=&(-1)^t\sum_{(\tau_1,\cdots,\tau_{t+1})\in S_{1n}^{t+1}}d_{\tau_1}\cdots d_{\tau_{t+1}}.
\end{align}
\end{lem}
Proof: We prove by induction. It is clear (\ref{vandermondedeterminantequ1}) holds for $n=1$. Suppose that (\ref{vandermondedeterminantequ1}) holds for $n=k-1$, we need to prove for $n=k$. We first prove (\ref{vandermondedeterminantequ1}) for $0\leq t\leq k-2$ and then for $t=k-1$. Indeed, when $n=k$, for $0\leq t\leq k-2$, the left hand side of (\ref{vandermondedeterminantequ1}) can be decomposed as
\begin{align*}
&M_{t+1}(d_1,\cdots,d_{k-t})-\Big(\sum_{p=1}^{k}d_p\Big)M_{t}(d_1,\cdots,d_{k-t})-(-1)\Big(\sum_{(\tau_1,\tau_2)\in S_{1k}^2}d_{\tau_1}d_{\tau_2}\Big)M_{t-1}(d_1,\cdots,d_{k-t})\nonumber \\
&-\cdots-(-1)^{t-1}\Big(\sum_{(\tau_1,\cdots,\tau_{t})\in S_{1k}^t}d_{\tau_1}\cdots d_{\tau_t}\Big)M_{1}(d_1,\cdots,d_{k-t})
=:I+II,
\end{align*}
where 
\begin{align}
I=&\Big[M_{t+1}(d_1,\cdots,d_{k-t})-M_{t+1}(d_2,\cdots,d_{k-t})\Big]-\Big[\Big(\sum_{p=1}^{k}d_p\Big)M_{t}(d_1,\cdots,d_{k-t})-\Big(\sum_{p=2}^{k}d_p\Big)M_{t}(d_2,\cdots,d_{k-t})\Big]\nonumber \\
&-(-1)\Big[\Big(\sum_{(\tau_1,\tau_2)\in S_{1k}^2}d_{\tau_1}d_{\tau_2}\Big)M_{t-1}(d_1,\cdots,d_{k-t})-\Big(\sum_{(\tau_1,\tau_2)\in S_{2k}^2}d_{\tau_1}d_{\tau_2}\Big)M_{t-1}(d_2,\cdots,d_{k-t})\Big]-\cdots\nonumber \\
-&(-1)^{t-1}\Big[\Big(\sum_{(\tau_1,\cdots,\tau_{t})\in S_{1k}^t}d_{\tau_1}\cdots d_{\tau_t}\Big)M_{1}(d_1,\cdots,d_{k-t})-\Big(\sum_{(\tau_1,\cdots,\tau_{t})\in S_{2k}^t}d_{\tau_1}\cdots d_{\tau_t}\Big)M_{1}(d_2,\cdots,d_{k-t})\Big]\nonumber 
\end{align}  
and 
\begin{align*}
II=&M_{t+1}(d_2,\cdots,d_{k-t})-\Big(\sum_{p=2}^{k}d_p\Big)M_{t}(d_2,\cdots,d_{k-t})-(-1)\Big(\sum_{(\tau_1,\tau_2)\in S_{2k}^2}d_{\tau_1}d_{\tau_2}\Big)M_{t-1}(d_2,\cdots,d_{k-t})\\
-&\cdots-(-1)^{t-1}\Big(\sum_{(\tau_1,\cdots,\tau_{t})\in S_{2k}^t}d_{\tau_1}\cdots d_{\tau_t}\Big)M_{1}(d_2,\cdots,d_{k-t}).
\end{align*}
We observe that the term $II$ corresponds to the left hand side of (\ref{vandermondedeterminantequ1}) with $n=k-1$. 
By the assumption in the induction argument, we have 
\begin{equation}
	II=(-1)^{t}\sum_{(\tau_1,\cdots,\tau_{t+1})\in S_{2k}^{t+1}}d_{\tau_1}\cdots d_{\tau_{t+1}}, \qquad  0\leq t \leq k-2.
\end{equation}
For the first term $I$, a direct calculation shows that
\begin{align}
I=&\Big[d_1M_{t}(d_1,\cdots,d_{k-t})\Big]-\Big[d_1M_{t}(d_1,\cdots,d_{k-t})+d_1\Big(\sum_{p=2}^{k}d_p\Big)M_{t-1}(d_1,\cdots,d_{k-t})\Big]\nonumber \\
&-(-1)\Big[d_1\Big(\sum_{p=2}^{k}d_p\Big)M_{t-1}(d_1,\cdots,d_{k-t})+d_1\Big(\sum_{(\tau_1,\tau_2)\in S_{2k}^2}d_{\tau_1}d_{\tau_2}\Big)M_{t-2}(d_1,\cdots,d_{k-t})\Big]-\cdots\nonumber \\
&-(-1)^{t-1}\Big[d_1\Big(\sum_{(\tau_1,\cdots,\tau_{t-1})\in S_{2k}^{t-1}}d_{\tau_1}\cdots d_{\tau_{t-1}}\Big)M_{1}(d_1,\cdots,d_{k-t})+d_1\Big(\sum_{(\tau_1,\cdots,\tau_{t})\in S_{2k}^{t}}d_{\tau_1}\cdots d_{\tau_{t}}\Big)M_{0}(d_1,\cdots,d_{k-t})\Big]\nonumber \\
=&(-1)^td_1\sum_{(\tau_1,\cdots,\tau_{t})\in S_{2k}^{t}}d_{\tau_1}\cdots d_{\tau_{t}}. 
\end{align}  
Therefore, 
\begin{align*}I+II=&(-1)^td_1\sum_{(\tau_1,\cdots,\tau_{t})\in S_{2k}^{t}}d_{\tau_1}\cdots d_{\tau_{t}}+(-1)^{t}\sum_{(\tau_1,\cdots,\tau_{t+1})\in S_{2k}^{t+1}}d_{\tau_1}\cdots d_{\tau_{t+1}}\\
=&(-1)^t\sum_{(\tau_1,\cdots,\tau_{t+1})\in S_{1k}^{t+1}}d_{\tau_1}\cdots d_{\tau_{t+1}}.
\end{align*}
This proves (\ref{vandermondedeterminantequ1}) for $0\leq t\leq k-2$. Finally, we prove (\ref{vandermondedeterminantequ1}) for $n=k$ and $t=k-1$. Indeed, 
for $t=k-1$ and $n=k$, the left hand side of (\ref{vandermondedeterminantequ1}) reads as 
\begin{align*}
	&M_{k}(d_1)-\Big(\sum_{p=1}^{k}d_p\Big)M_{k-1}(d_1)-(-1)\Big(\sum_{(\tau_1,\tau_2)\in    S_{1k}^2}d_{\tau_1}d_{\tau_2}\Big)M_{k-2}(d_1)\\
&-\cdots-(-1)^{k-2}\Big(\sum_{(\tau_1,\cdots,\tau_{t})\in S_{1k}^{k-1}}d_{\tau_1}\cdots d_{\tau_t}\Big)M_{1}(d_1)=: III.
\end{align*}
Using the decomposition
\[
\sum_{(\tau_1,\cdots, \tau_q)\in S_{1k}^q}d_{\tau_1} \cdots d_{\tau_q} = d_1\Big(\sum_{(\tau_1,\cdots, \tau_{q-1})\in S_{2k}^{q-1}}d_{\tau_1} \cdots d_{\tau_{q-1}}\Big) + \sum_{(\tau_1,\cdots, \tau_q)\in S_{2k}^q} d_{\tau_1} \cdots d_{\tau_q}, \quad q=1, 2, \cdots, k-1,
\]
we have 
\begin{align*}
III =&M_{k}(d_1)-\Big[d_1+\sum_{p=2}^{k}d_p\Big]M_{k-1}(d_1)-\Big[d_1\Big(\sum_{p=2}^{k}d_p\Big)+\sum_{(\tau_1,\tau_2)\in S_{2k}^2}d_{\tau_1}d_{\tau_2}\Big]M_{k-2}(d_1)\\
&-\cdots-(-1)^{k-2}\Big(\sum_{(\tau_1,\cdots,\tau_{t})\in S_{1k}^{k-1}}d_{\tau_1}\cdots d_{\tau_t}\Big)M_{1}(d_1).
\end{align*}
Therefore, 

\begin{align*}
	III=&\Big[M_{k}(d_1)-d_1M_{k-1}(d_1)\Big]-\Big[\Big(\sum_{p=2}^{k}d_p\Big)M_{k-1}(d_1)-d_1\Big(\sum_{p=2}^kd_{p}\Big)M_{k-2}(d_1)\Big]\\
	&-(-1)\Big[\Big(\sum_{(\tau_1,\tau_2)\in S_{2k}^2}d_{\tau_1}d_{\tau_2}\Big)M_{k-2}(d_1)-d_1\Big(\sum_{(\tau_1,\tau_2)\in S_{2k}^2}d_{\tau_1}d_{\tau_2}\Big)M_{k-3}(d_1)\Big]\\
	&-\cdots-(-1)^{k-2}\Big(\sum_{(\tau_1,\cdots,\tau_{k-1})\in S_{2k}^{k-1}}d_{\tau_1}\cdots d_{\tau_{k-1}}\Big)M_{1}(d_1)\\
	=&(-1)^{k-1}d_1d_2\cdots d_k.  \quad \Big(\text{using the fact that the value in each square bracket is $0$}\Big)
\end{align*}
This completes the proof for the case $n=k$ and $t=k-1$. This complete the induction argument and proves the lemma.

\subsection{Appendix C: Some inequalities}
In this section, we present some inequalities that are used in this paper.
We first recall the following Stirling approximation of factorial 
\begin{equation}\label{strilingformula}	\sqrt{2\pi}n^{n+\frac{1}{2}}e^{-n}\leq n!\leq en^{n+\frac{1}{2}}e^{-n},
\end{equation}
which will be used frequently in subsequent derivation.

\begin{lem}\label{sufficienthm2estimate1}
For $n\geq 2$,
\[	
\frac{(2n-1)\sqrt{\pi}(n-1)^{2n-1}2^{2n-2}}{(2n-2)!\sqrt{4n-3}}\leq 1.11^{2n-2}  e^{2n-2}.
\]
\end{lem}
Proof: By (\ref{strilingformula}), for $n\geq 2$,
\begin{align*}
&\frac{(2n-1)\sqrt{\pi}(n-1)^{2n-1}2^{2n-2}}{(2n-2)!\sqrt{4n-3}}\leq \frac{(2n-1)\sqrt{\pi}(n-1)^{2n-1}2^{2n-2}}{\sqrt{2\pi}(2n-2)^{2n-2+\frac{1}{2}}e^{-(2n-2)}\sqrt{4n-3}}\\
\leq&\frac{(2n-1)\sqrt{n-1}}{2\sqrt{4n-3}}e^{2n-2}\leq 1.11^{2n-2} e^{2n-2}.
\end{align*}


\medskip
\begin{lem}\label{spaceapproxcalculate1}
For $k=1,2,\cdots$, the following estimate holds
\begin{equation*}
	\frac{\zeta(k+1)\xi(k)}{(2k)!\sqrt{4k+1}}\geq \frac{1.15}{2^{4k}k},
\end{equation*}
where $\zeta(k+1),\xi(k)$ are defined in (\ref{equ:zetaxiformula1}).
\end{lem}
Proof: For $k=1,2,3$, the inequality holds. For odd $k\geq 4$, by (\ref{strilingformula}), 
\begin{align*}
&\frac{\zeta(k+1)\xi(k)}{(2k)!\sqrt{4k+1}}=\frac{(\frac{k+1}{2})!(\frac{k-1}{2})!(\frac{k-1}{2})!(\frac{k-3}{2})!}{4(2k)!\sqrt{4k+1}}\geq \frac{4\pi^2(\frac{k+1}{2})^{\frac{k+2}{2}}(\frac{k-1}{2})^{k}(\frac{k-3}{2})^{\frac{k-2}{2}}e^{-2k+2}}{4e(2k)^{2k+\frac{1}{2}}e^{-2k}\sqrt{4k+1}}\\
=&\frac{e\pi^2(\frac{1}{2})^{2k}}{2^{2k}\sqrt{2k}\sqrt{4k+1}}(\frac{k+1}{k})^{\frac{k+2}{2}}(\frac{k-1}{k})^{k}(\frac{k-3}{k})^{\frac{k-2}{2}}
\geq \frac{e\pi^2(\frac{1}{2})^{2k}}{2^{2k}\sqrt{2k}\sqrt{4k+1}}\frac{1}{e^2}\\
=&\frac{1}{2^{4k}k}\frac{\pi^2\sqrt{4k}}{e\sqrt{8}\sqrt{4k+1}} 
\geq \frac{1.15}{2^{4k}k}.\qquad \Big(\text{because $k\geq 4$}\Big)
\end{align*}
For even $k\geq 4$,  
\begin{align*}
&\frac{\zeta(k+1)\xi(k)}{(2k)!\sqrt{4k+1}}=\frac{(\frac{k}{2})^2(\frac{k-2}{2}!)^2!}{4(2k)!\sqrt{4k+1}}
\geq \frac{4\pi^2(\frac{k}{2})^{k+1}(\frac{k-2}{2})^{k-1}e^{-2k+2}}{4e(2k)^{2k+\frac{1}{2}}e^{-2k}\sqrt{4k+1}}\\
=&\frac{e\pi^2(\frac{1}{2})^{2k}}{2^{2k}\sqrt{2k}\sqrt{4k+1}}(\frac{k-2}{k})^{k-1}
\geq\frac{1}{2^{4k}k}\frac{e\pi^2\sqrt{k}}{8\sqrt{2}\sqrt{4k+1}}\geq\frac{1.15}{2^{4k}k}. \qquad \Big(\text{because $k\geq 4$}\Big)
\end{align*} 

\medskip
\begin{lem}\label{stability2calculation1}
	For $n=2,3,\cdots$
	\begin{equation}
\sqrt[2n-1]{\frac{4(1+d)^{2n-1}\sqrt{4n-1}(2n-1)!}{\zeta(n)\lambda(n)}}\leq 6.24(1+d),
	\end{equation}
	where $\zeta(n)$ is defined in (\ref{equ:zetaxiformula1}) and $\lambda(n)$ defined in (\ref{equ:lambda1}).
\end{lem}
Proof: For $n=2,3,4,5$, the inequality holds. For even $n\geq 6$, by (\ref{strilingformula}), 
\begin{align*}
&\frac{4\sqrt{4n-1}(2n-1)!}{\zeta(n)\lambda(n)}= \frac{16\sqrt{4n-1}(2n-1)!}{(\frac{n}{2})!(\frac{n-2}{2})!(\frac{n-4}{2}!)^2}\leq \frac{16\sqrt{4n-1}e (2n-1)^{2n-\frac{1}{2}}e^{-(2n-1)}}{(2\pi)^2(\frac{n}{2})^{\frac{n+1}{2}}(\frac{n-2}{2})^{\frac{n-1}{2}}(\frac{n-4}{2})^{n-3}e^{-(2n-5)}}\\
=&\frac{\sqrt{4n-1}(2n-1)^{\frac{5}{2}}2^{2n-1}}{4\pi^2e^3(\frac{1}{2})^{2n-1}} \frac{(n-\frac{1}{2})^{\frac{n+1}{2}}(n-\frac{1}{2})^{\frac{n-1}{2}}(n-\frac{1}{2})^{n-3}}{n^{\frac{n+1}{2}}(n-2)^{\frac{n-1}{2}}(n-4)^{n-3}}\\
\leq& \frac{e\sqrt{4n-1}(2n-1)^{\frac{5}{2}}\ 4^{2n-1}}{4\pi^2} \leq (6.24)^{2n-1}.
\end{align*}

For odd $n\geq 7$, 
\begin{align*}
&\frac{4\sqrt{4n-1}(2n-1)!}{\zeta(n)\lambda(n)}= \frac{16\sqrt{4n-1}(2n-1)!}{(\frac{n-1}{2}!)^2(\frac{n-3}{2})!(\frac{n-5}{2})!}
\leq \frac{16\sqrt{4n-1}e (2n-1)^{2n-\frac{1}{2}}e^{-(2n-1)}}{(2\pi)^2(\frac{n-1}{2})^{n}(\frac{n-3}{2})^{\frac{n-2}{2}}(\frac{n-5}{2})^{\frac{n-4}{2}}e^{-(2n-5)}}\\
=&\frac{\sqrt{4n-1}(2n-1)^{\frac{5}{2}}2^{2n-1}}{4\pi^2e^3(\frac{1}{2})^{2n-1}} \frac{(n-\frac{1}{2})^{n}(n-\frac{1}{2})^{\frac{n-2}{2}}(n-\frac{1}{2})^{\frac{n-4}{2}}}{(n-1)^{n}(n-3)^{\frac{n-2}{2}}(n-5)^{\frac{n-4}{2}}}\\
\leq&\frac{e\sqrt{4n-1}(2n-1)^{\frac{5}{2}}\ 4^{2n-1}}{4\pi^2}\leq (6.24)^{2n-1}.
\end{align*}
This completes the proof of the Lemma. 

\medskip
\begin{lem}\label{stability2calculation2} For $n=2,3,4,\cdots$
\[
\frac{2^{n-1}(2n-1)!\sqrt{4n-1}}{\zeta(n)(n-2)!}\leq \frac{7.73\sqrt{4n-1}(2n-1)4^{2n-1}}{4e\pi^{\frac{3}{2}}},
\]
where $\zeta(n)$ is defined in (\ref{equ:zetaxiformula1}).
\end{lem}
Proof: 
For $n=2$, the inequality holds. For odd $n\geq 3$, by (\ref{strilingformula}), 
\begin{align*}
&\frac{2^{n-1}(2n-1)!\sqrt{4n-1}}{\zeta(n)(n-2)!}= \frac{2^{n-1}(2n-1)!\sqrt{4n-1}}{(\frac{n-1}{2}!)^2(n-2)!}\leq \frac{2^{n-1}\sqrt{4n-1}e(2n-1)^{2n-\frac{1}{2}}e^{-(2n-1)}}{(\sqrt{2\pi})^3(\frac{n-1}{2})^{n}(n-2)^{n-2+\frac{1}{2}}e^{-(2n-3)}}\\
\leq & \frac{\sqrt{4n-1}(2n-1)4^{2n-1}}{4e\pi^{\frac{3}{2}}}\frac{(n-\frac{1}{2})^{n}(n-\frac{1}{2})^{n-\frac{3}{2}}}{(n-1)^n (n-2)^{n-\frac{3}{2}}}\leq \frac{7.73\sqrt{4n-1}(2n-1)4^{2n-1}}{4e\pi^{\frac{3}{2}}}. \quad \Big(\text{since $n\geq 3$}\Big)
\end{align*}
For even $n\geq 4$,  
\begin{align*}
&\frac{2^{n-1}(2n-1)!\sqrt{4n-1}}{\zeta(n)(n-2)!}=\frac{2^{n-1}(2n-1)!\sqrt{4n-1}}{(\frac{n}{2})!(\frac{n-2}{2})!(n-2)!}\leq \frac{2^{n-1}\sqrt{4n-1}e(2n-1)^{2n-\frac{1}{2}}e^{-(2n-1)}}{(\sqrt{2\pi})^3(\frac{n}{2})^{\frac{n+1}{2}}(\frac{n-2}{2})^{\frac{n-1}{2}}(n-2)^{n-2+\frac{1}{2}}e^{-(2n-3)}}\\
\leq & \frac{\sqrt{4n-1}(2n-1)4^{2n-1}}{4e\pi^{\frac{3}{2}}}\frac{(n-\frac{1}{2})^{\frac{n+1}{2}}(n-\frac{1}{2})^{\frac{n-1}{2}}(n-\frac{1}{2})^{n-\frac{3}{2}}}{n^{\frac{n+1}{2}}(n-2)^{\frac{n-1}{2}} (n-2)^{n-\frac{3}{2}}}\leq  \frac{7.73\sqrt{4n-1}(2n-1)4^{2n-1}}{4e \pi^{\frac{3}{2}}}.
\end{align*}

\begin{lem}\label{equ:sigmaminsestimate1}
For integer $s\geq 2$,
\[
\sqrt{2}\zeta(s)(\frac{1}{s})^{s} \geq \frac{8\sqrt{2}\pi }{27}\frac{1}{(2e)^{s-1}} .
\]	
\end{lem}
Proof: When $s=2$, by (\ref{equ:zetaxiformula1}), the Lemma holds. For odd $s\geq 3$, by (\ref{equ:zetaxiformula1}), $\zeta(s)= (\frac{s-1}{2}!)^2$. By (\ref{strilingformula}), 
\begin{align*}
&\sqrt{2}\zeta(s)(\frac{1}{s})^{s}=\sqrt{2}(\frac{s-1}{2}!)^2(\frac{1}{s})^{s}\geq \sqrt{2}\Big(\sqrt{2\pi}(\frac{s-1}{2})^{\frac{s}{2}}e^{-\frac{s-1}{2}}\Big)^2(\frac{1}{s})^{s}\\
\geq& \frac{\sqrt{2}\pi}{2^{s-1}e^{s-1}}(\frac{s-1}{s})^s \geq \frac{8\sqrt{2}\pi}{ 2^{s-1}27e^{s-1}}= \frac{8\sqrt{2}\pi}{27}\frac{1}{(2e)^{s-1}}.
\end{align*}
For even $s\geq 4$, by (\ref{equ:zetaxiformula1}), $\zeta(s)=(\frac{s}{2})!(\frac{s-2}{2})!$. By (\ref{strilingformula}), 
\begin{align*}
&\sqrt{2}\zeta(s)(\frac{1}{s})^{s}=\sqrt{2}(\frac{s}{2})!(\frac{s-2}{2})!(\frac{1}{s})^{s} \geq 2\sqrt{2}\pi(\frac{s}{2})^{\frac{s+1}{2}}(\frac{s-2}{2})^{\frac{s-1}{2}}e^{-\frac{s}{2}}e^{-\frac{s-2}{2}}(\frac{1}{s})^{s} \\
\geq&  \frac{\sqrt{2}\pi}{2^{s-1}e^{s-1}}(\frac{s-2}{s})^{\frac{s-1}{2}}\geq \frac{8\sqrt{2}\pi }{27}\frac{1}{(2e)^{s-1}}.
\end{align*}

\bibliography{references}

\end{document}